\documentclass[10pt]{article}
\pdfoutput=1
\usepackage[utf8]{inputenc}
\usepackage{cite}
\usepackage{amsmath,amssymb,amsbsy,amstext,amsthm,simplewick,amsfonts}
\usepackage{graphicx}
\usepackage{subcaption}
\usepackage{lscape}
\usepackage{cancel}
\usepackage{wrapfig}
\usepackage{upgreek}
\usepackage{bm} 
\usepackage{framed}
\usepackage{bbm}
\usepackage{textcomp}
\usepackage{tikz}
\usepackage{pifont}
\usetikzlibrary{matrix,shapes,fit,tikzmark,calc}
\usepackage{adjustbox}
\usepackage{makecell}
\usepackage{tcolorbox}
\usepackage{physics}
\usepackage{empheq}
\usepackage[normalem]{ulem}
\usepackage{enumitem}
\usepackage{braket} 
\usepackage{array}
\usepackage{dsfont}
\usepackage{ulem}
\usepackage{titlesec}
\titleformat{\section}{\normalfont\fontsize{12}{16}\bfseries}{\thesection}{1em}{}

\numberwithin{equation}{section}

\def\be{\begin{equation}}
\def\ee{\end{equation}}

\def\e{\epsilon}

\def\ba{\begin{eqnarray}}
\def\ea{\end{eqnarray}}

\def\bfx{\textbf{x}}

\def\bfk{\textbf{k}}

\def\bfx{\textbf{x}}
\def\bfq{\textbf{q}}

\newcommand{\Psioff}{\Psi^{\text{off-shell}}}

\definecolor{blue3}{RGB}{31,119,180}
\definecolor{red3}{RGB}{214,39,40}
\definecolor{orange3}{RGB}{255,127,14}
\definecolor{green3}{RGB}{44,160,44}

\usepackage{colortbl}
\definecolor{lightgreen}{cmyk}{0.2, 0, 0.2, 0.2}
\definecolor{lightgray}{cmyk}{0.1,0.2,0,0.1}
\definecolor{lightgray2}{cmyk}{0.1,0.1,0,0.1}

\usepackage{hyperref}
\hypersetup{colorlinks=true,linkcolor=teal,citecolor=orange3,urlcolor=green3,pdfencoding=auto}

\makeatletter
\newlength{\apb@width}
\newcommand{\autoparbox}[2][c]{\settowidth{\apb@width}{#2}\parbox[#1]{\apb@width}{#2}}

\makeatother


\def\bfp{\textbf{p}}

\def\beq{\begin{equation}}
\def\eeq{\end{equation}}

\newcommand{\psit}{\psi^{\text{tree}}}
\newcommand{\psil}[1]{\psi^{#1\text{-loop}}}
\newcommand{\Il}[1]{\mathcal{I}^{#1\text{-loop}}}
\newcommand{\dt}{\tilde \delta_{D}^{(3)}}
\newcommand{\phix}{\Phi_\text{ex}}
\newcommand{\bco}{\bar X^{\text{out}}}

\allowdisplaybreaks[1]
\begin{document}


\begin{titlepage}
\setcounter{page}{1} \baselineskip=15.5pt

\thispagestyle{empty}

\renewcommand*{\thefootnote}{\fnsymbol{footnote}}

\begin{center}

{\fontsize
{20}{20} \bf The Analytic Wavefunction } \\
\end{center}

\vskip 18pt
\begin{center}
\noindent
{\fontsize{12}{18}\selectfont Santiago Agüí Salcedo\footnote{\tt  santiago.agui@protonmail.com}, Mang Hei Gordon Lee\footnote{\tt  mhgl2@cam.ac.uk}, Scott Melville\footnote{\tt  scott.melville@damtp.cam.ac.uk} and Enrico Pajer\footnote{\tt enrico.pajer@gmail.com}}
\end{center}

\begin{center}
\vskip 8pt
\textit{Department of Applied Mathematics and Theoretical Physics, University of Cambridge, Wilberforce Road, Cambridge, CB3 0WA, UK} 
\end{center}


\vspace{1.4cm}

\noindent The wavefunction in quantum field theory is an invaluable tool for tackling a variety of problems, including probing the interior of Minkowski spacetime and modelling boundary observables in de Sitter spacetime. 
Here we study the analytic structure of wavefunction coefficients in Minkowski as a function of their kinematics.
We introduce an \emph{off-shell} wavefunction in terms of amputated time-ordered correlation functions and show that it is analytic in the complex energy plane except for possible singularities on the negative real axis. 
These singularities are determined to all loop orders by a simple energy-conservation condition. 
We confirm this picture by developing a Landau analysis of wavefunction loop integrals and corroborate our findings with several explicit calculations in scalar field theories. 
This analytic structure allows us to derive new UV/IR sum rules for the wavefunction that fix the coefficients in its low-energy expansion in terms of integrals of discontinuities in the corresponding UV-completion. 
In contrast to the analogous sum rules for scattering amplitudes, the wavefunction sum rules can also constrain total-derivative interactions. 
We explicitly verify these new relations at one-loop order in simple UV models of a light and a heavy scalar. 
Our results, which apply to both Lorentz invariant and boost-breaking theories, pave the way towards deriving wavefunction positivity bounds in flat and cosmological spacetimes.


\end{titlepage}


\setcounter{tocdepth}{2}
{
\hypersetup{linkcolor=black}
\tableofcontents
}

\renewcommand*{\thefootnote}{\arabic{footnote}}
\setcounter{footnote}{0} 

\newpage

\section{Introduction}

Complexification often leads to simplification.
While physical observables are always real, their extension to the complex plane can reveal strikingly simple structures that encode a variety of physical features.
This is because quantities that vary smoothly along the real axis (such as most observables) can be extended to functions that are \emph{analytic} almost everywhere in the complex plane. 
Such functions feature a rigid structure and are completely specified by their singularities. Once these singularities are connected to physical properties of the system, the picture that emerges in the complex plane is both physically transparent and mathematically simple.

\paragraph{The analytic $S$-matrix.}
The importance of the complex plane is perhaps greatest in particle physics,
where the probability amplitude to scatter between different asymptotic states can be analytically continued to complex values of the particle momenta.
This analyticity, which is deeply rooted into causality,
was the cornerstone of the $S$-matrix programme of the 60's and 70's \cite{Eden:1966dnq,Martin:1969ina}, and now underpins most modern amplitude methods \cite{Elvang:2015rqa,Benincasa:2013faa,Cheung:2017pzi}. Analyticity has led to two particularly important results:
\begin{itemize}
	
	\item[(i)] a straightforward map between physical properties of states (e.g. their mass and spin) and the analytic structure of off-shell scattering amplitudes extended to the complex plane (e.g. their pole residues and branch cut discontinuities),
	
	\item[(ii)] a variety of \emph{sum rules} that connect low-energy observables to particular integrals over the underlying high-energy theory, providing a concrete connection between Effective Field Theory (EFT) Wilson coefficients and its ultraviolet (UV) completion.  
	
\end{itemize}
In practice, (i) is what allows us to empirically determine the properties of the particles that make up our Universe---for instance, the mass of the Higgs boson is measured by determining the location of a pole in the complex energy plane, via the Breit-Wigner formula \cite{ATLAS:2012yve, CMS:2012qbp, CMS:2013btf}. 
Increasingly, (ii) is the most promising way to search for new physics at the precision frontier---for instance by using an accurate determination of the low-energy coefficients in the Standard Model EFT \cite{Brivio:2017vri} to place constraints on physics beyond the Standard Model  \cite{Zhang:2018shp, Bi:2019phv, Remmen:2019cyz, Gu:2020ldn, Bonnefoy:2020yee, Remmen:2020vts, Zhang:2020jyn, Gu:2020thj, Fuks:2020ujk, Yamashita:2020gtt, Remmen:2020uze, Chala:2021wpj, Li:2022rag}. The sum rules that follow from analyticity can also be combined with fundamental properties such as unitarity and locality \cite{Pham:1985cr, Ananthanarayan:1994hf, Pennington:1994kc}, which has led to the recent explosion of different ``positivity bounds'' on low-energy EFTs \cite{Adams:2006sv, Nicolis:2009qm, Bellazzini:2016xrt,deRham:2017avq,deRham:2017zjm,Bellazzini:2020cot,Tolley:2020gtv,Caron-Huot:2020cmc, Sinha:2020win, Trott:2020ebl, Li:2021cjv, Arkani-Hamed:2020blm, Chiang:2021ziz, Chiang:2022ltp, Davighi:2021osh, Henriksson:2021ymi, Haring:2022sdp, Chen:2022nym}.

\paragraph{The analytic wavefunction.}
In this work, our goal is to develop an equivalent understanding of analyticity for the field-theoretic wavefunction\footnote{While the wavefunction is not a standard topic in QFT textbooks, see \cite{Jackiw:1987ar,Guven:1987bx,Benincasa:2022gtd} for several useful pedagogical references.}. 
Unlike scattering amplitudes, which are defined in terms of asymptotic states in the far past/future, the wavefunction is instead defined at a fixed time in the bulk. 
Because the choice of a time hypersurface obscures Lorentz invariance, the field-theoretic wavefunction has received relatively little attention. 
However it is a particularly useful object since it allows one to capture a wide variety of phenomena that cannot be probed by amplitudes, for instance finite-time effects. 
Since it does not rely on having well-defined particle eigenstates in the past/future, the wavefunction also generalises more readily to the curved spacetime backgrounds relevant for cosmology. 
In cosmological models, Lorentz invariance is always broken by the gravitational background and so using the wavefunction does not obscure any relevant symmetry. 
Already starting with \cite{Maldacena:2002vr,Maldacena:2011nz,Raju:2012zr,Raju:2012zs,Raju:2011mp,Mata:2012bx,Bzowski:2011ab,Bzowski:2012ih,Bzowski:2013sza,Ghosh:2014kba,Anninos:2014lwa,Kundu:2014gxa,Kundu:2015xta}, the wavefunction has therefore played a prominent role in theoretical cosmology, where it can describe the quantum fluctuations produced in the early inflationary Universe. 
This has driven much recent progress in importing amplitude techniques to the wavefunction \cite{Arkani-Hamed:2015bza,Shukla:2016bnu,Arkani-Hamed:2018kmz,Green:2020ebl,BBBB,Baumann:2019oyu,Arkani-Hamed:2017fdk,Albayrak:2018tam,Albayrak:2019yve,COT,Cespedes:2020xqq,Baumann:2020dch,sCOTt,MLT,Goodhew:2021oqg,Baumann:2021fxj,Bonifacio:2021azc,Goodhew:2022ayb,Cabass:2022rhr,Bonifacio:2022vwa} and to the calculation of cosmological correlators \cite{Bzowski:2019kwd,Baumgart:2019clc,Gorbenko:2019rza,Mirbabayi:2020vyt,Cohen:2020php,Green:2020whw,Sleight:2020obc,Sleight:2021iix,Albayrak:2020fyp,Albayrak:2020isk,Armstrong:2020woi,Meltzer:2021zin,Hogervorst:2021uvp,DiPietro:2021sjt,Sleight:2021plv,Premkumar:2021mlz,Bzowski:2022rlz,Pethybridge:2021rwf, Green:2022slj,Pimentel:2022fsc,Jazayeri:2022kjy,Armstrong:2022csc,Armstrong:2022jsa,Armstrong:2022vgl,Mirbabayi:2022gnl}.

Here, we develop the crucial ingredient of analyticity for the wavefunction.
Guided by the analogous results for amplitudes, we establish:
\begin{itemize}

\item[(o)] an analytic \emph{off-shell wavefunction} using a particular analytic continuation of the wavefunction coefficients to the complex energy plane, 

\item[(i)] relations between its singularities (pole/branch cuts) and the properties of the underlying fields exchanged in perturbation theory,

\item[(ii)] UV/IR sum rules which relate the coefficients in the finite-time EFT action to its underlying UV completion. 

\end{itemize}
We focus throughout on scalar fields interacting on a fixed Minkowski spacetime background, a setting in which we are able to completely characterise all possible non-analyticities that can arise in the wavefunction in perturbation theory. 
This lays down a framework for future applications of the wavefunction, e.g. to cosmological perturbations on quasi-de Sitter spacetimes.

\paragraph{Causality and analyticity.}
Analyticity in the energy domain is closely connected with causality in the time domain. 
If the application of a localised source only alters the future, then the Fourier transform of the response function is analytic in the lower-half of the complex energy/frequency plane (a mathematical result known as Titchmarsh's theorem). 
This is the analyticity used by Kramers and Kronig in their seminal work \cite{Kroning,Kramers:1924wxa}, which derived relations between the real and imaginary part of the refractive index as a consequence of causality.
Scattering amplitudes respect a stronger causality condition, namely \emph{microcausality}, which requires that all local operators commute at space-like separations. 
Together with crossing symmetry, this implies the analyticity of the amplitude in the complex $s$-plane in several cases\footnote{
Rigorous proofs of analyticity from causality assumptions are mostly restricted to the scattering of real scalar fields, and usually require additional assumptions such as the adiabatic hypothesis \cite{Gell-Mann:1954ttj, Bremermann:1958zz, Hepp_1964, Bros:1964iho}. 
}.   

The analyticity that we exploit in this work is the basic non-relativistic version of causality.
Concretely, the wavefunction $\Psi_{t=0} [ \phi ]$ that we consider is defined by evolving the vacuum in the far past until some later time $t=0$ using an interacting Hamiltonian $\mathcal{H} (t)$, and then projecting onto the field eigenstate at that time. 
Causality in this picture corresponds to the condition that $\Psi_{t=0}$ is insensitive to any interactions that take place at times $t > 0$, i.e. the wavefunction at time $t=0$ is determined only by its past. 
This means that $\mathcal{H} (t)$ could be replaced with $\mathcal{H} (t) \Theta (t)$, turning all interactions off after $t=0$, and the wavefunction $\Psi_{t=0}$ would be unchanged.
The wavefunction can therefore be viewed as the response to the interactions that turn off after a fixed time.
This is the intuitive reason why we find a simple analytic structure: the Fourier transform of such a causal response is generally analytic in (at least) half of the complex frequency plane (the lower half in our Fourier conventions).
Our sum rules for the wavefunction are therefore the analogue of the Kramers-Kronig relations. Pioneering ideas and results on the analytic structure of wavefunction coefficients can be found in \cite{Arkani-Hamed:2017fdk} and \cite{MLT,Baumann:2021fxj}.

\paragraph{Outline.}
Below we provide a summary of our conventions and a technical statement of our main results ((o), (i) and (ii) above).
The remainder of the paper then demonstrates these results as follows:
\begin{itemize}

\item[(o)] In Sec.~\ref{ssec:analytic_review} we review the recursion relations of \cite{Arkani-Hamed:2017fdk} for the Minkowski wavefunction, which allow us to express arbitrary Feynman-Witten diagrams in terms of a rational integrand. 
By extending this recursion relation to complex values of the energy, we define a new off-shell wavefunction in Sec.~\ref{offshell}. 

\item[(i)]  
Based on this rational integral representation, in Sec.~\ref{ssec:analytic_heuristic} we provide a simple physical criterion for determining the singular points of any given diagram, and we show in Sec.~\ref{ssec:analytic_examples} that this criterion indeed produces a complete list of all singular points in a number of examples. 
In Sec.~\ref{ssec:analytic_landau} we develop systematic Landau conditions for determining the singular points of any given wavefunction integral. 

\item[(ii)] In Sec.~\ref{sec:dispersion} we use analyticity to derive UV/IR sum rules, and apply them to a simple scalar EFT with interactions up to mass-dimension-8. 
In Sec.~\ref{sec:UV_examples}, we consider two different  explicit UV completions and confirm that they satisfy the sum rules. 

\end{itemize}
We then conclude in Sec.~\ref{sec:conclusions} with a discussion of future directions.
In Appendix~\ref{App:A} we describe in detail how to set up and perform loop integrals for the wavefunction coefficients.
In Appendix~\ref{A:rules} we briefly review the standard Feynman rules for computing Feynman-Witten diagrams for the wavefunction.
Throughout the main text we focus on the analytic structure of the off-shell wavefunction in perturbation theory, and in Appendix~\ref{App:C} describe how this generalises to a fully non-perturbative object.

\paragraph{Conventions.}
Our Fourier conventions are
\begin{align}
f(\bfx)&=\int\frac{d^{3}k}{(2\pi)^{3}}e^{i\bfk\bfx}f(\bfk)\equiv \int_{\bfk}e^{i\bfk\bfx}f(\bfk)\,, &f(\bfk)&=\int d^{3}x\,e^{-i\bfk\bfx}f(\bfx)\equiv \int_{\bfx}e^{-i\bfk\bfx}f(\bfx)\,.
\end{align}
To avoid cluttering our equations with factors of $(2\pi)^3$ from Fourier transform we employ the notation
\begin{align}
\dt(\bfk)\equiv (2\pi)^{3} \delta_D^{(3)}\left(  \bfk \right)\,.
\end{align}

In our convention the wavefunction is parameterized as
\begin{align}\label{psin}
\Psi[\phi;t]=\exp\left[   +\sum_{n}^{\infty}\frac{1}{n!} \int_{\bfk_{1},\dots\bfk_{n}}\,  \dt  \left( \sum_{a}^{n} \bfk_{a} \right) \psi_{n}(\{\bfk\};t)  \phi(\bfk_{1})\dots \phi(\bfk_{n})\right]\,,
\end{align}
where $\{\bfk\}$ is shorthand for the set of external momenta, $\{\bfk_1,\bfk_2,\dots,\bfk_n\}$. This is the same as for example in \cite{Maldacena:2002vr,Benincasa:2019vqr,Melville:2021lst}, but it is the opposite of that used in \cite{COT,MLT}, where $\Psi\sim e^{-\psi_n \phi^n}$ was chosen. Because of time-translation invariance, the wavefunction coefficients have a very simple time dependence:
\begin{align}
    \psi_n ( \{ \bfk \}; t ) = e^{i \sum_{a=1}^n\Omega_{k_a} t} \psi_n ( \{ \bfk \}; 0 )\,.
\end{align}
Therefore, we simply focus on $t=0$, without loss of generality, and omit specifying the time dependence.  

We work perturbatively and adopt a power-counting in which $\psi_n$ starts at $\mathcal{O} ( g_*^{n-2} )$ for some small expansion parameter $g_*$. 
Expanding a given $\psi_n$ in powers of this coupling corresponds to a diagrammatic expansion in the number of loops,
\begin{align}
    \psi_n  =  \psit_n + \psil{1}_n + \psil{2}_n + \mathcal{O} \left( g_*^{n+1} \right) \,,
\end{align}
where $\psil{L}_n \sim \mathcal{O} \left( g_*^{n -2 + L} \right)$. 
It will often be convenient to work with the loop integrand, which we write as, 
\begin{align}
    \psil{L}_n ( \{ \bfk \}  ) = \int_{\bfp_1,\dots,\bfp_L} \Il{L}_n ( \{ \bfk \} ,\{ \bfp \} ) \,,
\end{align}
Momenta of external legs are denoted by $\bfk_a$ (where $a$ runs over the number of external lines) and loop momenta are denoted by $\bfp_a$ (where $a$ runs over the number of loops). The momenta of internal lines are denoted by $\bfq_a$, and can always be fixed by momentum conservation in terms of the external momenta and the loop momenta. 
When discussing 4-point coefficients, we denote the total 3-momentum in each of the channels as 
\begin{align}
    \bfk_s &= \bfk_1 + \bfk_2\,, &\bfk_t &= \bfk_1 + \bfk_3\,, &\bfk_u &= \bfk_2 + \bfk_3\,.
\end{align}
We use $\Omega_{\bfk}$ to denote the ``on-shell" value of the energy when the spatial momentum is $\bfk$ (and $k= | \bfk|$ is its magnitude). For each field of mass $m$ in the theory, this is given by
\begin{align}
    \Omega_{k} = \sqrt{ k^2 + m^2 }\,.
\end{align}
Our main object of study will be ``off-shell" wavefunction coefficients (defined in Sec.~\ref{offshell}). These depend on additional variables $\omega_a$ representing the off-shell energy of each external leg, which is independent of the mass and momentum of the field. It is only when we want to evaluate a wavefunction coefficient on-shell that we impose $\omega_a \overset{!}{=}\Omega_{k_a}$. In the context of 4-point coefficients we sometimes use:
\begin{align}
    \omega_s &= \omega_1 + \omega_2\,, & \omega_t &= \omega_1 + \omega_3\,, & \omega_u  &= \omega_2 + \omega_3\,.
\end{align}
The total energy flowing into a diagram will be denoted by $\omega_T = \sum_{a} \omega_a$. 
Off-shell wavefunction coefficients will be denoted with the same symbol $\psi_n$ as their on-shell cousins, but depend on $n$ additional variables:
\begin{align}
  \text{Off-shell:} \qquad \psil{L}_n (\{\omega\}, \{ \bfk \}  ) = \int_{\bfp_1,\dots,\bfp_L} \Il{L}_n (\{\omega\}, \{ \bfk \} , \{\bfp \}) \,.
\end{align}

\subsection{Summary of main results}
For the convenience of the busy reader, we summarize here our main findings.

\paragraph{Introducing the off-shell wavefunction.}
The central object that we study in this work is a new off-shell extension of the wavefunction coefficients at $t=0$. 
They can be defined as the following half-interval Fourier transform, 
\begin{align}
   \psi_n \left( \{ \omega \} , \{ \bfk \} \right) 
   =
   \left[ \prod_{j=1}^n \int_{-\infty}^0 d t_j \, e^{ i \omega_j t_j}  \right]\;  G_{\bfk_1 ... \bfk_{n-1}}^{\rm amp., con.} ( t_1, ... , t_n )
\; ,
\end{align}
where $G_{\bfk_1 ... \bfk_{n-1}}^{\rm amp., con.} ( t_1, ... , t_n )$ is the amputated, connected part of the time-ordered Green's function (descussed in detail in App.~\ref{App:C}),
\begin{align}
 \langle \phi ( 0 ) = 0 |  \; T \;  \hat{\Phi}_{\bfk_1} ( t_1 ) ... \hat{\Phi}_{\bfk_n} ( t_n ) \;  | \Omega_{\rm in} \rangle 
 =
 G_{\bfk_1 ...  \bfk_{n-1} } (t_1, ... , t_{n} ) \; \tilde{\delta}_D^{(3)} \left( \sum_{a}^n\bfk_a \right) ,
\end{align}
where $| \Omega_{\rm in} \rangle$ is the vacuum in the far past and $| \phi (0) = 0 \rangle$ is the field eigenstate $\hat{\Phi}_{\bfk} (0) | \phi (0) = 0 \rangle = 0$ at time $t=0$.
These off-shell coefficients coincide with the usual wavefunction coefficients (the $\psi_n ( \{ \bfk \} )$ appearing in \eqref{psin}) in the limit $\omega \to \sqrt{k^2 + m^2}$, which we refer to as ``going on-shell''. 
In perturbation theory, the off-shell $\psi_n ( \{ \omega \} , \{ \bfk \} )$ can be represented as a series of Feynman-Witten diagrams in which $n$ external edges are each labelled by an energy and a spatial momentum, and each vertex is labelled by a bulk time which is integrated from $-\infty$ to $0$. 
Internal lines carry spatial momenta only, which are partially fixed by imposing momentum conservation at each vertex (any undetermined internal momenta due to loops are integrated over). 
The Feynman rules for evaluating each diagram are:
\begin{itemize}

\item An external edge with energy $\omega$ connected to a bulk vertex at time $t$ gives an off-shell bulk-boundary propagator, $K_\omega (t) = e^{i \omega t}$,

\item An internal edge carrying momentum $\bfk$ between vertices at times $t_1$ and $t_2$ gives a bulk-bulk propagator, namely the two-point $G_{\bfk} (t_1 ,t_2)$ associated with the field being exchanged, 

\item Each $n$-point vertex gives a factor of $i \delta^n S[\Phi;0]/\delta \Phi^n |_{\Phi = 0}$, where $S[\Phi;0]$ is the action from $t=-\infty$ to $t=0$. Note that each temporal (or spatial) derivative becomes a factor of $i \omega$ (or $i \bfk$) when acting on an external line.

\end{itemize} 
We give a number of explicit examples throughout. 

\paragraph{Signatures of new heavy physics.}
With these definitions, the analytic structure of any particular Feynman diagrams is completely determined by the properties of the virtual fields being exchanged. 
This is familiar in the context of scattering amplitudes, where for instance the $s$-channel $2 \to 2$ scattering amplitude at one-loop receives contributions from the following diagrams (where $\omega_s$ and $\bfk_s$ are the total ingoing energy and momentum), 
\begin{align}
	\mathcal{A}_{2 \to 2}^{(s)} 
	&= \;\; 
	&&\vcenter{\hbox{\includegraphics[width=0.2\textwidth]{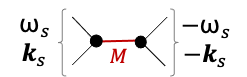}}}
	\;\; &&+ \;\;
	&&\vcenter{\hbox{\includegraphics[width=0.2\textwidth]{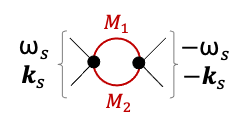}}} 
	\;\; &&+ \;\; &&\cdots  \nonumber \\ 
	&\sim &&\qquad\;\; \frac{1}{M^2- s} &&+ &&\;\; \sqrt{ (M_1 + M_2 )^2 - s} &&+ &&... 
\end{align}
and, as a result, develops poles and branch cuts at particular \emph{thresholds} which are determined by the masses of the heavy internal fields (in this case at $s= M^2$ and $s = (M_1+ M_2 )^2$, where $s= \omega_s^2 - k_s^2$ is the usual Mandelstam invariant).

Bringing this perspective to the wavefunction, we show that the location of similar thresholds in $\psi_n$ are tied to the masses of the heavy internal fields.
For instance, the one-loop four-point wavefunction coefficient receives contributions from the following diagrams, where $\omega_1$ and $\omega_2$ are the total energies flowing into each interaction vertex,  
\begin{align}  \label{eqn:intro_psi4s}
	\psi_4^{(s)} 
	&= \;\; 
	&&\vcenter{\hbox{\includegraphics[width=0.15\textwidth]{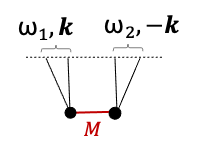}}}
	\;\; &&+ \;\;
	&&\qquad\;\; \vcenter{\hbox{\includegraphics[width=0.15\textwidth]{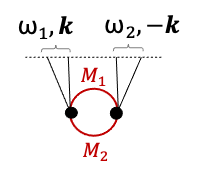}}} 
 	\;\; &&+ \;\; &&\cdots  \nonumber \\ 
	&\sim && \frac{1}{  \sqrt{M^2 + k^2 }  + \omega_1 } &&+ && \sqrt{ \sqrt{  (M_1 + M_2 )^2 + k^2 } + \omega_1 } &&+ &&... 
\end{align}
and, consequently, develops analogous poles and branch cuts at thresholds determined by the heavy internal fields (in this case at $\omega_1 = - \sqrt{k^2 + M^2}$ and $\omega_1 = - \sqrt{k^2 + (M_1 + M_2)^2}$). 

However, since energy is no longer conserved in the wavefunction, there can be additional non-analyticities that have no amplitude counterpart. For instance, the one-loop diagram,
\begin{align}
   \psi_4
	&\supset \;\; \qquad	\vcenter{\hbox{\includegraphics[width=0.15\textwidth]{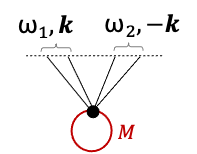}}}  \nonumber \\ 
&\sim \qquad  \sqrt{ 2M + \omega_1 + \omega_2 } 
 \end{align}
produces a branch point at $\omega_1 = - \omega_2 - 2M$ which depends on the value of the other particle's energies.
These additional non-analyticities necessarily do not appear in the energy-conserving limit $\omega_T \to 0$ (in this example when $\omega_1 + \omega_2 \to 0$ is fixed), since in that limit the wavefunction coefficient $\omega_T \psi_n$ coincides with the $n$-particle scattering amplitude amplitude. 

We show that a given diagram will develop singular points whenever the external energies are such that one or more of the bulk vertices become energy-conserving. 
This energy-conservation condition can be applied at both tree and at loop level, and we have verified in a variety of examples that the resulting list of singular points is exhaustive. 
In \eqref{eqn:intro_psi4s} above, for instance, the thresholds in $\omega_1$ can be recognised as the values at which the left-most vertex can carry zero energy (in the case of the loop diagram, a branch cut develops because for every $\omega_1 < -\sqrt{k^2 + ( M_1 + M_2)^2}$ there is at least one value of the loop momentum at which the left-most vertex carries zero energy).

\begin{figure}
    \centering
     \begin{subfigure}[b]{0.47\textwidth}
         \centering
         \includegraphics[width=\textwidth]{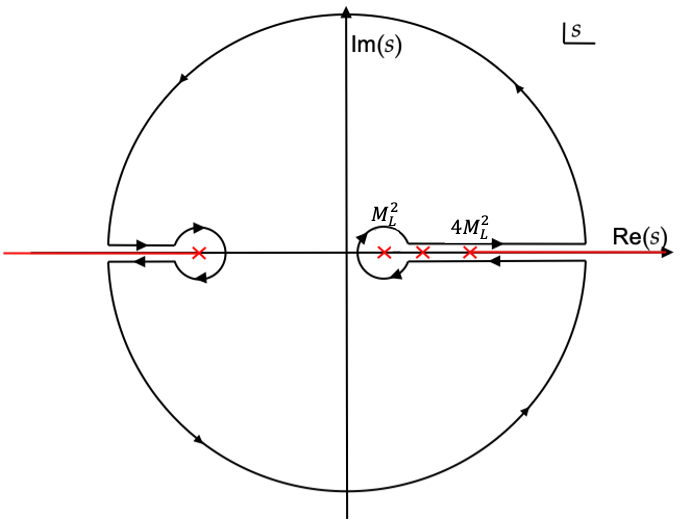}
         \caption{$A(s,t)$ in complex $s$-plane for a fixed $t<0$. The pole at $M^{2}_{L}$ corresponds to the lightest tree-level exchange in the $s$-channel. }
     \end{subfigure}
     \hfill
     \begin{subfigure}[b]{0.47\textwidth}
         \centering
         \includegraphics[width=\textwidth]{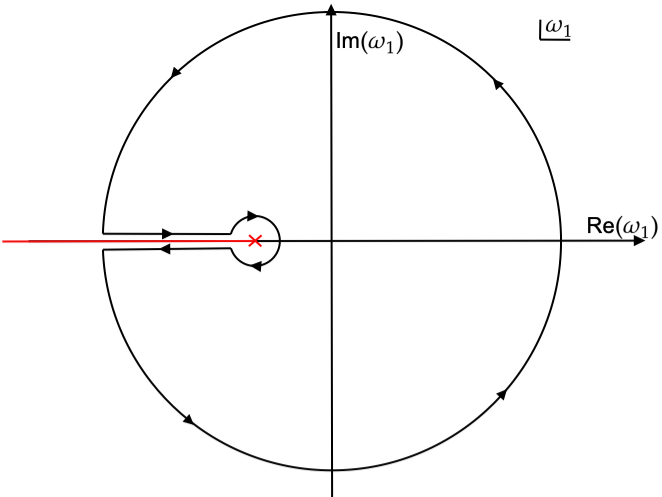}
         \caption{$\psi_{n}( \{ \omega \} , \{ \bfk \} )$ in the complex $\omega_1$-plane for fixed $\omega_{a\neq1}>0$ and real $\bfk_a$. All tree-level poles/loop-level branch cuts lie on the negative real axis.}
     \end{subfigure}
        \caption{Analytic structure of Minkowski scattering amplitude (left) and wavefunction coefficients (right).}
\end{figure}

\paragraph{New UV/IR sum rules.}
At low centre-of-mass energies, the $2 \to 2$ scattering amplitude between four scalar particles can be expanded\footnote{
In theories without a mass gap, there can also be non-analytic contributions to \eqref{eqn:A_EFT_exp} generated by loops of massless particles. 
In this case, we must additionally assume a weak coupling which allows these loop diagrams to be computed perturbatively (to some fixed order in the coupling) and subtracted from \eqref{eqn:A_disp_intro}.  
}, 
\begin{align}
	\mathcal{A} (s, t) = \sum_{n=0}^{\infty} s^n \; c_{n} (t)   ,
	\label{eqn:A_EFT_exp}
\end{align}
where the EFT coefficients $c_{n}$ can be straightforwardly related to the couplings which appear in the EFT action. 
Exploiting the analytic structure of the amplitude in the complex $s$ plane at fixed $t$ leads to UV/IR sum rules,
\begin{align}
	c_{n} (t) =  \int_{M^2_L}^{\infty} \frac{ds}{2 \pi i}  \frac{ \text{disc}_s \, \mathcal{A} (s,t ) }{ s^{n+1} }
	+ \int_{M^2_L}^{\infty} \frac{du}{2 \pi i}  \frac{  \text{disc}_u \, \mathcal{A} ( s , t ) }{ s^{n+1} } + \mathcal{C}_{\infty} \left[ \mathcal{A} \right],
	\label{eqn:A_disp_intro}
\end{align}
where $M_L$ represents the mass of the lightest particle which couples to the external fields, $s(u) = 4m^2 - t - u$ in the second term (where $m$ is the mass of the external fields), and $\text{disc}_z f(z) = \lim_{\epsilon \to 0} \left[ f(z + i \epsilon) - f (z - i \epsilon ) \right]$ is the usual discontinuity of a complex function. $\mathcal{C}_{\infty}$ is a particular contour integral at asymptotically large values of $s$. 
See \cite{Eden:1966dnq, Martin:1969ina} for a standard textbook presentation of these ideas, or any of the more recent literature on the $S$-matrix bootstrap \cite{Paulos:2016but, Paulos:2017fhb, Correia:2020xtr}. 

The off-shell wavefunction coefficients can similarly be expanded at low energies,
\begin{align}
 \omega_T \psi_n ( \{ \omega \} , \{ \bfk \} ) = \sum_{n=0}^{\infty} \omega_1^n  \; \alpha_n ( \{ \omega_{a \neq 1} \} , \{ \bfk \} )  ,
\end{align}
where the factor of $\omega_T = \sum_{a=1}^n \omega_a$ is included for convenience. The EFT coefficients $\alpha_n$ can be related to the couplings which appear in the finite-time action $S[ \Phi ; 0 ]$.
Utilising their analytic structure, we are able to derive an analogous UV/IR sum rule,
\begin{align}
\alpha_n ( \{ \omega_{a \neq 1} \} , \{ \bfk \} ) = \int_{-\infty}^{-M_L} \frac{d \omega_1}{2 \pi i}  \frac{ \text{disc}_{\omega_1}  \left[ \omega_T \psi_n \left( \{ \omega \} , \{ \bfk \} \right) \right] }{ \omega_1^{n+1} }
 + \mathcal{C}_{\infty} \left[ \omega_T \psi_n \right],
\end{align}
where the $\{ \bfk \}$ and $\{ \omega_{a \neq 1} \}$ are held fixed at positive real values. This can be used to connect each EFT Wilson coefficients to a particular integral over the discontinuity of the wavefunction in the underlying UV theory.

 
\section{The analytic structure of the wavefunction}
\label{sec:analytic}

In this section we describe the analytic structure of the coefficients of the field-theoretic wavefunction in Minkowski spacetime. We begin with a brief review of the useful recursion relations for the wavefunction coefficients derived  in \cite{Benincasa:2019vqr}. Then we present a conjecture for the location of all poles and branch points in perturbation theory. We substantiate this conjecture with a series of tree-level and one-loop calculations. Finally, we conclude with a more formal Landau-like analysis of the singularities of wavefunction coefficients from their integral representation. 

 
\subsection{A review of recursion relations for the Minkowski wavefunction}
\label{ssec:analytic_review}

The Minkowski wavefunction $\Psi$ is a functional of all the fields in theory that describes the pure quantum state of a system. It can be thought of as the projection of an abstract state vector $\ket{\Psi}$ onto the set $\{\ket{\phi ; t }\}$ of eigenstates of the field operators $\hat \Phi_{\bfk}(t)$. For example, in Fourier space the basis obeys $\hat\Phi_{\bfk}(t)\ket{\phi ; t }=\phi_{\bfk} \ket{\phi ; t }$, where we take $\phi$ to represent the set of all possible fields in a theory with Lorentz indices omitted. The wavefunction is formally defined by the following $(3+1)$-dimensional path integral
\begin{align}
    \Psi[\phi;t]=\int_{\text{BD}}^{\Phi(t)=\phi} [d\Phi] e^{iS[\Phi ; t]}\,,
\end{align}
where $\text{BD}$ refers to the Bunch-Davies initial state in the infinite past and $S$ is the action functional of a given theory. It is often convenient to parameterize the wavefunction as\footnote{This parameterization is not the most general but it is sufficient for perturbation theory and for certain classes of non-perturbative toy models.}
\begin{align}
\Psi[\phi;t]=\exp\left[ +  \sum_{n}^{\infty}\frac{1}{n!} \int_{\bfk_{1},\dots\bfk_{n}}\, \tilde{\delta}^{(3)} \left( \sum_{a}^{n} \bfk_{a} \right) \psi_{n}(\{\bfk\};t)  \phi(\bfk_{1})\dots \phi(\bfk_{n})\right]\,,
\end{align}
where $\psi_n$ are momentum- and time-dependent \textit{wavefunction coefficients}. These can be computed in perturbation theory to any desired order from a set of diagrammatic rules analogous to Feynman diagrams. We briefly review these rules in App.~\ref{A:rules}. For amplitude Feynman diagrams, the integrals over the time at which an interaction can happen become delta functions in frequency space and are readily accounted for by imposing energy conservation at each vertex. Conversely, since the wavefunction $\Psi[\phi;t]$ singles out a particular time $t$, which we will choose to be $t=0$, the invariance under time translations (and Lorentz boosts) enjoyed by Minkowski amplitudes is spontaneously broken and the wavefunction Feynman rules now involve a number $V$ of nested time integrals for a diagram with $V$ interactions. While each integral is straightforward to evaluate, calculations become rather lengthy for all but the simplest diagrams. 

Fortunately, a purely algebraic set of recursion relations can be derived for time-translation invariant theories that involve polynomial interactions\footnote{This can be extended to derivative interactions by ``dressing" the interaction vertices along the lines of \cite{Hillman:2021bnk}.} \cite{Benincasa:2019vqr}. The solutions in turn admit an elegant representation in terms of canonical forms of polytopes. The relations work diagram by diagram. More in detail, take a diagram contributing to a given wavefunction coefficient $\psi_n$, as discussed in App.~\ref{A:rules}, and remove all external lines. This give a ``skeleton" diagram with $V$ vertices and $I$ internal lines. Associate to each vertex a total vertex energy $x_A$, with $A=1,\dots,V$. Also, to each internal line associate an energy $y_m$, with $m=1,\dots,I$. For tree diagrams all $y_m$'s are fixed in terms of external spatial momenta by momentum conservation at each vertex, but at loop level this is not the case. The dependence of wavefunction coefficients on vertex energies $x$ and internal-line energies $y$ can now be written as
\begin{align}
    \psi_n=\psi_n(x_1,x_2,\dots,x_V;y_1,y_2,\dots,y_I)\,.
\end{align}
This dependence can be determined by the following recursion relation
\begin{align}
 \left(\sum\limits_A^{V} x_A \right)\psi_{n}(\{x_A\})=\sum_{m}^{I}\text{Cut}_m \psi_n(\dots,x_B+y_m,\dots,x_{B'}+y_m,\dots)\,,
 \end{align}
 where the operation Cut$_m$ means that one should remove the $m$-th internal line and add its energy $y_m$ to each of the vertex energies that that line connected. If after the line is cut the diagram becomes disconnected one should interpret $\text{Cut}_m \psi_n$ as the product of the wavefunction coefficients shifted by $y_m$ of the disconnected parts, $\psi_{n'}\times\psi_{n-n'}$ with $n'<n$. Notice that in the recursion relation all coupling constants are omitted, but can be easily re-inserted if desired. The recursion relation can be represented graphically as
\begin{equation} \label{eqn:recursion}
	\begin{tikzpicture}[baseline=(current  bounding  box.center)]
		\coordinate (psi) at (-3.7, 0);
		\coordinate (psiL) at (-0.5, 0);
		\coordinate (psiR) at (2.3, 0);
		\coordinate (loop) at (5, 0);
		\draw[thick] (psi) circle (7mm);
		\draw[thick] (psiL) circle (7mm);
		\draw[thick] (psiR) circle (7mm);
		\node at (-2.65, 0) {$=$};
		\node at (-1.75, -0.1) {\large $\sum\limits_m $};
		\node at (-5.5, 0) {\large $\left(\sum\limits_A x_A \right)$};
		\node at (-3.7, 0) {\large $\psi_{n}$};
		\node at (psiL) {\large $\psi_{n'}$};
		\node at (psiR) {\large $\psi_{n-n'}$};
		\draw[dashed] (.2, 0) -- (1.6, 0);
		\node at (0.5, -.2) {\tiny $+y_m$};
		\node at (1.35, -.2) {\tiny $ +y_m$};
		\node at (3.5, 0) {$+$};
		\draw (loop) circle (7mm);
		\node at (5, 0) {\large$\psi_{n}$};
		\draw[dashed] (4.5, -.5) to [out=240, in=180] (5, -1.5) to [out=0, in=310] (5.5, -.5);
		\node at (4.15, -.5) {\tiny $+y_m$};
		\node at (5.85, -.5) {\tiny $ +y_m$};
	\end{tikzpicture}\,.	
\end{equation}
Using this relation over and over again, one can reduce any diagram to a diagram with one vertex and no internal lines, for which the initial condition of the recursion is
\begin{align} \label{eqn:recursion_initial}
\psit_{1}(x)=\bullet_{x}\hspace{3mm} = \hspace{3mm} \frac{1}{x}\,.
\end{align}
Note that \eqref{eqn:recursion} is the perturbative version of the Hamilton-Jacobi equation which determines the time evolution of the wavefunction, and \eqref{eqn:recursion_initial} corresponds to a Bunch-Davies initial condition (see \cite{Cespedes:2020xqq} for a recent review of this Schr\"{o}dinger picture\footnote{In particular, compare \eqref{eqn:recursion} with Figure 2 of \cite{Cespedes:2020xqq}.}).

Sometimes an example is worth a thousand words. Tree-level examples of the two- and three-site chains are  
\begin{align} \label{psi1}
 \psit_{2}(x_1,x_2;y) &= \frac{\psit_{1} (x_{1}+y)\psit_{1}(x_{2}+y)} {x_{T}} =\frac{1}{(x_1+x_2)(x_1+y)(x_2+y)}\,, \\ 
 \psit_{3}(x_1,x_2,x_3;y_1,y_2) &=\frac{ \psit_{2}(x_{1},x_{2}+y_{2})\psit_{1}(x_{3}+y_{2})+\psit_{1}(x_{1}+y_{1})\psit_{2}(x_{2}+y_{1},x_{3})}{ \sum_{A}^{3} x_{A}}  \nonumber \\  \label{psi3}
 &=\frac{\left( \frac{1}{x_1+x_2+y_2}+ \frac{1}{x_2+x_3+y_1} \right)}{(x_1+x_2+x_3)(x_1 + y_1)(x_2 + y_1 + y_2)(x_3+y_2)} \,. 
 \end{align}
 For loop diagrams, the recursion relation produce the loop \textit{integrand}, as opposed to the integral. To make the distinction clear, we introduce the following notation
 \begin{align}
     \psil{L}_n ( \{ \bfk \}  ) = \int_{\bfp_1,\dots,\bfp_L} \Il{L}_n ( \{ \bfk \} , \{\bfp\} ) \,,
 \end{align}
 where the set of external momenta $\{\bfk\}$ and internal momenta $\{\bfp\}$ will be connected to the recursion relations shortly.
Examples of a one-loop diagram with one or two vertices are
\begin{align}\label{psiloop}
    \Il{1}_1(x;y) &=\frac{1}{x}\psit_1(x+2y)=\frac{1}{x(x+2y)}\,,\\
     \Il{1}_2(x_1,x_2;y_1,y_2)  &=\frac{1}{x_1+x_2}\left[\psit_2(x_1+y_1,x_2+y_1)+\psit_2(x_1+y_2,x_2+y_2)\right] \label{eqn:I1_1loop} \\
     &=\frac{1}{(x_1+x_2)(x_1+y_1+y_2)(x_2+y_1+y_2)} \left[\frac{1}{(x_1+x_2+2y_2)}+\frac{1}{(x_1+x_2+2y_1)}\right]\,.\nonumber
\end{align}
Notice that, loosely speaking, the recursion relation is giving us the result of the integrand expanded in partial fractions. 

With the solution of this recursion relations one can easily write down any desired wavefunction coefficient. We will write that the standard wavefunction coefficients are ``\textit{on-shell}" to emphasize the difference from an off-shell generalization to be introduced shortly. On-shell wavefunction coefficients are obtained from the solution of the recursion relations by performing the following identification:
\begin{align}\label{onshell}
 \text{On-shell $\psi_n$:}  &&  x_A & = \sum_{a\in A} \Omega_{k_a}=\sum_{a\in A} \sqrt{k_a^2+m_a^2}\,, & y_m &= \Omega_{q_m} =\sqrt{q_m^2+m_m^2}\,,
\end{align}
where $\bfk_a$ are the momenta of the external fields connected to the vertex $A$ by a bulk-boundary propagator (an external leg) and $\bfq_m$ are the momenta of the internal lines, which might or might not be integrated over. The label on the mass reminds us that that these results are valid for fields with arbitrary unequal masses. On-shell wavefunction coefficients depend on the external momenta $\bfk_a$, subject to the constraint of momentum conservation, i.e. $
\sum_a \bfk_a=0$ and rotational invariance,
\begin{align}
\text{On-shell:} \qquad \psi_n(\{\bfk\})=\psi_n(\bfk_1,\bfk_2,\dots,\bfk_n) = \int_{\bfp_1,\dots,\bfp_L} \Il{L}_n ( \{ \bfk \} , \{\bfp\} )  \,.
\end{align}
For example, the tree-level contact contribution to $\psi_n$ is found to be\footnote{Here we are abusing our notation. The function $\psi_n(\{x\},\{y\})$ produced by the recursion relations is not exactly the same as the wavefunction coefficients $\psi_n(\{\bfk\},\{\bfp\})$ we want to compute. In particular the label $n$ on the former refers to the number of vertices, while on the latter it refers to the number of external legs, which for a given diagram can be larger if more than one external leg is connected to a vertex. We chose to keep our notation simple and tolerate this nuisance.}
\begin{align}\label{contact}
    \psi_n^\text{contact}(\bfk_1,\bfk_2,\dots , \bfk_n)=\psit_1(\sum \Omega_{k_a})=\frac{1}{\sum_{a} \Omega_{k_a}}\,,
\end{align}
up to an overall coupling constant. Similarly, the quartic wavefunction from particle exchange in the $s$-channel is
\begin{align}\label{on4}
\psi_4(\bfk_1,\bfk_2,\bfk_3,\bfk_4)&=\includegraphics[scale=0.9]{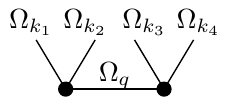}\nonumber \\
    &=\frac{1}{(\Omega_{k_1}+\Omega_{k_2}+\Omega_{k_3}+\Omega_{k_4})(\Omega_{k_1}+\Omega_{k_2}+\Omega_{q_s})(\Omega_{k_3}+\Omega_{k_4}+\Omega_{q_s})}\,,
\end{align}
where $q_s=|\bfq_s|$ and $\bfq_s=\bfk_1+\bfk_2$ is the momentum of the exchanged field.

\paragraph{Distributional terms and the $i\epsilon$ prescription.} There is a technical but important point to be made about the singularities appearing in the wavefunction coefficients when the sum of energies $\Omega_k$ of a given subdiagram (a.k.a. partial energies) vanish: the poles should be shifted into the upper-half complex plane by an infinitesimal amount. To understand this better, let's consider the simplest example of a contact interaction, whose wavefunction coefficient is given in \eqref{contact}. When computing this term from the Feynman-Witten rules (reviewed in App.~\ref{A:rules}), one finds the following ``bulk time integral" representation
\begin{align}
\psi_n^\text{contact}=i\int_{-\infty(1-i\epsilon)}^0 dt\, e^{i\Omega_T t}\,,
\end{align}
where we introduced the total energy, $\Omega_T\equiv \sum_a \Omega_{k_a}$. Notice that the lower boundary of integration is shifted into the complex plane by an infinitesimal amount $i\epsilon$, which should be taken to zero at the end of the calculation. At the mathematical level, this makes the integral well defined. At the physical level this projects the wavefunction in the infinite past onto the Fock vacuum of the free theory. This is the wavefunction $i\epsilon$ prescription. A more colorful way to state the same fact is that the $i\epsilon$ deformation of the time contour turns the oscillatory mode functions into an exponential damping that turns off all interactions in the asymptotic past, effectively turning the interacting theory into a free one. An alternative way to achieve the same result, which was heavily used in \cite{COT}, is to shift $\Omega_T$ by a small imaginary part,
\begin{align}
\Omega_T \to \Omega_T -i\epsilon\,.
\end{align}
For positive and infinitesimal $\e$, the integral converges to
\begin{align}
    \psi_n^\text{contact}=i\int_{-\infty}^0 dt\, e^{i (\Omega_T-i\epsilon) t}=\frac1{\Omega_T-i\epsilon}\,.
\end{align}
Notice that the pole has been pushed into the upper-half complex plane by an infinitesimal amount. For non-vanishing $\Omega_T$, this is again $1/\Omega_T$, as before. However, for vanishing $\Omega_T$ we discover an additional delta function (e.g. by integrating this against a test function)
\begin{align}\label{distrib}
    \frac1{\Omega_T-i\epsilon}=\frac{1}{\Omega_T}+i\pi\delta(\Omega_T)\,.
\end{align}
The wavefunction coefficient has acquired a (distributional) imaginary part. Analogues of this distributional terms appear in wavefunction coefficients to all orders, an in general one should think that all total and partial energy poles, such as those in \eqref{on4}, are shifted by an appropriate $-i\epsilon$, where in principle there is a different $\e$ for each partial energy singularity. We will not study these distributional terms in detail here, but we would like to make a few remarks. First, it might not have escaped the attentive reader that the distributional imaginary term in \eqref{distrib} is very similar to the one found when taking the imaginary part of the Feynman propagator. In that context, the imaginary part appears only when an off-shell ``virtual" particle goes on-shell and becomes a real particle. These terms are particularly important when studying unitarity and are responsible for ensuring that the optical theorem is satisfied order by order, in the form of Cutkosky cutting rules \cite{Cutkosky:1960sp} (see e.g. \cite{Veltman:1994wz,tHooft:1973wag,Schwartz:2014sze}). Indeed, the analog for the exchange four-point function of the delta function in \eqref{distrib} is essential to ensure that the cosmological optical theorem of \cite{COT} reproduces the standard amplitude optical theorem\footnote{We are aware of progress in this direction, which will appear soon \cite{Carlos}.}. Second, the presence of delta functions might seem at odds with analyticity, since they are invisible to any contour integral on the complex plane as appearing for example in Cauchy's theorem. Conversely, an imaginary shift in the poles is a convenient way to maintain some analytical properties as well as an effective prescription to handle singularities on the real axis. Finally, notice that the location of the singularities of wavefunction coefficients depends crucially on the choice of initial state. Starting from the Minkowski vacuum we only encounter singularities that cannot be reached for real, physical momenta. However, for modified initial states delta functions appear at the boundary of the physical kinematical region, for example on folded triangles in the bispectrum.

 
\subsection{Off-shell wavefunction coefficients}
\label{offshell}

In this subsection, we will define the main object of our study: ``off-shell" wavefunction coefficients.

It has long been known that scattering amplitudes enjoy analytic properties in the kinematical variables as a consequence of causality \cite{Eden:1966dnq}. 
Such a connection has proven challenging to establish in full generality and much work has been devoted to deducing the analytic structure by studying the result of explicit perturbative calculations. One might hope that something similar happens for the wavefunction and the goal of this work is to make a first step in this direction. The very first question that we have to ask when tackling this problem is: analyticity in what variable(s)? A first guess might be that the right variable is the energy $\Omega_{k_a}$ of an external field or the associated norm of the momentum $k^2=\Omega_k^2-m^2$. However, already from the tree-level expression in \eqref{on4} one can see that the dependence on this variable will feature branch points and cuts in the complex plane. The reason is that the internal energies $\Omega_q$, which appear analytically in these expressions, are themselves non-analytic functions of the external energies $\Omega_{k_a}$ because of momentum conservation. For example, for the four-point exchange diagram, the internal energy is
\begin{align}
    \Omega_{q_s} &= \sqrt{|\bfk_1+\bfk_2|^2+m^2}=\sqrt{k_1^2+k_2^2+2\bfk_1\cdot\bfk_2+m^2}  \nonumber \\
    &=\sqrt{\Omega_{k_1}^2-m_1^2+2\bfk_1\cdot\bfk_2+\Omega_{k_2}^2-m_2^2+m^2}\,,
\end{align}
where $m$ here is the mass of the exchanged particle. This function has branch points both in the upper and lower complex plane and the location depends on $m$ and hence on the spectrum of the theory. At loop order and in the presence of many particles it quickly becomes hard to keep track of all the associated discontinuities. Another conceptual issue of studying analyticity in $\Omega_{k_a}$ is that the connection to causality is obscured since this variable is not obviously related to time by a Fourier transform. While it might be possible to make progress in this direction, here we choose a different path. 

Let's define a new set of objects, which we will call \textit{off-shell wavefunction coefficients}. There are many equivalent ways to think about these objects, as we will now discuss. The first way is to consider the solution of the recursion relations reviewed in the last subsection, but where the energies of external legs are new variables that are not a priori related in any way to the external three-momenta. This is obtained by performing the following identification
\begin{align}
    \text{Off-shell $\psi_n$:}&&  x_A &= \sum_{a\in A} \omega_a \,, & y_m &=\Omega_{q_m}=\sqrt{q_m^2+m_m^2}\,.
\end{align}
Contrasting these expressions with \eqref{onshell} we see that the only difference is that now energies of external particles are parameterized by $\omega_a$ and are not fixed by the corresponding $k_a$. Conversely, internal energies are still on-shell and are defined as before. Therefore, off-shell wavefunction coefficients for $n$ fields now depend on $n$ more variables as compared to their on-shell cousins. We will denote off-shell wavefunction coefficient with the same symbol $\psi_n$ as the on-shell ones, but with additional dependence on the $\omega_a$'s
\begin{align}
\text{Off-shell:} && \psi_n(\{\omega\},\{\bfk\})=\psi_n(\omega_1,\omega_2,\dots,\omega_n;\bfk_1,\bfk_2,\dots,\bfk_n)  = \int_{\bfp_1,\dots,\bfp_L} \Il{L}_n (\{\omega\}, \{ \bfk \} , \{\bfp\} )  \,.
\end{align}
Since the rest of this paper discusses exclusively off-shell coefficients this should not lead to confusion. Moreover, an off-shell wavefunction coefficient can always be put on-shell simply by specifying 
\begin{align}
\psi_n(\bfk_1,\bfk_2,\dots,\bfk_n)=\psi_n(\Omega_{k_1},\Omega_{k_2},\dots,\Omega_{k_n};\bfk_1,\bfk_2,\dots,\bfk_n)\,.
\end{align}

There is an equivalent and closely related way to define off-shell wavefunction coefficients. Recall that the wavefunction coefficients are determined to all orders in perturbation theory by diagrammatic Feynman-Witten-like rules. In particular, all perturbative contributions to the on-shell wavefunction coefficient can be written as (for $V\geq 1$)
\begin{align}\label{seed}
\text{On-shell:} \qquad	 \psi_n(\bfk_1,\bfk_2,\dots,\bfk_n) &= i \int_{\bfp_1,\dots \bfp_L}\int d^V t \left[  \prod_{a=1}^{n}K(\Omega_{k_{a}})\right] \left[ \prod_m^{I}  G(\Omega_{p_m}) \right]\,,
\end{align}
where $K$ and $G$ are bulk-boundary and bulk-bulk propagators
\begin{align}\label{Kflat}
K(\omega,t) &=e^{i \omega t}\,.\\
G(\omega,t, t' ) &=iP(\omega)\left[  \theta(t-t') K^{\ast}(\omega,t)K(\omega,t')+\theta(t'-t) K^{\ast}(\omega,t')K(\omega,t)-K(\omega,t)K(\omega,t')\right] \\ \nonumber
&= \frac{i}{2\omega}\left[e^{-i\omega(t-t')}\theta(t-t')+e^{-i\omega(t'-t)}\theta(t'-t)-e^{i\omega(t+t')} \right]	\,,
\end{align}
and for simplicity we have assumed only polynomial interactions and omitted the coupling constants. The off-shell wavefunction coefficients are defined using the same propagators (for\footnote{There is a single zero-vertex contribution ($V=0$), which appears already in the free theory. The on-shell result is $\psi_2(\bfk_1,\bfk_2)=-\Omega_k$, while off-shell this becomes $\psi_2(\omega_1,\omega_2)=-(\omega_1+\omega_2)/2$.} $V\geq 1$) as follows:
\begin{align}
\text{Off-shell:} \qquad	 \psi_n(\{\omega\},\{\bfk\}) &= i \int_{\bfp_1,\dots \bfp_L}\int d^V t \left[  \prod_{a=1}^{n}K(\omega_a)\right] \left[ \prod_m^{I}  G(\Omega_{p_m}) \right]\,,
\end{align}
where the $\omega_a$'s are independent variables. 

There are two more ways to define off-shell wavefunction coefficients, which are also valid beyond perturbation theory and highlight a different perspective. We briefly quote here these results because they might help familiarizing oneself with these new objects, but we will not make use of these formulae in the rest of this paper. First, as we show in App.~\ref{App:C}, the off-shell extension $\psi_n \left( \{ \omega \}, \{ \bfk \} \right)$ of the equal-time wavefunction coefficients can be thought of as amputated in-out Green's functions in frequency space. More concretely, the following relation holds
\begin{align}
 \left[ \prod_{j=1}^n \int_{-\infty}^0 d t_j \, e^{ i \omega_j t_j} \mathcal{E}_j \right]\;  \langle \phi ( 0 ) = 0 |  \; T \;  \hat{\Phi}_{\bfk} ( t_1 ) ...  \hat{\Phi}_{\bfk_n} ( t_n ) \;  | \Omega_{\rm in} \rangle_c
 =
 \psi_n \left( \{ \omega \}, \{ \bfk \} \right) \, \tilde{\delta}_D^{(3)} \left( \sum_{a}^n\bfk_a \right) \; , 
\label{eqn:off_shell_psi_def}
\end{align}
where,
\begin{itemize}

\item $| \Omega_{\rm in} \rangle$ is the state in the interacting theory that asymptotes to the vacuum in the far past ($t \to -\infty$), and the label $c$ reminds us to only consider connected contributions (i.e. featuring a single momentum-conserving delta function),

\item $| \phi ( 0 ) = 0 \rangle$ is the field eigenstate annihilated by $\hat{\Phi}_{\bfk}(t)$ at $t=0$, 

\item $T$ represents time-ordering of the subsequent operators, 

\item $\mathcal{E}_j \hat{\Phi}_{\bfk_j} (t_j ) = i \left( \partial_{t_j}^2 + k_j^2 + m^2 \right) \hat{\Phi}_{\bfk_j} (t_j)$ is the free equation of motion acting on the $\hat{\Phi}_{\bfk_j} (t_j )$ operator, 

\item $\int_{-\infty}^0 dt \, e^{i \omega t}$ is the half-interval Fourier transform from the time domain to the frequency domain. 
\end{itemize}

\noindent Note that the use of the half-interval Fourier transform in \eqref{eqn:off_shell_psi_def} shows that the off-shell wavefunction coefficients are analytic in half of the complex $\omega$-plane.
This is because \eqref{eqn:off_shell_psi_def} provides a convergent representation of $\psi_n$ for any $\text{Im} \, \omega < 0$ (for which $e^{i \omega t} \to 0$ at $t \to - \infty$), providing the correlator does not grow exponentially in the far past. 
An equivalent argument, which is closer to the non-relativistic Kramers-Kronig dispersion relation, is to write this time integral as a two-sided Fourier transform at the expense of introducing a step function in time,
\begin{align}
    \int_{-\infty}^0 dt \, e^{i \omega t} \; \mathcal{E} \hat{\Phi}_{\bfk} (t) = \int_{-\infty}^{\infty} dt \, \Theta ( t ) \, e^{i \omega t} \; \mathcal{E} \hat{\Phi}_{\bfk} (t) \; . 
\end{align}
This definition of $\psi_n$ then takes the form of a response to sources which are only turned on the past, and such a response is analytic in (half of) the complex frequency domain. The fact that the amputated in-out Green's function in \eqref{eqn:off_shell_psi_def} reduce to the on-shell wavefunction coefficients when we take $\omega_a=\Omega_{k_a}$ can be thought of as the wavefunction analog of the Lehmann-Symanzik-Zimmermann (LSZ) reduction for amplitudes\footnote{
The connection between wavefunction coefficients, time-ordered correlators and the LSZ procedure for extracting $S$-matrix elements will be discussed in more detail in the upcoming~\cite{MelvillePimentel}.
}.

The last representation of off-shell wavefunction coefficients highlights the relation between analyticity and causality by re-interpreting $\psi_n$ as a response to an appropriate classical external source. Define the off-shell wavefunction by the following path integral,
\begin{align}\label{entry2}
\Psioff[\phix;t_{0}]=e^{iS_{\partial}[\phix ; t_0]} \int_{BD}^{\delta\Phi(t_{0})=0}[d\delta\Phi] e^{iS_{2}[\delta\Phi ; t_0]+S_{\rm int}[\delta\Phi+\phix ; t_0]}\,,
\end{align}
where $\delta \Phi$ is the integration variable, $\phix=\phix(t,\bfk)$ is a spacetime-dependent classical external source and the boundary term is
\begin{align}
iS_{\partial}[\phix; t_0]&=  i \frac{1}{2}\int_{\bfk} \phix(t_{0},\bfk) \dot \Phi_\text{ex}(t_{0},-\bfk)\,.
\end{align}
$S_2 + S_{\rm int}$ is the usual split of the classical action into free (quadratic) and interacting (non-linear) parts.
Then the coefficients of the expansion of $\log \Psioff$ in powers of $\phix(\omega)$ are the desired off-shell wavefunction coefficients
\begin{align}
\Psioff_{n}(\{\omega\},\{\bfk\})\equiv \prod_{a=1}^{n} \frac{\delta }{\delta \phix(\omega_{a},\bfk_{a})} \log \Psioff[\phix]\Big|_{\phix=0}\,.
\end{align}
From this point of view, analyticity in $\omega$ is rooted in the causal response of $\Psioff$ to the external source $\phix$.

To conclude, we notice that so far we have discussed only polynomial interactions of the form $\lambda \Phi^n/n!$ when setting up the recursion relation for $\psi_n ( \{ \omega \} , \{ \bfk \} )$. For derivative interactions it suffices to decorate each vertex with an appropriate vertex factor, accounting for permutations. This leads to the schematic structure
\begin{align}
    \psi_n ( \{\omega\}, \{ \bfk \} ) =  \int_{\bfp_1 ... \bfp_L}  F \left( \{ \bfk \} , \{ \bfp \} \right) \, \mathcal{I}_n \left( \{\omega\}, \{ \bfk \} , \{ \bfp \}  \right)\,,
\end{align}
where $F$ is determined by the Feynman rules for each vertex in the diagram (which can include both time and spatial derivatives), and $\mathcal{I}_n$ is given by the recursion relation above. For example, each spatial derivative leads to a factor of $i \bfk$. When time derivatives act on one or more bulk-bulk propagator the form of $F$ can be more complicated (see \cite{Hillman:2021bnk} for a dedicated discussion). However, our study of the singularities of the wavefunction relies on the structure of the denominator and so still applies, albeit with the possible non-generic cancellation of some singularity.

 
\subsection{On the location of poles and branch points}
\label{ssec:analytic_heuristic}

We are now in a position to state the central claim of this paper:

\begin{tcolorbox}\label{Analyticity}
\begin{quote}

The off-shell wavefunction coefficient $\psi_n ( \omega_a, \bfk_a)$ at any order in perturbation theory is \emph{analytic} in the complex $\omega_1$-plane at fixed real, positive values of $(\omega_{a \neq 1}, \bfk_a)$, \emph{except} for singularities along the negative real axis, $\omega_1 \leq 0$. The location of singularities corresponds to the vanishing of the partial energy of a connected sub-diagram (the \textit{energy-conservation condition}).

\end{quote}
\end{tcolorbox}

\noindent Note that a trivial relabelling of the arguments implies that $\psi_n ( \omega_a , \bfk_a)$ is also analytic in the other $\omega_b$ planes (at fixed $\omega_{a \neq b} >0$ and $\bfk_a > 0$) with singularities at $\omega_b \leq 0$.
As for amplitudes, these non-analyticities have a natural interpretation in terms of the external kinematics crossing various \emph{thresholds} at which new internal processes become important. 
We will first describe these thresholds at a heuristic level (which captures the relevant physics), then in SubSec.~\ref{ssec:analytic_examples} we present various explicit computations of $\psi_n$ that indeed contain the corresponding singularities, and finally in SubSec.~\ref{ssec:analytic_landau} we describe a more formal derivation of the analytic structure in terms of Landau conditions.    

\paragraph{The physical picture.}
In perturbation theory, $\psi_n ( \omega_a, \bfk_a)$ can be represented as a sum over Feynman-Witten diagrams in which each interaction vertex represent an integral of the schematic form
\begin{align}
\int_{-\infty}^0 d t \;  f_{\omega_1}^* ( t ) ... f_{\omega_n}^* ( t ) = \int_{-\infty}^0 d t \; e^{+i \omega_T t}\,, 
\end{align}
where $\omega_T = \sum_{j=1}^n \omega_j$ is the total energy flowing into the vertex from its $n$ legs. Evaluating these integrals requires a prescription to handle the limit $t \to -\infty$. This comes from imposing that the infinite past the system is in the Minkowski ground state. This ensures that the effect of interactions become small in the far past and the integral converges. This is the precise analog of the choice of the Bunch-Davies initial state in accelerating FLRW spacetimes. 
In practice, this physical picture is achieved by deforming the integration contour in the far past to $t \to - \infty ( 1 - i \epsilon)$, such that $e^{i \omega_T t}$ provides an exponential suppression for each interaction vertex in the infinite past. 
However, as already recognized in \cite{Arkani-Hamed:2017fdk}, if there is an \emph{energy-conserving} vertex at which $\omega_T = 0$, then this exponential suppression is removed and such an infinitely long-lived interactions can produce singularities in the Bunch-Davies wavefunction.

This is precisely analogous to a long-lived (on-shell) internal state producing divergences in a scattering amplitude. In the amplitude context, the tree-level exchange of a single on-shell line produces a simple pole and the loop-level exchange of multiple on-shell lines produces a branch cut.
For the wavefunction, the integral representation introduced above makes it clear that tree-level wavefunction coefficients also possess simple poles, while branch cuts are produced only at loop level. 
The conceptual difference is that, rather than being determined by where intermediate lines go on-shell, the non-analyticities in the wavefunction are determined by where interaction vertices become energy-conserving and hence the factors in the denominators of \eqref{psi1}-\eqref{psi3} or \eqref{psiloop}-\eqref{eqn:I1_1loop} vanish. 
 
\begin{figure}
\centering
 \includegraphics[width=0.9\textwidth]{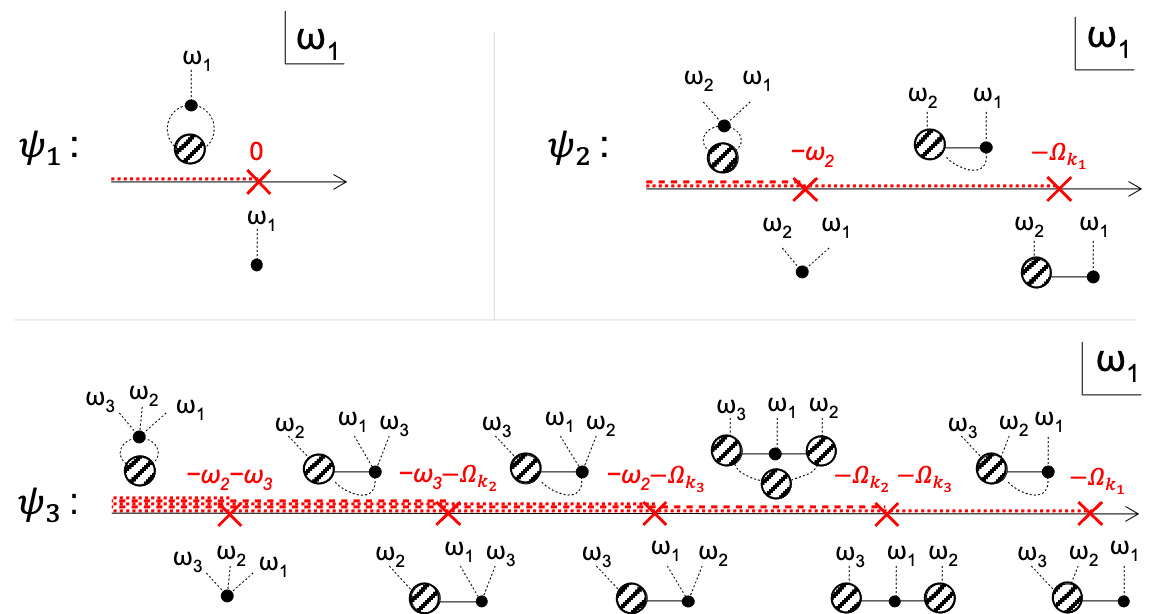}
 \caption{
 The analytic structure in the complex $\omega_1$-plane for the off-shell wavefunction coefficients, $( \psi_1, \psi_2, \psi_3 )$, in a theory with massless particles. In all cases we analytically continue $\omega_1$ with the other $\omega_j$ and $k_j$ held fixed at real positive values, 
 and to provide a concrete order for the singularities we assume that $\omega_j \geq k_j$ and $\omega_j > \omega_{j'}$ if $j > j'$. 
 Red crosses/lines indicate poles/branch cuts, and for each the diagram responsible is shown.
 Solid/dashed lines denote on/off-shell legs. 
 \label{fig:branch_cut_diagrams}}
\end{figure}

This heuristic argument is illustrated in Fig.~\ref{fig:branch_cut_diagrams}, where we summarise our conjectured analytic structure for the off-shell $\psi_1, \psi_2$ and $\psi_3$ by considering the values of $\omega_1$ for which there exists a diagram with an energy-conserving vertex. 
Before attempting to prove that this simple rule indeed captures all of the non-analyticities in the perturbative wavefunction coefficients, we will show how it can be used to systematically generate a list of singular points for any off-shell $\psi_n$.

\paragraph{Tree-level poles.}
The simplest way to enumerate all possible poles in the perturbative wavefunction is to proceed inductively, beginning with the off-shell $\psi_{n=1}$ with a single external leg and then adding further external legs one at a time. 
This is useful because an off-shell diagram with $n$ external legs entering the same bulk vertex with energies $(\omega_1, ... , \omega_n)$ and momenta $( \bfk_1, ... , \bfk_n )$ is identical to the same off-shell diagram with a single external leg entering that vertex with energy $\sum_{a=1}^n \omega_a$ and momenta $\sum_{a=1}^n \bfk_a$, and therefore will give poles that are analogous to those of a lower-point diagram. Because of this we find it convenient to also include quadratic vertices in our analysis, corresponding to perturbative correction from the linear mixing of fields. So for the first three wavefunction coefficients,

\begin{itemize}
    
    \item[$\psi_1$:] The only tree-level diagram for a single (off-shell) external line carrying energy $\omega_1$ is: 
    \begin{equation}
    \centering
    \includegraphics[height=24pt]{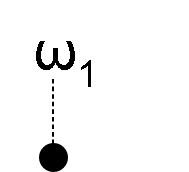}
    \end{equation}
    For an interaction with no derivatives, the corresponding wavefunction coefficient is simply $\psi_1 \propto 1/\omega_1$ and contains a simple pole at $\omega_1 = 0$. 
    Adding derivatives only produces positive powers of $\omega_1$ or $\bfk_1$, and so cannot lead to any additional singularities.

    \item[$\psi_2$:] When two lines carry energies $( \omega_1, \omega_2 )$ and momenta $( \bfk_1 , \bfk_2 = -\bfk_1)$ into the bulk, there are now two possibilities. Either $(a)$ the two lines both terminate on the same interaction vertex, 
    \begin{equation}
        \centering
        \includegraphics[height=24pt]{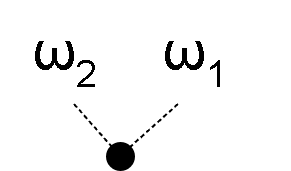}
    \end{equation}
    in which case we find the same result as for $\psi_1$ (with $\omega_1 \to \omega_1 + \omega_2$), or $(b)$ the two lines end on different vertices,
    \begin{equation}
    \centering
        \includegraphics[height=24pt]{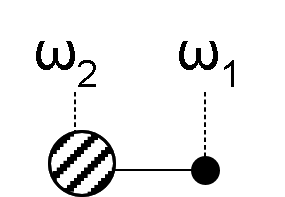}
    \end{equation} 
    where we have used a solid internal line to denote that this is on-shell (i.e. carries an energy $\Omega_{\bfk_1} = \sqrt{k_1 ^2 + m^2}$,  where $m$ is the mass of the field being exchanged). 
    The black vertex carries zero energy when $\omega_1 = - \Omega_{k_1}$, and this is a qualitatively new threshold that develops when there is more than one external line. We have used hatched blob to indicate that the details of the $\omega_2$ coupling are unimportant for this threshold---of course one could relabel the external arguments and similarly conclude that there is a pole at $\omega_2 = - \Omega_{ k_2}$ independently of the coupling to the $\omega_1$ external line. So up to this permutation, there are 2 simple poles which can appear in $\psi_2$, at
    \begin{align}
        \omega_1 + \omega_2 &= 0 \; ,   \nonumber \\ 
        \omega_1 + \Omega_{k_1} &= 0 \; .  
    \end{align}

    \item[$\psi_3$:] With three lines carrying energy and momentum into the bulk, there are now three options.
    The first possibility is that all of the external lines terminate on the same vertex, 
    \begin{equation}
    \centering
    \includegraphics[height=24pt]{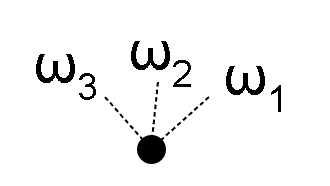} 
    \end{equation}        
    which produces a simple pole at $\omega_1 + \omega_2 + \omega_3 = 0$ just like in $\psi_1$ above (with $\omega_1 \to \omega_1 + \omega_2 + \omega_3$). 
    The second possibility is that just two of the external lines terminate on the same vertex: this can happen \emph{either} as, 
    \begin{equation}
    \centering
    \includegraphics[height=24pt]{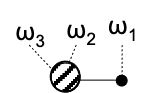}
    \end{equation}     
    which produces a simple pole at $\omega_1 = - \Omega_{k_1}$ just like in $\psi_2$ above, \emph{or} as,
    \begin{equation}
    \centering
    \includegraphics[height=24pt]{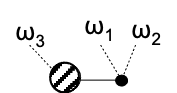} 
    \end{equation}
    which produces a simple pole at $\omega_1 + \omega_2 = - \Omega_{k_3}$, again like in $\psi_2$ above (with $\omega_1 \to \omega_1 + \omega_2$ and $\bfk_1 \to \bfk_1 + \bfk_2$) and up to permutations of the external legs. 
    Finally, there is a qualitatively new threshold which corresponds to the three external legs terminating on different vertices,
    \begin{equation}
    \centering
    \includegraphics[height=24pt]{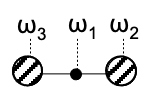} 
    \end{equation}        
    which produces a simple pole at $\omega_1 = - \Omega_{k_2} - \Omega_{ k_3}$. Again, permuting the labels of the external energies implies analogous poles also in $\omega_2$ and $\omega_3$.  
    Overall, up to this permutation of the external leg labels, $\psi_3$ can therefore have simple poles at 4 locations:
    \begin{align}
        \omega_1 + \omega_2 + \omega_3 &= 0 \; , \nonumber \\ 
        \omega_1 + \omega_2 + \Omega_{ k_3} &= 0 \; ,  \\
        \omega_1 + \Omega_{k_2} + \Omega_{ k_3} &= 0 \; , \nonumber \\ 
        \omega_1 + \Omega_{ k_1} &= 0  \; .
    \end{align}

\end{itemize}
Based on this recursive pattern, we see that each time $n$ is increased a qualitatively new kind of threshold appears (in addition to the thresholds which exist for all lower-point coefficients). 
A simple algorithm for explicitly listing all of these poles in a given $\psi_n$ at tree level is the following,

\begin{tcolorbox}
\begin{quote}
\emph{Energy-conservation condition (at tree-level):} \\
For each partition of the $n$ external legs into $q$ subsets, each with a total energy $\omega_a$ and total momentum $\bfk_a$ (for $a = 1, ... , q$), there can be a pole in $\psi_n$ whenever,
\begin{align}
    \omega_1 + \sum_{a = 2}^q \sqrt{ | \bfk_a |^2 + m_a^2 } = 0 
\end{align}
where $m_a$ is the mass of any field that can couple to the external legs in subset $a$.
\end{quote}
\end{tcolorbox}
\noindent As an illustration, consider the off-shell four-point coefficient, $\psi_4$. 
Up to permutations of the particle labels, this algorithm produces a list of 7 possible poles for every set of masses $m_j$ which the exchanged fields may have:
\begin{align}
    \begin{array}{c | r}
    \text{Partition}  & \text{Pole conditions} \qquad ~ \\
    \hline
    \{ 1,2,3,4 \} &  \omega_1 + \omega_{2} + \omega_{3} + \omega_{4} =0 \\ 
    %
    \{1 , 2 , 3 \} , \{ 4 \} & \omega_1 + \omega_2 + \omega_3 + \Omega_{k_4 } = 0 \\
    %
    \{1 ,2 \} ,  \{ 3 , 4 \} & \omega_1 + \omega_2 + \Omega_{| \bfk_3 + \bfk_4 |} = 0 \\
    %
    \{1 \} , \{ 2,3,4 \} & \omega_1 + \Omega_{ |\bfk_2 + \bfk_3 + \bfk_4 |} = 0 \\
    %
    \{ 1, 2 \} , \{ 3 \} , \{ 4 \}  & \omega_1 + \omega_{2} + \Omega_{ k_3} + \Omega_{ k_4 } = 0 \\
    \{1 \} , \{ 2 \} ,  \{ 3 , 4 \} & \omega_1 + \Omega_{ k_2 } + \Omega_{ |\bfk_3 + \bfk_4 |} = 0 \\
    %
    \{1 \} , \{ 2 \} , \{ 3 \} , \{ 4 \} & \omega_1 + \Omega_{ k_2 } + \Omega_{ k_3 } + \Omega_{k_4 } = 0  
    \end{array}
\end{align}
where $\Omega_{ k_a} = \sqrt{ k_a^2 + m_a^2}$ is the energy associated with any of the massive fields that can be exchanged in that channel. 
This list is indeed exhaustive of all of the poles we will find in explicit examples below.

\paragraph{Loop-level branch cuts.}
Beyond tree level, the wavefunction is no longer a rational function and can develop branch cuts in the complex $\omega$-planes. 
These cuts can be viewed as a continuum of poles which arise from integrating a rational integrand (determined by the recursion relations reviewed in Sec.~\ref{ssec:analytic_review}) over continuous loop momenta.
To systematically enumerate all possible branch points that can be generated by loops, it is again useful to proceed inductively starting from $\psi_{n=1}$, 
\begin{itemize}
    
    \item[$\psi_1$:] When a single line carries energy $\omega_1$ and momentum $\bfk_1 = 0$ into the bulk, it must terminate on an interaction vertex.
    At loop level, this vertex may also be connect to internal lines, each of which carries a momentum $\bfq_a$ which is related to the momenta flowing in the loop (and hence integrated over). For instance, in the diagram,
    \begin{equation}
    \centering
    \includegraphics[height=32pt]{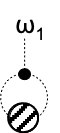}
    \end{equation}
    the black vertex conserves energy when,
    \begin{align}
        \omega_1 = - \Omega_{ q_1} - \Omega_{q_2},
    \end{align}
    where $\Omega_{ q_a } = \sqrt{ q_a^2 + m_a^2 }$ is the energy of the internal lines (which have masses $m_a$). Note that momentum-conservation requires $\bfq_1 + \bfq_2 = 0$. 
    Integrating over all values of $\bfq_1$ therefore creates a continuum of poles on the negative $\omega_1$ axis, which begins at the value,
    \begin{align}
        - \min_{ \substack{ \bfq_1 \\ (\bfq_1+\bfq_2 = 0)} } \left( \Omega_{ q_1} + \Omega_{ q_2}  \right) = - m_1 - m_2 .
    \end{align}
    In general, allowing for an arbitrary number $I$ of internal lines to be connected to the black interaction vertex, there will be a continuum of poles on the negative real axis beginning at, 
    \begin{align}
        \omega_1 = - \min_{ \substack{ \bfq_a \\  ( \sum_{a=1}^I \bfq_a = 0 ) } } \left( \sum_{a=1}^{I} \Omega_{ q_a } \right) = -\sum_{a=1}^I m_a ,
    \end{align}
    where the $m_a$ are the masses of the internal lines.
    Note that when the theory includes massless particles this branch cut threshold coincides with the tree-level pole, but for gapped theories these non-analyticities are separated.

    \item[$\psi_2$:] With two external lines, there are again two possibilities. The first is that the two lines both terminate on the same interaction vertex, e.g. 
    \begin{equation}
    \centering
    \includegraphics[height=32pt]{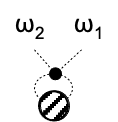}
    \end{equation}     
    in which case we have a branch cut (i.e. a continuum of poles) on the negative real axis which begins at $\omega_1 + \omega_2 = - \sum_{a=1}^I m_a$, just as for $\psi_1$ (with $\omega_1 \to \omega_1 + \omega_2$). 
    The qualitatively new threshold is when the two lines end on different vertices,
    \begin{equation}
    \centering
    \includegraphics[height=32pt]{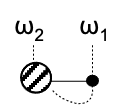}
    \end{equation}     
    in which case the internal momenta are now constrained as $\sum_{a=1}^I \bfq_a = \bfk_1$ by momentum conservation.
    Consequently, the branch cuts from diagrams of this kind begin at
    \begin{align}
        \omega_1 = - \min_{ \substack{ \bfq_a \\ ( \sum_{a=1}^I \bfq_a = \bfk_1) } } \left( \sum_{a=1}^I \Omega_{ q_a} \right) 
        = - \sqrt{ k_1^2 + \left( \sum_{a=1}^I m_a \right)^2 }.
        \label{eqn:psi2_branch_point}
    \end{align} 
    Note that when all of the internal lines carry the same mass, the minimum is achieved at $\bfq_a = \bfk_1/I$ for every $a$ and this threshold is simply $\omega_1 = -\sqrt{ k_1^2 + (I m)^2 }$, and again would coincide with the tree-level pole in any theory which contains massless exchange. This threshold is analogous to the $I$-particle threshold for scattering amplitudes, which comes about because with relativistic energy $\omega_1^2 - | \bfk_1 |^2 = (I m )^2$ the off-shell particle 1 can decay into $I$ on-shell particles of mass $m$. 
    As usual, the freedom to relabel the external leg arguments implies an analogous branch cut in $\omega_2$.

    \item[$\psi_3$:] With three external legs, there are again three possibilities. 
    They could all terminate on the same vertex,
    \begin{equation}
    \centering
    \includegraphics[height=32pt]{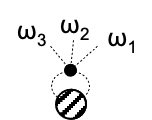}
    \end{equation}     
    which reproduces the same $\omega_1 + \omega_2 + \omega_3 = -\sum_{a=1}^I m_a$ branch cuts as in $\psi_1$ above (with $\omega_1 \to \omega_1 + \omega_2 + \omega_3$). 
    Two could terminate on the same vertex, \emph{either} as
    \begin{equation}
    \centering
    \includegraphics[height=32pt]{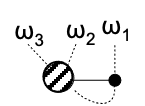}
    \end{equation}     
    which reproduces the same $\omega_1 = -\sqrt{k_1^2 + ( I m)^2}$ type branch cuts as in $\psi_2$ above, \emph{or} as,
    \begin{equation}
    \includegraphics[height=32pt]{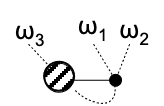}
    \end{equation}     
    which produces a branch cut with threshold $\omega_1 + \omega_2 = - \sqrt{k_3^2 + ( I m)^2}$ following the same $\psi_2$ argument (with $\omega_1 \to \omega_1 + \omega_2$ and $\bfk_1 \to \bfk_1 + \bfk_2 = - \bfk_3$).
    The third possibility is that all three legs terminate on different vertices,
    \begin{equation}
    \centering
    \includegraphics[height=32pt]{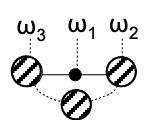}
    \end{equation}     
    This produces a qualitatively new threshold due to the different momentum conservation conditions for the $\bfq_a$. For instance, for $I=2$ internal lines connected to the black vertex and considering a single loop momentum $\bfp$, this threshold occurs at
    \begin{align}
        \omega_1 = - \min_{\bfp} \left( \Omega_{ |\bfk_3 + \bfp|} + \Omega_{| \bfk_2 - \bfp|}    \right) \; .
    \end{align}
    The precise value of this minimum depends on the relative size of $\bfk_3$ and $\bfk_2$, but again we note that in the limit of massless internal lines the branch cut extends all the way to the tree-level pole (which corresponds to $\bfp = 0$). 
    
\end{itemize}
We see that for every tree-level diagram leading to a simple pole there is a corresponding series of loop-level diagrams (labelled by $I$) that create a branch cut at related thresholds. When the exchanged fields are massless all thresholds approach the location of a corresponding tree-level pole. The general conclusion is therefore:
\begin{tcolorbox}
\begin{quote}
    \emph{Energy-conservation condition (at loop level):} \\
    When loops of massless fields are included, every pole in $\psi_n$ becomes a branch point. In massive theories, for each pole there is an infinite series of branch points at successively lower negative values of $\omega_1$ (with a separation determined by the mass gap).   
\end{quote}
\end{tcolorbox}
This closely parallels the analytic structure of scattering amplitudes, for which each tree-level channel produces a corresponding pole at the single-particle threshold (e.g. $s= m^2$), and then loops in each of these channels produce branch cuts at the multi-particle thresholds (e.g. $s = 4m^2, 9m^2, ...$). ~\\

Altogether, we have shown how a simple heuristic argument that links singularities in the wavefunction to long-lived interactions in the bulk (i.e. those that have vanishing total energy) can be used to generate a systematic list of where we expect to find poles and branch cuts in the complex $\omega$-planes.
Next, we will confirm that these lists are indeed an exhaustive classification of the singularities in some concrete wavefunction coefficients computed in perturbation theory.

 
\subsection{Examples}
\label{ssec:analytic_examples}

In this subsection, we present examples of tree-level and one-loop wavefunction coefficients with up to three external legs and confirm the location of singularities predicted by our general argument in the previous section. 
We focus on the singularities in the complexified $\omega_1$ variable, but similar results apply to the other off-shell energies $\omega_a$.

\subsubsection{Tree-level examples} 

Since the Minkowski wavefunction coefficients are particularly simple at tree-level (they are given directly by the recursion relation of Sec.~\ref{ssec:analytic_review}), for the following examples we allow for arbitrary interaction vertices. 

\paragraph{One vertex.}
Let's start by considering tree-level diagrams with a single vertex. These are all related to the starting solution of the recursion relation $\psi_{1}^\text{tree}(x)=1/x$. For one, two and three external legs respectively these are given by
\begin{align}
 \psi_1(\omega_1)&=\includegraphics[scale=0.5]{psi_1_tree.png}=\frac{F_1(\omega_1)}{\omega_1} \,, \\  \psi_2(\omega_1,\omega_2;\bfk_1)&=\includegraphics[scale=0.5]{psi_2_tree_A.png}=\frac{F_2(\omega_1,\omega_2,\bfk_1)}{\omega_1+\omega_2} \,, \\
 \psi_3(\omega_1,\omega_2,\omega_3;\bfk_1,\bfk_2)&=\includegraphics[scale=0.5]{psi_3_tree_A.png}=\frac{F_3(\omega_1,\omega_2,\omega_3;\bfk_1,\bfk_2)}{\omega_1+\omega_2+\omega_3} \,,
\end{align}
where $F_1$, $F_2$ and $F_3$ are vertex factors. We find poles at $\omega_1=0$, $\omega_1=-\omega_2$ and $\omega_1=-\omega_2-\omega_3$, in agreement with the energy-conservation condition of the previous section.

\paragraph{Two vertices.}
Diagrams with two vertices are a bit more interesting. They are all related to the second term in the recursion relation, $\psi_2(x_1,x_2;y)$ in \eqref{psi1}, and they only appear for two or more external legs.  For $\psi_2$ with 2 vertices, we find
\begin{align}
 \psi_2(\omega_1,\omega_2; \bfk)&=\includegraphics[scale=0.5]{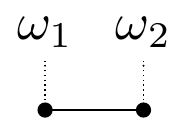}=\frac{F_{L}(\omega_1; \bfk)F_R(\omega_2; \bfk)}{(\omega_1+\Omega_k)(\omega_2+\Omega_k)(\omega_1+\omega_2)} \,.  
\end{align}
There are poles at $\omega_1=-\omega_2$, as well as $\omega_1=-\Omega_k$, which are predicted by the energy-conservation condition. For $\psi_3$ with 2 vertices, there are two possibilities. We can have $\omega_1$ alone on one of the vertices, which gives:
\begin{align}
    \quad\psi_3=\includegraphics[scale=0.5]{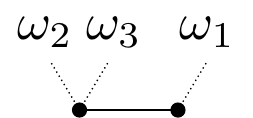}=\frac{\tilde{F}_L(\omega_2,\omega_3;\bfk_1)\tilde{F}_R(\omega_1;\bfk_1)}{(\omega_1+\Omega_{k_1})(\omega_2+\omega_3+\Omega_{k_1})(\omega_1+\omega_2+\omega_3)},
\end{align}
and so we find poles at $\omega_1=-\omega_2-\omega_3$ and $\omega_1=-\Omega_{k_1}$. We can also have $\omega_1$ and another external leg on the same vertex, which gives:
\begin{align}
    \psi_3&=\includegraphics[scale=0.5]{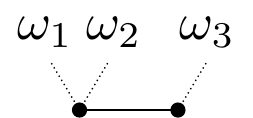}+(2\leftrightarrow 3) \nonumber \\
    &=\frac{\tilde{F}_L(\omega_1,\omega_2;\bfk_3)\tilde{F}_R(\omega_3;\bfk_3)}{(\omega_3+\Omega_{k_3})(\omega_1+\omega_2+\Omega_{k_3})(\omega_1+\omega_2+\omega_3)}+(2\leftrightarrow 3).
\end{align}
In addition to $\omega_1=-\omega_2-\omega_3$ we also find $\omega_1=-\omega_2-\Omega_{k_3}$ and $\omega_1=-\Omega_{k_2}-\omega_3$. 

\paragraph{Three vertices.}
For $\psi_3$ with three vertices there are 3 different permutations for the location of external leg. If $\omega_1$ is attached to the vertex on the side we have:
\begin{align}
       \psi_3&=\includegraphics[scale=0.5]{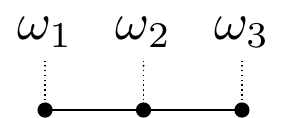}+(2\leftrightarrow 3)\nonumber\\
       &=\frac{F_A(\omega_1;\bfk_1)F_B(\omega_2;\bfk_1,\bfk_2)F_C(\omega_3;\bfk_3)\left(\frac{1}{\omega_1+\omega_2+\Omega_{k_3}}+\frac{1}{\omega_2+\omega_3+\Omega_{k_1}}\right)}{(\omega_1+\omega_2+\omega_3)(\omega_1+\Omega_{k_1})(\omega_2+\Omega_{k_1}+\Omega_{k_3})(\omega_3+\Omega_{k_3})}+(2\leftrightarrow 3).
\end{align}
Poles are located at $\omega_1=-\omega_2-\omega_3$, $\omega_1=-\Omega_{k_1}$, $\omega_1=-\omega_2-\Omega_{k_3}$ and $\omega_1=-\Omega_{k_2}-\omega_3$. 

If $\omega_1$ is attached to the middle vertex we have:
\begin{align}
    \psi_3 &=\includegraphics[scale=0.5]{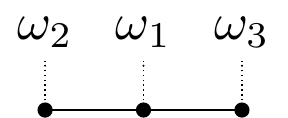} 
 \nonumber \\
&=\frac{F_A(\omega_2;\bfk_2)F_B(\omega_1;\bfk_1,\bfk_2)F_C(\omega_3;\bfk_3) \left(\frac{1}{\omega_1+\omega_3+\Omega_{k_2}}+\frac{1}{\omega_1+\omega_2+\Omega_{k_3}}\right) }{(\omega_1+\omega_2+\omega_3)(\omega_3+\Omega_{k_3})(\omega_1+\Omega_{k_2}+\Omega_{k_3})(\omega_3+\Omega_{k_3})} \, .
\end{align}
 Here we find a pole at $\omega_1=-\Omega_{k_2}-\Omega_{k_3}$ as well as $\omega_1=-\omega_2-\omega_3$, $\omega_1=-\omega_2-\Omega_{k_3}$ and $\omega_1=-\Omega_{k_2}-\omega_3$. 
 
 All these poles correspond precisely to the list of tree-level singularities predicted by the energy-conservation condition for $\psi_n$ with $n=1,2$ and $3$. Now we move on to consider loop diagrams.
 

\subsubsection{One-loop examples} 

At one-loop, the computation of wavefunction coefficients becomes more involved due to the integration over the loop momentum. 
To streamline our presentation, we will therefore now focus on polynomial interactions. We also give only the final results here in the main text, and describe the technical details of the computations in App.~\ref{App:A}.

\paragraph{One vertex.} 
Consider the following diagram:
\begin{equation}
\centering
    \includegraphics[scale=0.4]{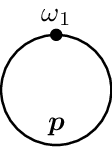} 
\end{equation}
in which all external legs are to be attached to the single vertex. 
Define $\omega_1$ to be the total energy entering the vertex. Since there is only one vertex, the energy-conservation condition predicts a branch point at the threshold,
\begin{equation}
    \omega_1=-\min_{\bfp}\left(2\Omega_p\right)=-2M ,
\end{equation}
where $M$ is the mass of the internal line forming the loop.
This diagram corresponds to the integral,
\begin{equation}
   \omega_1 \psil{1}_{1} =\int_{\bfp}\frac{1}{ \omega_1 + 2 \Omega_p } \; , 
\end{equation}
and is evaluated explicitly in App.~\ref{sec:1edge}.
The result is,
\begin{align}
\omega_1 \psil{1}_{1} = \frac{2 \omega_1 }{16 \pi^2 } \sqrt{4 M^2-\omega_1^2} \text{arcsin} \left( \sqrt{ \frac{2M-\omega_1}{4M} } \right)  + \text{analytic} \; , 
\label{eqn:psi1site}
\end{align}
see \eqref{md3}. 
Note that the UV divergence is analytic in $\omega_1$ (and can therefore be absorbed into local counter-terms). 
There is a branch point at $\omega_1= - 2M$ due to the argument of the $\text{arcsin}$ exceeding unity, but otherwise $\omega_1 \psil{1}_1$ is analytic in the complex $\omega_1$ plane.  

If the internal field is massless, the branch point is located at $\omega_1=0$, so the branch cut starts at the location of the tree-level pole. 
Indeed, taking the massless limit of \eqref{eqn:psi1site} gives,
\begin{equation}\label{psi1site}
    \psil{1}_{1}= - \frac{\omega_1}{16\pi^2}\log\left( \omega_1 \right) + \text{analytic} \; , 
\end{equation}
which has a logarithmic branch point at $\omega_1 = 0$ (with the conventional branch cut running along the negative real axis, $\omega_1<0$). 
So this simple example agrees with our energy-conservation condition.


\paragraph{Two vertices.} 
Now consider the following two-vertex one-loop diagram:
\begin{align}\label{2V}
    \includegraphics{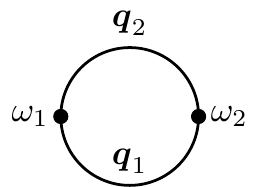}
\end{align}
which contributes to $\psil{1}_{2}$. 
Define $\bfk$ to be the momentum entering the left vertex and exiting the right vertex, and $\omega_1$, $\omega_2$ to be the energies entering each vertex. 
The energy-conservation condition predicts the following singularities in the complex $\omega_1$ plane:
\begin{itemize}

    \item[(i)] $\omega_1=- \sqrt{k^2 + (M_1 + M_2)^2}$, 
    
    \item[(ii)] $\omega_1=-\omega_2- M_1 - M_2$. 

\end{itemize}
The diagram in \eqref{2V} corresponds to the integral:
\begin{align}
 \omega_{12} \psil{1}_2 ( \omega_1, \omega_2 ; k )  &=  \int_{\bfp}   \, \frac{ 1 }{ ( \omega_1 + \Omega_{q_1} + \Omega_{q_2} ) ( \omega_2 + \Omega_{q_2} ) } \left[ \frac{1}{ \omega_{12} + 2 \Omega_{q_1}} + \frac{1}{\omega_{12} + 2 \Omega_{q_2}} \right] \; . 
 \label{eqn:psi2_int_main}
\end{align}
where $q_1=|\bfp|$ and $q_2=|\bfk-\bfp|$ are the momenta of the internal lines and $\omega_{12} = \omega_1 + \omega_2$ is the total energy. 
This integral is evaluated in detail in App.~\ref{sec:2edge}, and given in \eqref{eqn:psi2_ans} in terms of incomplete elliptic integrals. 
For finite values of $M$ and $k$, it has a branch point at $\omega_1 = - \sqrt{k^2 + 4M^2}$ in the complex $\omega_1$ plane, as predicted by the energy-conservation condition. 

The second singularity predicted by the energy-conservation condition appears when either $M$ or $k$ vanish, as shown in \eqref{psi2site} and \eqref{eqn:Jk_soft}. 
For the case of massless internal edges, the incomplete elliptic integrals simplify to dilogarithms, and \eqref{eqn:psi2_int_main} can be written as,
\begin{align}
    \omega_{12} \psil{1}_{2} =\frac{1}{8\pi^2}\Bigg[ 
    & \frac{\omega_2\log\left( \omega_1+k \right)-\omega_1\log\left( \omega_2+k \right)}{\omega_1-\omega_2} 
      \nonumber \\
  &- \frac{\omega_{12}}{2k}\left(\frac{1}{2}\log^2\left(\frac{\omega_1+k}{\omega_2+k}\right) +\text{Li}_2\left(\frac{k-\omega_2}{k+\omega_1}\right)+\text{Li}_2\left(\frac{k-\omega_1}{k+\omega_2}\right)\right) + \text{analytic} \Bigg] \; . \label{psi2site_main}
\end{align}
The singularities in the complex $\omega_1$ plane are\footnote{Recall that the dilogarithm $\text{Li}_2(z)$ has a branch point at $z=1$ and the conventional branch cut goes from $z=1$ to $z=\infty$ along the real axis.}
\begin{itemize}

    \item[(i)] $\omega_1=-k$, from both $\log(\omega_1+k)$ and $\text{Li}_2\left(\frac{k-\omega_2}{k+\omega_1}\right)$.
    
    \item[(ii)] $\omega_1=-\omega_2$, from both $\text{Li}_2\left(\frac{k-\omega_1}{k+\omega_2}\right)$ and $\text{Li}_2\left(\frac{k-\omega_2}{k+\omega_1}\right)$ (while the dilogarithm is finite at that point, it is not smooth).

\end{itemize}
This list of singularities matches exactly the predictions of the energy-conservation condition. It is also worth mentioning that the first line of (\ref{psi2site}) is not singular at $\omega_1=\omega_2$ (at fixed $k \neq -\omega_2$), since this apparent pole has zero residue. 


\paragraph{Three vertices.} 
Let's move on to the most complicated one-loop diagram we will consider, involving three vertices:
\begin{equation}
    \includegraphics[scale=0.4]{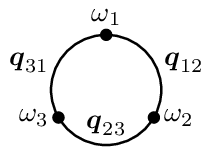}
\end{equation}
Momentum conservation at each vertex fixes all but one of the internal momenta (which we denote by $\bfp$), and also sets $\bfk_1 + \bfk_2 + \bfk_3 = 0$. 
This diagram then corresponds to the integral,
\begin{align}
    \omega_{123} \psil{1}_3 = \int_{\bfp} \frac{1}{ (\omega_1 + \Omega_{q_{12}} + \Omega_{q_{31}} ) ( \omega_2 + \Omega_{q_{12}} + \Omega_{q_{23}} ) ( \omega_3 + \Omega_{q_{23}} + \Omega_{q_{31}} ) } \sum_{\text{perm.}}^6  \frac{1}{ ( \omega_{123} + 2 \Omega_{q_{12}} )(\omega_{23} + \Omega_{q_{12}} + \Omega_{q_{31}}  )} \; . 
    \label{eqn:psi3_int_main}
\end{align}
which is discussed in App.~\ref{sec:3edge}. 
The singularities expected from the energy-conservation condition are:
\begin{itemize}

    \item[(i)] $\omega_1=-\min_{\bfp}(\Omega_{q_{31}}+\Omega_{q_{12}})$. This gives $\omega_1=-|\bfk_1|$ for massless internal lines.

    \item[(ii)] $\omega_1=-\omega_2-\min_{\bfp}(\Omega_{q_{23}}+\Omega_{q_{31}})$. This gives $\omega_1 = -\omega_{2}-|\bfk_3|$ for massless internal lines.

    \item[(iii)] $\omega_1=-\omega_3-\min_{\bfp}(\Omega_{q_{12}}+\Omega_{q_{23}})$. This gives $\omega_1=-\omega_3 -|\bfk_2|$ for massless internal lines.

    \item[(iv)] $\omega_1=-\omega_2-\omega_3-
    \sum_{a=1}^{2} M_a$. This gives $\omega_1=-\omega_2-\omega_3$ for massless internal lines.

\end{itemize}

Evaluating the integral \eqref{eqn:psi3_int_main} in full generality is a difficult task in $d=3$ dimensions: the main complication is that the boundary of the integration region for the $\{ \Omega_{q_{12}} , \Omega_{q_{23}} , \Omega_{q_{31}} \}$ internal energies is a non-trivial surface (defined by a hyperelliptic curve). 

For simplicity, let us consider here the case where all internal fields are massless and let us further suppose that one of the external fields carries zero spatial momentum, say $\bfk_3 = 0$ (though note that we are not fixing $\omega_3$). 
In this limit $\bfk_1=-\bfk_2$ (and so we denote their common magnitude as $k$), and the integration region degenerates to the same region encountered in the two-vertex diagram above.
Consequently $\psil{1}_3$ can be written in a closed form in terms of dilogarithms. 
The full expression is left in App.~\ref{sec:3edge} (equation \eqref{eqn:psi3_soft}), however we notice that it is analytic in the complex $\omega_1$ plane (at fixed $\{\omega_2, \omega_3, k \}$) modulo branch points at:
\begin{itemize}

    \item[(i)] $\omega_1=- k$, where $\psil{1}_{3}\sim \log \left( \omega_1 + k \right)$,

    \item[(ii)] $\omega_1=-\omega_2$, where $\psil{1}_{3}\sim \text{Li}_2\left(-\frac{\omega_2-k}{\omega_1+k}\right)$ and $\text{Li}_2\left(-\frac{\omega_1-k}{\omega_2+k}\right)$,
    
    \item[(iii)] $\omega_1 = - \omega_3 - k$, 
    where $\psil{1}_{3}\sim \log \left( \omega_{13} + k \right)$, 
         
    \item[(iv)] $\omega_1=-\omega_{23}$, where $\psil{1}_{3}\sim\text{Li}_2\left(-\frac{\omega_{23} -k}{\omega_1+k}\right)$ and $\text{Li}_2\left(-\frac{\omega_{13} -k}{\omega_2+k}\right)$ .

\end{itemize}
This precisely saturates the list of singularities expected from the energy-conservation condition.

Having established the validity of the energy-conservation condition in a number of examples, we now turn to a robust ``proof'' that wavefunction integrals generically possess singularities in these locations. 

 
\subsection{Landau analysis for the wavefunction coefficients}
\label{ssec:analytic_landau}

In this section, we develop the analogue of the Landau analysis commonly used in amplitude literature. This provides a list of necessary conditions for a point in kinematic space to be singular. We will see that the list of singularities presented in Sec.~\ref{ssec:analytic_heuristic} is contained within the list of singularities from the Landau analysis. 

From the recursion relations we know that $\psi_n$ can always be written in the following form:

\begin{equation}
\psi_n(\{\omega\},\{\bfk\})=\int_{\bfp_1,\dots,\bfp_L}\frac{F(\{\omega\},\{\bfk\},\{\bfp\})}{\prod_{j=1}^{2V+L-2} S_j(\{\omega\},\{\bfk\},\{\bfp\})}.\label{plytope}
\end{equation}
Here $S_j$ are linear functions of the internal and external energies, and $F(\{\omega\},\{\bfk\},\{\bfp\})$ can always be expressed in terms of a sum of products of $S_j$ times analytic functions of $\bfk$ from derivative interactions. 

We would like to find all singular points of $\psi_n$ without explicitly computing the integral. Similar technology has been developed in the amplitude literature, leading to a set of conditions for singularity known as the Landau equations (for a review on Landau conditions for amplitudes, see \cite{Eden:1966dnq}, \cite{Zwicky:2016lka},\cite{Sashanotes}). We will review some key ideas used to derive the Landau equations, and show how they can be used to find singularities for wavefunction coefficients as well.

\paragraph{Singularities: one integral variable.}
Consider the following expression:
\begin{equation}
f(z_1,\dots,z_n)=\int_{C}dw\,g(z_1,\dots,z_n,w).
\end{equation}

Here $C$ denotes a contour in the complex $w$ plane. $g(z_1,\dots,z_n,w)$ contains singularities, and their positions in the complex $w$ plane are determined by an algebraic equation $S(z_1,\dots,z_n,w)=0$. Changing $z_1,\dots,z_n$ corresponds to changing the position of poles in the complex $w$ plane. 

Singular points in $g(z_1,\dots,z_n,w)$ can be avoided by deforming the contour $C$, and this prevents singularities from developing in $f(z_1,\dots,z_n)$. However, contour deformation cannot avoid the following singularities:

\begin{itemize}
\item When a singularity approaches the endpoint of the contour $C$, which is fixed by the boundary conditions of the integral. This is known as an endpoint singularity.
\item When two different singularities approach the contour from opposite sides and pinch the contour in between. This is known as a pinch singularity.
\end{itemize}

\begin{figure}
    \centering
    \includegraphics[scale=0.5]{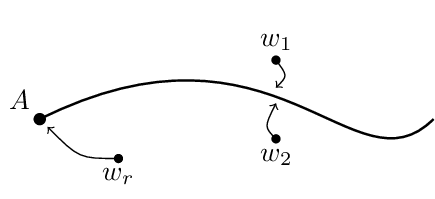}
    \caption{Usual picture for singularities in the case with one integration variable. Here $w_r$ gives an endpoint singularity, while $w_1$ and $w_2$ gives a pinch singularity.}
\end{figure}

\paragraph{Singularities: multiple integral variables.}
In order to illustrate how singularities can develop in cases with multiple integral variables, let us consider the case of two complex integral variables:
\begin{equation}
f(z_1,\dots,z_n)=\int_C dw_1 dw_2\,g(w_1,w_2,z_1,\dots,z_n).
\end{equation}
The hypercontour $C$ is a two (real) dimensional surface in a four (real) dimensional space. In general the hypercontour would have a set of boundaries, and each of them would be described by an equation:
\begin{equation}
    \tilde{S}_i=0.
\end{equation}
For instance, suppose the integration region is given by:
\begin{equation}
    f(z_1,\dots,z_n)=\int_{1}^{\infty} dw_1 \int_{-1}^{1}dw_2\,g(w_1,w_2,z_1,\dots,z_n). \label{exampleint}
\end{equation}
Then the boundaries of the hypercontour would be given by:
\begin{align}
    &\tilde{S}_{1}=w_1-1=0,\\
    &\tilde{S}_{2+}=w_2+1=0,\\
    &\tilde{S}_{2-}=w_2-1=0.
\end{align}

\begin{figure}
    \centering
    \includegraphics[scale=0.35]{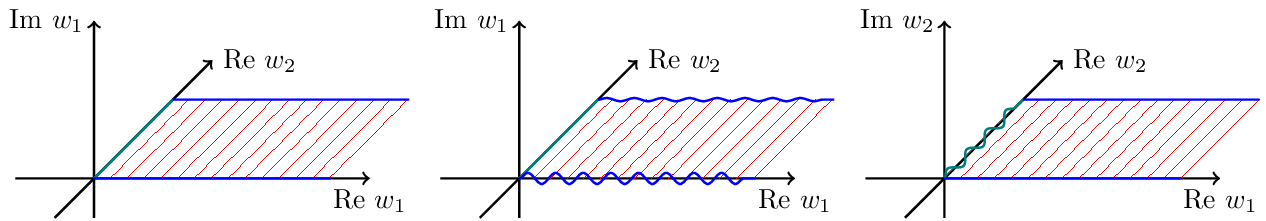}
    \caption{The hypercontour for the integral (\ref{exampleint}), sketched in three of the four (real) directions. The left figure shows the undistorted integration contour. The boundary $\tilde{S}_{2+}$ and $\tilde{S}_{2-}$ is indicated by blue lines, while the boundary $\tilde{S}_1$ is indicated by a teal line. Since the equation for $\tilde{S}_{2+}$ and $\tilde{S}_{2-}$ only fixes $w_2$, they are allowed to deform in the imaginary $w_1$ direction, as shown in the middle figure. The right figure shows the allowed deformation for the boundary $\tilde{S}_1$, which is in the imaginary $w_2$ direction.}
    \label{fig:contour}
\end{figure}

Notice that each of the equation $\tilde{S}_i=0$ describe a two (real) dimensional surface in a four (real) dimensional space. This is less constraining than the single integral variable case: since the hypercontour $C$ is two dimensional, its boundary should be one dimensional. This implies that the boundary of the contour is not rigidly fixed: we are allowed to deform the boundary as long as it remains on the surface described by the equation $\tilde{S}$. See Fig.~\ref{fig:contour}.

Since the allowed deformations are all constrained on a surface $\tilde{S}_i=0$, the boundary of the contour cannot be deformed in the normal direction of the surface, which is described by the vector with components:
\begin{equation}
    \frac{\partial\tilde{S}_i}{\partial \boldsymbol{w}}=\left(\frac{\partial\tilde{S}_i}{\partial w_1},\frac{\partial\tilde{S}_i}{\partial w_2}\right).
\end{equation}

Similarly, the singularities for the function $g(w_1,w_2,z_1,\dots,z_n)$ are described by algebraic equations of the form:
\begin{equation}
    S_i(w_1,w_2,z_1,\dots,z_n)=0.
\end{equation}
Once again these are two dimensional surfaces, and their normal vectors are $\frac{\partial S_i}{\partial \boldsymbol{w}}$. The singular surface in general may not be planar, since the equation $S_i=0$ may not be linear. 

Given a set of singular surfaces and boundary constraint surfaces, there are three ways where singularities can emerge from the integral $I(z_1,\dots,z_n)$:
\begin{itemize}
\item A singular surface approaches the boundary of the hypercontour in the normal direction of a constraint surface, such that no deformation can be carried out to avoid the singular surface. This is analogous to the endpoint singularities in the one dimensional case.
\item Two surfaces approaches each other from opposite sides of the hypercontour, and they pinch the hypercontour in between. This is analogous to the pinch singularities in the one dimensional case.
\item A single surface may become locally cone-like and pinch the contour in the vertex of the cone, see Fig.~\ref{fig:pinch}. It is also a pinch singularity, but unlike the single integral variable case, only one singularity surface is involved. In this case, the normal vector near the vertex satisfies the following:
\begin{equation}
\frac{\partial S_i}{\partial \boldsymbol{w}}=0 .
\end{equation}

\end{itemize}
\begin{figure}[t]
    \centering
    \includegraphics[scale=0.5]{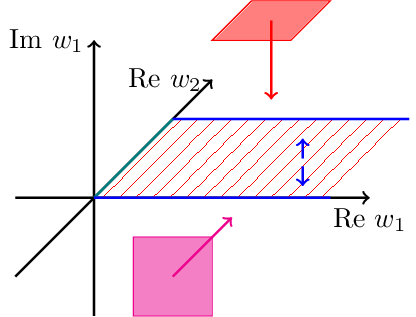}
    \caption{As the boundary $\tilde{S}_{2+}$ and $\tilde{S}_{2-}$ can be deformed in the imaginary $w_1$ direction (indicated by the blue arrows in the figure), the contour can be deformed downwards to avoid the red surface approaching from above (along $\Im w_1$). However, the boundary is fixed along $\Re w_2$, so the contour cannot be deformed to avoid the magenta surface approaching from that direction, which results in an endpoint singularity.}
    \label{fig:endpt}
\end{figure}
\begin{figure}[t]
    \centering
    \includegraphics[scale=0.35]{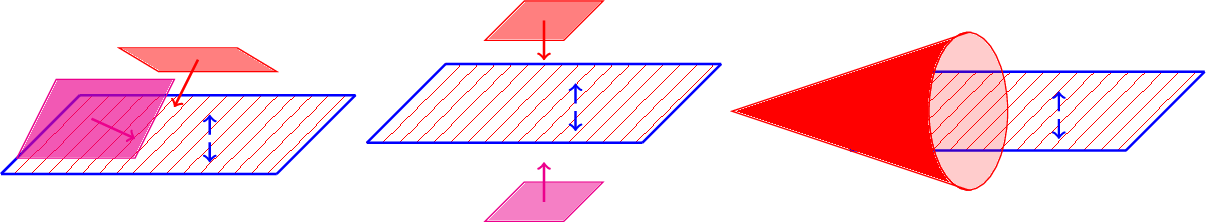}
    \caption{Consider a contour that can be deformed in the vertical direction (indicated by the blue arrows). In the left panel, the contour can deform downwards to avoid colliding with the red and magenta surfaces. However, if the surfaces approach from the opposite side as in the middle panel, a contour deformation cannot avoid the pinch. Finally, a cone like surfaces, such as that in the right panel, can pinch a contour on its own.}
    \label{fig:pinch}
\end{figure}
It is also possible to have multiple surfaces pinching the hypercontour, or multiple surfaces approaching the boundary of the hypercontour. In general, given a function of the form (\ref{exampleint}), a singularity can form if:
\begin{itemize}
\item For a subset $I$ of the singularity surfaces and a subset $\tilde{I}$ of boundary constraints, the following is satisfied:
\begin{equation}
    S_i=\tilde{S}_j=0\, (i\in I,j\in \tilde{I}).
\end{equation}
For amplitudes the analogous condition gives $p_i^2=m_i^2$. Here $I$ cannot be an empty set, but $\tilde{I}$ can be empty.
\item For some real and non-zero choice of $a_i$ and $\tilde{a}_j$, the normal vectors of the singularity surfaces in subset $I$ and boundary constraints in $\tilde{I}$ must satisfy:
\begin{equation}
\sum_{i\in I}a_i \frac{\partial S_i}{\partial\boldsymbol{w}}+\sum_{j\in \tilde{I}}\tilde{a}_j \frac{\partial\tilde{S}_j}{\partial\boldsymbol{w}}=0.\label{nvec}
\end{equation}
In other words, the normal vectors are linearly dependent. For amplitudes, this gives $\sum_{i}\alpha_i p_i=0$.
\end{itemize}
These are the necessary conditions for the formation of singularities. However, these conditions are not sufficient: one needs to check whether the singularities actually appear. This is similar to the pseudo thresholds in amplitudes: the Landau analysis may predict singular points that are not present for physical configurations. We will see that something similar also occurs for wavefunction coefficients. 


\paragraph{Feynman parameters.}
The constants $a_i$ in the normal vector condition looks suspiciously like Feynman parameters in the usual Landau analysis. Indeed, if we consider the following integral:
\begin{equation}
\psi_n(\omega_1,\dots,\omega_n)=\int_{\bfp_1,\dots,\bfp_L} \left[ \prod_{n=1}^{2V+L-2} \int_0^1d\alpha_n \right]\frac{\delta(1-\sum\alpha_n)F}{(\sum_{j=1}^{2V+L-2} \alpha_j S_j)^{2V+L-2}} \label{plytopef},
\end{equation}
we find the condition (\ref{nvec}) again, but with $a_i$ replaced by the Feynman parameters $\alpha_i$. Just like for amplitudes, this does not introduce new singularities \cite{Eden:1966dnq}. In contrast to $a_i$, which is just restricted to be non-zero, we have $\alpha_i\in[0,1]$. This provides a stricter criterion for singularities to arise in $\psi_n$.

\paragraph{Example: massless two-vertex integral.}
As a first example let's consider the massless two-vertex integral (\ref{eqn:psi2_int_main}), which can be written in the following form (see App.~\ref{sec:2edge}):
\begin{equation}
\psil{1}_{2}(\omega_1,\omega_2)=\frac{1}{8\pi^2 (\omega_{1}+\omega_2)k}\int_{k}^{\infty}dp_+\int_{-k}^{k} dp_-\frac{(p_++p_-)(p_+-p_-)}{(\omega_1+\omega_2+p_++p_-)(\omega_1+p_+)(\omega_2+p_+)}.
\label{eqn:psi2_int_p+p-}
\end{equation}
The singular surfaces are:
\begin{align}
    &S_1=\omega_1+p_+=0,\\
    &S_2=\omega_2+p_+=0,\\
    &S_3=\omega_1+\omega_2+p_++p_-=0.
\end{align}
In addition, the boundary is described by the following equations:
\begin{align}
&\tilde{S}_1=p_+-k=0,\\
&\tilde{S}_{2+}=p_-+k=0,\\
&\tilde{S}_{2-}=p_--k=0.
\end{align}

As an example, consider:
\begin{equation}
    S_3=\tilde{S}_{2+}=\tilde{S}_1=0.
\end{equation}
Solving this gives $\omega_1+\omega_2=0$. Now we check the normal-vector condition:
\begin{align}
    &a_3\frac{\partial S_3}{\partial p_+}+\tilde{a}_1\frac{\partial \tilde{S}_1}{\partial p_+}+\tilde{a}_{2+}\frac{\partial \tilde{S}_{2+}}{\partial p_+}=a_3+\tilde{a}_1=0,\\
    &a_3\frac{\partial S_3}{\partial p_-}+\tilde{a}_1\frac{\partial \tilde{S}_1}{\partial p_-}+\tilde{a}_{2+}\frac{\partial \tilde{S}_{2+}}{\partial p_-}=a_3+\tilde{a}_{2+}=0.
\end{align}
Clearly this can be satisfied if $a_3=-\tilde{a}_1=-\tilde{a}_{2+}$. Going through the procedure for all combinations of the surfaces, we eventually find the following list of potential singularities:
\begin{align}
    &S_1=\tilde{S}_1=0\Rightarrow\omega_1=-k,\\
    &S_2=\tilde{S}_1=0\Rightarrow\omega_2=-k,\\
    &S_3=\tilde{S}_{2+}=\tilde{S}_1=0\Rightarrow\omega_1+\omega_2=0,\\
    &S_3=\tilde{S}_{2-}=\tilde{S}_1=0\Rightarrow\omega_1+\omega_2=-2k,\\
    &S_1=S_2=0\Rightarrow\omega_1=\omega_2\,(\text{if }\,\omega_2<-k).\label{pin2site}
\end{align}

Notice this list is larger than the list of physical singularities. Let's examine these extra singularities:
\begin{itemize}
    \item The $\omega_1+\omega_2=-2k$ singularity is not found in the expression (\ref{psi2site}). This suggests that the residue of the pole is vanishing. In App.~\ref{App:A} we explore this further, and show that it vanishes due to the cancellation between two dilogarithms.
    \item For $S_2=0$, we must have $\omega_2<-k$. Since we restrict ourselves to positive $\omega_2$, the pinch singularity from (\ref{pin2site}) is not visible in the complex $\omega_1$ plane. This is linked to the fact that the integral has a finite value at $\omega_1=\omega_2$ when $\omega_2>-k$. 
\end{itemize}


\paragraph{Example: Massive two-vertex integral.}
The boundary of integration changes when the internal lines become massive. Consider the case where both internal lines have the same mass. The integral becomes:

\begin{equation}
    \psil{1}_{2}(\omega_1,\omega_2)=\frac{1}{8\pi^2(\omega_1+\omega_2)k}\int_{\sqrt{k^2+4m^2}}^{\infty}d\Omega_+\int_{-k\delta}^{k\delta}d\Omega_-\frac{\Omega_+^2-\Omega_-^2}{(\omega_1+\omega_2+\Omega_++\Omega_-)(\omega_1+\Omega_+)(\omega_2+\Omega_+)}.
\end{equation}
Here we have:
\begin{equation}
    \delta=\frac{\sqrt{\Omega_+^2-k^2-4m^2}}{\sqrt{\Omega_+^2-k^2}}.
\end{equation}
The details on how to obtain and evaluate this integral are given in App.~\ref{App:A}. Despite appearances, the boundary of the integration contour is described by only one equation:
\begin{equation}
    \tilde{S}=(\Omega_+^2-k^2)(\Omega_--k^2)+4m^2k^2=0.
\end{equation}
Here we have three singularity surfaces:
\begin{align}
    &S_1=\omega_1+\omega_2+\Omega_++\Omega_-=0,\\
    &S_2=\omega_1+\Omega_+,\\
    &S_3=\omega_2+\Omega_+.
\end{align}
Going through the Landau analysis again gives us the end-point singularity from a single surface:
\begin{align}
    &S_2=\tilde{S}=0\Rightarrow\omega_1=-\sqrt{k^2+4m^2},\\
    &S_1=\tilde{S}=0\Rightarrow\omega_1+\omega_2=-2m.
\end{align}
We also have an end-point singularity from two surfaces:
\begin{equation}
    S_1=S_2=\tilde{S}=0\Rightarrow\omega_1=-k\frac{\sqrt{\omega_2^2-k^2-4m^2}}{\sqrt{\omega_2^2-k^2}}.\label{2pinch}
\end{equation}
Here I have picked the negative solution so that $S_1$ can be satisfied, since $\Omega_+$ is positive. However, this singularity is in fact spurious. To see that this is indeed the case, notice that the constraints $S_1=S_2=0$ imply
\begin{equation}
    \omega_2+\Omega_-=0.
\end{equation}
Since $|\Omega_-|\leq k\delta\leq k$, we have $\omega_2\leq k$ in order for the singularity to appear. In fact one can check that $\omega_2<k$ unless we take $\omega_1^2\rightarrow \infty$. However, from the kinematics of the system, $\omega_2\geq k$. Therefore, as long as we restrict ourselves to values of $\omega_2$ that are physically allowed, we will not encounter this singularity. In App.~\ref{App:A} we will see that this pole is indeed spurious.


\paragraph{Example: Three vertex integral.}
For more complicated graphs it may prove difficult to write down variables like $\Omega_+$ and $\Omega_-$. Therefore it is instructive to understand how to carry out Landau analysis with the loop momentum $\bfp$, and derive the singularities from our general arguments. We will use the three vertex integral as an example. The integral is:
\begin{multline}
        \psil{1}_{3}=\int_\bfp \frac{1}{(\omega_2+\Omega_{q_{12}}+\Omega_{q_{23}})(\omega_2+\omega_3+\Omega_{q_{12}}+\Omega_{q_{31}})}\\
        \times \frac{1}{(\omega_1+\omega_2+\omega_3+2\Omega_{q_{31}})(\omega_1+\Omega_{q_{12}}+\Omega_{q_{31}})(\omega_3+\Omega_{q_{23}}+\Omega_{q_{31}})},
\end{multline}
Here I assume the masses of the internal lines are $m$. Now we will analytically continue in $\bfp$. Since we are integrating over all $\bfp$ there is no boundary to the integration contour. The singularity surfaces are:
\begin{align}
    &S_1=\omega_2+\Omega_{q_{12}}+\Omega_{q_{23}}=0,\\
    &S_2=\omega_2+\omega_3+\Omega_{q_{12}}+\Omega_{q_{31}}=0,\\
    &S_3=\omega_1+\omega_2+\omega_3+2\Omega_{q_{31}}=0,\\
    &S_4=\omega_1+\Omega_{q_{12}}+\Omega_{q_{31}}=0,\\
    &S_5=\omega_3+\Omega_{q_{23}}+\Omega_{q_{31}}=0.
\end{align}
Since $\Omega_{p}=\sqrt{|\bfp|^2+m^2}$, the singularity surfaces are not linear anymore. As a result a single surface can pinch a contour. As an example consider $S_3=0$. The normal vector condition gives:
\begin{equation}
    \frac{\textbf{q}_{31}}{\Omega_{q_{31}}}=\frac{\bfp-\bfk_1}{\Omega_{q_{31}}}=0.
\end{equation}
This is solved by $\bfp=\bfk_1$, and $\Omega_{q_{31}}=m$. Putting this back into $S_3=0$ gives:
\begin{equation}
    \omega_1+\omega_2+\omega_3=-2m.
\end{equation}
Similarly, let us write down the rest of the single pinch singularities:
\begin{align}
    &S_1=0\Rightarrow\omega_2+\sqrt{|\bfk_1|^2+4m^2}=0,\\
    &S_2=0\Rightarrow\omega_2+\omega_3+\sqrt{|\bfk_1|^2+4m^2}=0,\\
    &S_4=0\Rightarrow\omega_1+\sqrt{|\bfk_1|^2+4m^2}=0,\\
    &S_5=0\Rightarrow\omega_3+\sqrt{|\bfk_3|^2+4m^2}=0.
\end{align}
This list of singularity is related to the list produced from our general argument, up to some permutation. Naturally, when we take the massless limit, this simply reproduces the list of singularities from the expression we computed.

\paragraph{Thresholds for massive fields.}

From the recursive relations for the wavefunction (\ref{plytope}), the equations for the singularity surfaces have the following form:
\begin{equation}
S_i=\omega_1+\sum_{e\in E}\omega_e+\sum_{i\in I}c_i\Omega_{i},\label{Si}
\end{equation}
where $E$ is a subset of external legs and $I$ is a subset of internal legs and $c_i$ being either $1$ or $2$. The form of these singularity surfaces comes from the recursion relations in Sec.~\ref{ssec:analytic_review}: each external energy can only appear once within the expression. Internal energy can only appear at most with a factor of $2$, coming from cutting a loop diagram. 

We will now show the following for massive fields:
\begin{tcolorbox}
    \begin{quote}
        \emph{Landau conditions for wavefunction coefficients}
        
        Given (\ref{Si}), the singularities corresponding to the energy-conservation condition are found by solving:
        \begin{align}
             &\quad\quad\quad S_i=0,\\
             &\sum_{i\in I}c_i\frac{\partial \Omega_{i}}{\partial \bfp_l}=0, \quad\forall \bfp_l\in \{\bfp\}.
        \end{align}
    \end{quote}
\end{tcolorbox}

We will analytically continue in the loop momentum $\bfp_l\in\{\bfp\}$ and carry out the Landau analysis. Since the numerator in the expression (\ref{plytope}) is a sum of products of $S_i$ and possibly powers of the momenta from spatial derivative interactions, to find singularities we can simply focus on the denominators,
i.e., we simplify $\psi_n$ until the numerator no longer contains any factor of $S_i$, we carry out the Landau analysis term by term and finally we sum over all the singularities of each individual term.

The full list of singularities from the Landau analysis includes:
\begin{itemize}
\item Endpoint singularities. Since we are integrating over all $\bfp_l$, there is no boundary for the hypercontour. As a result we cannot have endpoint singularities, in these integration variables.
\item Pinching the contour with a single surface. By considering the cutting procedure in (\ref{ssec:analytic_review}), it is clear that each $S_i$ corresponds to the total energy entering a subgraph of the Feynman diagram, and so solving $S_i=0$ corresponds to energy conservation. In terms of the components of the loop momentum $\bfp_l$, the normal vector condition reads:
\begin{equation}
\frac{\partial S}{\partial \bfp_l}=0\Rightarrow \frac{\partial }{\partial \bfp_l}\sum_{i\in I}c_i\Omega_i=0. \label{extcondition}
\end{equation}
Unlike the amplitude case, this equation can be consistent with setting only one $S_i=0$ \footnote{For amplitudes $S_i=p_i^2-m_i^2=0$, and $\frac{\partial S_i}{\partial \bfp_l}=2p_i=0$. The normal vector condition requires $p_i=0$ which is not possible if $S_i=0$ as well.}. In fact this equation is equivalent to extremizing $\sum_{i\in I}c_i\Omega_i$ with respect to the loop momentum $\boldsymbol{p}_l$. Observe that for an arbitrary $\bfk$,
\begin{equation}
    \frac{\partial^2 \Omega_{|\bfp+\bfk|}}{\partial p_i\partial p_j}=\frac{\delta_{ij}}{\Omega_{|\bfp+\bfk|}}-\frac{(p+k)_i (p+k)_j}{\Omega_{|\bfp+\bfk|}^3}.
\end{equation}
Since $\Omega_p>p^2\geq 0$ this is a positive definite matrix. When we take the second derivative of $\sum_{i\in I}c_i\Omega_i$ we simply get a sum of positive definite matrices (with positive coefficients), and the resulting matrix is also positive definite. Therefore when we solve (\ref{extcondition}), the solution corresponds to a minimum.

Therefore, $S_i=0$ satisfies:
\begin{equation}
\omega_1=-\sum_{e\in E}\omega_e-\min_{\boldsymbol{p}_l}\sum_{i\in I} c_i\Omega_i,
\end{equation}
where the minimization is with respect to all $\bfp_l\in\{\bfp\}$. This is exactly the type of singularities obtained by the energy-conservation condition in Sec.~\ref{ssec:analytic_heuristic}. 
\item Pinches from multiple surfaces. Given two singularity surfaces $S_1$ and $S_2$ with subset of external legs $E_1$ and $E_2$, either $E_1\subseteq E_2$ or $E_2\subseteq E_1$. This comes from the cutting procedure described in Sec.~\ref{ssec:analytic_review}: the set of external vertices in a cut diagram must be smaller after every cut. This property of $S_i$ gives us a nice picture of what a multiple surface pinch would mean: the energy going into part of a diagram vanishes, and simultaneously the energy going into a subset of the diagram also vanishes.

At tree level, this type of singularity doesn't give us new poles in $\omega_1$. Instead, it tells us about non-analyticity in the other $\omega_e$. This is because we can just take one of the singularity surfaces, say $S_1$, to write down:
\begin{equation}
    \omega_1=-\sum_{e\in E_1}\omega_e-\sum_{i\in I_1}c_i\Omega_i.
\end{equation}
This can then be used to remove any $\omega_1$ dependence from the rest of the expression. Any further non-analyticities of the expression are a result of analytically continuing the other $\omega_e$ from their physically allowed values.

In the case of a two-vertex loop with massive fields, we used the Landau analysis and found that it can only occur for unphysical values of $\omega_2$, and by explicitly computing the integral we found that the singularity is indeed spurious. 

However, it is difficult to prove that these multiple surface pinches are always spurious. In amplitudes, when we look at more complicated graphs, such as the three-vertex graph, we discover that for certain external kinematics there are singularities known as anomalous thresholds. It might be possible that by looking at multiple surface pinches for the three-vertex graph we may discover new singularities similar to these anomalous thresholds, however we leave this for future work.

\end{itemize}

Using Landau analysis, we have successfully derived the list of singular points from our physical argument in (\ref{ssec:analytic_heuristic}). Once again, the full list of singularities from Landau analysis is over-complete, but these extra poles are (likely) removable by considering the external kinematics.

\paragraph{Thresholds for massless fields.}
If we attempt to directly extend the proof above to the case of massless particles, we run into the following issues even for single pinch singularities:
\begin{itemize}
    \item Since $\Omega_{p_l}=|\bfp_l|$, when we take derivative to obtain the normal vector condition, we get $\frac{\partial}{\partial \bfp_l}\Omega_{|p_l|}=\frac{\bfp_l}{|\bfp_l|}$. For $\bfp_l=0$ this is ill-defined. To deal with this problem one needs to regulate $\Omega_{p_l}$ properly. An example would be introducing artificial boundaries of integration so that $|\bfp_l|$ never reaches zero, for example:
    \begin{equation}
        \tilde{S}=|\bfp_l|-\epsilon=0,
    \end{equation}
    then take $\epsilon$ to zero. However, doing this procedure also introduces spurious singularities, such as the $\omega_1+\omega_2=-2k$ pole found in the two-vertex example.
    \item When we take the second derivative of $\Omega_{|\bfp+\bfk|}$, the result is:
    \begin{equation}
         \frac{\partial^2 |\bfp+\bfk|}{\partial p_i\partial p_j}=\frac{\delta_{ij}}{|\bfp+\bfk|}-\frac{(p+k)_i (p+k)_j}{|\bfp+\bfk|^3}.
    \end{equation}
    This is only positive semi-definite, rather than positive definite. Therefore, the solution may not be a minimum.
\end{itemize}
For the massless case, the Landau analysis would provide us with a list that matches our physical intuition in (\ref{ssec:analytic_heuristic}), with some extra singularities. This is analogous to the case in amplitudes, where we get extra soft/collinear singularities for massless particles. The $\omega_1+\omega_2=-2k$ pole in the two-vertex example above is one such singularity. In that case the residue is zero, so it does not give rise to new poles or branch cuts.

In the wavefunction coefficients we computed explicitly, these extra singularities always cancels. We believe this to be the case for any wavefunction coefficients for massless particle, however we will leave the proof of this statement for future research.

 
\section{UV/IR sum rules}
\label{sec:dispersion}

In this section, we derive a \emph{dispersion relation} for the (off-shell) wavefunction coefficients by analytically continuing one of the energies while keeping all remaining kinematics fixed. This dispersion relation can be used to fix the low-energy expansion of the wavefunction in terms of precise integrals of (the discontinuity of) the complete high-energy wavefunction, hence providing a set of UV/IR sum rules for the wavefunction. 
We provide explicit expressions for these sum rules for all Wilson coefficients up to mass-dimension-8 in the effective field theory of a single scalar field. 
Finally, we contrast and compare these results with the sum rules that one would obtain from the study of scattering amplitudes. 


\subsection{The effective field theory wavefunction}
 
Our goal is to use analyticity as a bridge between the low-energy effective field theory (EFT) and its underlying UV-completion. 
As a first step then, we should specify precisely what we mean by a low-energy EFT in the language of wavefunction coefficients. 

Recall that for amplitudes, given a bulk light field $\Phi$ and a bulk heavy field $X$, amplitudes for $\Phi$ can be computed using the generating functional:
\begin{equation}
    Z[J]=\int [d\Phi] [dX]\, e^{i S_{\text{UV}}[\Phi,X]+i\int_{x} J(x)\Phi(x)}.
\end{equation}
The heavy field can be integrated out by the following procedure to obtain action for the EFT:
\begin{equation}
    e^{iS_{\text{EFT}}[\Phi]}=\int [dX]\, e^{i S_{\text{UV}}[\Phi,X]}.
\end{equation}
We can carry out a similar procedure for the wavefunction. Consider the wavefunction specified by the path integral for the fields $\Phi$ and $X$. The boundary conditions to the past correspond to the Bunch-Davies vacuum. The path integral is a functional of the field boundary conditions to the future $\Phi(t_\ast)=\phi$ and $X(t_\ast)=\chi$: 
\begin{equation}
    \Psi[\phi,\chi;t_\ast]=\int_{BD}^{\Phi(t_\ast)=\phi}[d\Phi]\,\int_{BD}^{X(t_\ast)=\chi}[dX]e^{S_{\text{UV}}[\Phi,X;t_\ast]},
\end{equation}
where for some Lagrangian $\mathcal{L}$ we defined:
\begin{align}
    S_{\text{UV}}[\Phi,X;t_\ast]=\int_{-\infty}^{t_\ast} dt \, \mathcal{L}[\Phi,X]\,.
\end{align}
The path integral is then the transition amplitude between the Bunch-Davies vacuum $\ket{\text{BD}}_{\Phi}\otimes\ket{\text{BD}}_{X}$ and the field eigenstate $\ket{\phi}_{\Phi}\otimes\ket{\chi}_{X}$. To define the EFT wavefunction we focus on this wavefunction:
\begin{equation}
    \Psi[\phi,0;t_\ast]=\int_{BD}^{\Phi(t_\ast)=\phi}[d\Phi]\,\int_{BD}^{X(t_\ast)=0}[dX]e^{S_{\text{UV}}[\Phi,X;t_\ast]},\nonumber
\end{equation}
i.e. the wavefunction with the heavy field set to zero at $t=t_\ast$ \footnote{One may ask whether this is the most physically relevant quantity to compute. For example, in an EFT with cutoff $\Lambda$ we might have to average the value of wavefunction in an interval $t^\ast\pm \Lambda^{-1}$. We postpone this issue to the conclusions and to future work.}, and compute its wavefunction coefficients in powers of $\phi$. The coefficients of the perturbative expansion of $\Psi[\phi,0;t_\ast]$ in powers of $\phi$ computed with the interactions of the UV action $S_{\text{UV}}[\Phi,X;t_\ast]$ is what we call the UV wavefunction coefficients $\psi_{\text{UV}}$: 
\begin{equation}
\begin{split}
\Psi[\phi,0;t_\ast]&=\int_{BD}^{\Phi(t_\ast)=\phi}[d\Phi]\,\int_{BD}^{X(t_\ast)=0}[dX]e^{S_{\text{UV}}[\Phi,X;t_\ast]}\\
&=\exp\left[ +  \sum_{n}^{\infty}\frac{1}{n!} \int_{\bfk_{1},\dots\bfk_{n}}\,(2\pi)^3 \delta^{(3)} \left( \sum_{a}^{n} \bfk_{a} \right) \psi_{\text{UV}}^{(n)}(\{\bfk\};t_\ast)  \phi(\bfk_{1})\dots \phi(\bfk_{n})\right]\,.
\end{split}
\end{equation}
For this wavefunction, we can write down an EFT just like for amplitudes:
\begin{equation}
    e^{iS_{\text{EFT}}[\Phi;t_\ast]}=\int_{BD}^{X(t_\ast)=0} [dX]\,e^{S_{\text{UV}}[\Phi,X;t_\ast]}.
\end{equation}
Crucially, this $S_{\rm EFT} [ \Phi , t_* ]$ can be expanded as a series of local interactions for $\Phi$ and its derivatives, which at low energies (small derivatives) can be 
 truncated at a finite nuber of terms. 
The EFT wavefunction coefficients $\psi_{\text{EFT}}(\omega_{i},\{\textbf{k}_{j}\})$ are then obtained from expanding in powers of $\phi$ the wavefunction $\Psi[\phi,0;t_\ast]$ computed from the truncated EFT action:
\begin{equation}
\begin{split}
\Psi[\phi,0;t_\ast]&=\int_{BD}^{\Phi(t_\ast)=\phi}[d\Phi]\, e^{iS_{\text{EFT}}[\Phi;t_\ast]}\\
    &=\exp\left[ +  \sum_{n}^{\infty}\frac{1}{n!} \int_{\bfk_{1},\dots\bfk_{n}}\,(2\pi)^3 \delta^{(3)} \left( \sum_{a}^{n} \bfk_{a} \right) \psi_{\text{EFT}}^{(n)}(\{\bfk\};t_\ast)  \phi(\bfk_{1})\dots \phi(\bfk_{n})\right]\,.
\end{split}
\end{equation}
At low energies, where the truncation made in $S_{\rm EFT}$ is valid, the $\psi_{\rm EFT}$ coefficients coincide with the true $\psi_{\rm UV}$ coefficients. 
As the energy is increased, eventually the derivative expansion in $S_{\rm EFT} [ \Phi ; t_*]$ breaks down and one must ``UV complete'' the EFT by returning to the $S_{\rm UV} [ \Phi , X ; t_* ]$---physically this corresponds to having enough energy to excite heavy $X$ fluctuations (which are then not faithfully captured by an $S_{\rm EFT}$ involving only light degrees of freedom).  

There is a crucial difference between this EFT for the wavefunction and the EFT for amplitudes: the boundary condition for the fields are different. This comes from the new $t=t_\ast$ boundary we introduced for the wavefunction. 
As a result we need to keep track of both total derivatives in time and of terms proportional to the equations of motion. To this end we separate the EFT action into a bulk and a boundary term localized at $t_\ast$ :

\begin{equation}\label{EFTaction}
    S_{\text{EFT}}[\Phi;t_{\ast}]=\int_{-\infty}^{t_{*}} dt \int d^{3}\textbf{x}(\mathcal{L}_{\text{EFT}}^{\text{bulk}}[\Phi,\partial_{\mu}]+\partial_t  \mathcal{L}_{\text{EFT}}^{\text{boundary}}[\phi,\partial_{\mu},\partial_{t}]).
\end{equation}
All total derivatives in time have been collected into $\partial_t \mathcal{L}_{\text{EFT}}^{\text{boundary}}$, so that $\mathcal{L}_{\rm EFT}^{\rm bulk}$ is constructed in the usual way (with the freedom to integrate by parts). 
Notice that our formalism applies to both Lorentz-invariant theories as well as to theories that break boosts, explicitly or spontaneously. 
For concreteness our examples will include Lorentz invariant interactions in the bulk and non-boost invariant time and spatial derivatives will appear only in $ \mathcal{L}_{\text{EFT}}^{\text{boundary}}$.


\subsection{A wavefunction dispersion relation}

We have now defined the low-energy EFT approximation, $\psi_{\rm EFT}$, to the full wavefunction coefficient. 
The question which we wish to address next is: what information about the underlying UV physics can be gleaned from a measurement/calculation of this EFT object?

\paragraph{Analyticity and the dispersion relation.}
The analytic properties that we developed in~\ref{sec:analytic} can give a very concrete answer to this question.
The reason is that complex analytic functions are very constrained: they are all but completely fixed by their singularities and asymptotics via Cauchy's theorem. One can use said theorem to recover the value of a function $f(z)$ at a point $z=z_{0}$ using a closed contour integral:
\begin{equation}\label{Cauchy}
    f(z_{0})=\frac{1}{2\pi i}\oint_{C}\frac{f(z)}{z-z_{0}}.
\end{equation}
where $C$ is a counter-clockwise contour around the pole at $z=z_0$ (and contains no further singularities). 

\begin{figure}[ht]
    \centering
    \includegraphics[width=0.7\textwidth]{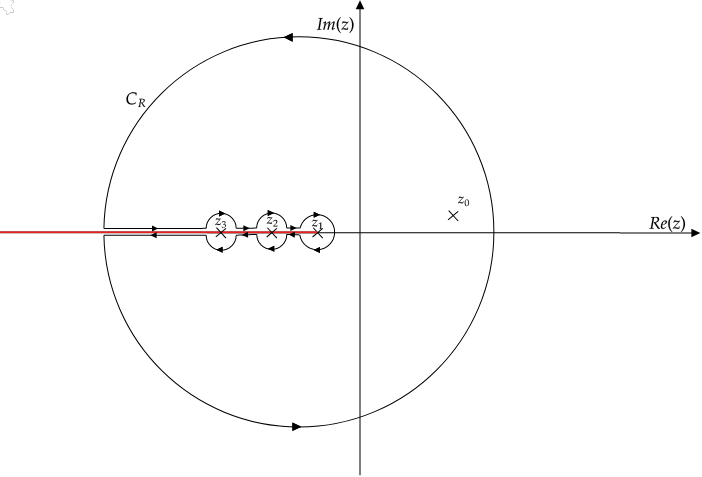}
    \caption{Deformed contour for dispersion relations. There is a branch point at $z=z_{1}$ and a branch cut on the negative real axis. There are isolated poles at $z=z_{2}$ and $z=z_{3}$. }
    \label{ComplexPlane}
\end{figure}

Now imagine we expand $C$ until we start intersecting poles and branch cuts of $f(z)$. We need to deform the contour to properly defined the integral (\ref{Cauchy}). In particular, we have three different contributions to the deformed contour $C_{R}$ as shown in Fig.~\ref{ComplexPlane}:
\begin{itemize}

    \item[(i)] The isolated poles $z_{i}$ of $f(z)$. The deformed contour wraps clockwise around the poles of $f(z)$. 

    \item[(ii)] The branch cuts of $f(z)$. The contour runs above and below the branch cut, and is therefore proportional to the \emph{discontinuity} of the function along the cut, where,
    \begin{align}
\text{disc} \, f (z ) = \lim_{\epsilon \to 0} \left[ f (z + i \epsilon ) - f (z - i \epsilon )  \right] \; . 
    \end{align}
    
    \item[(iii)] the arc at infinity $C_{R}$. Once the contour is made arbitrarily large, we can identify this contribution with the residue of the pole at infinity.
\end{itemize}
This expresses the right-hand side of (\ref{Cauchy}) as a sum of three terms:
\begin{equation}\label{disrel}
    f(z_{0})= - \sum_{z_i} \underbrace{\underset{z=z_{i}}{\text{Res}}\left(\frac{f(z)}{z-z_{0}}\right)}_{\text{isolated poles}}+ \underbrace{\int\frac{dz}{2\pi i}\frac{\text{disc}(f(z))}{z-z_{0}}}_{\text{branch cut}} 
    + \underbrace{\underset{\infty}{\text{Res}}\left(\frac{f(z)}{z-z_{0}} \right)}_{\text{Pole at infinity}}.
\end{equation}
Note that if one defines the ``discontinuity'' of an isolated pole as $\text{disc} \left( 1/ (z - z_i ) \right) = - 2 \pi i \delta ( z - z_i )$, then the first term can be absorbed into the second. 

\paragraph{Wavefunction dispersion relation.}
We can apply these considerations to the off-shell wavefunction coefficients $\psi_n ( \{ \omega \} , \{\textbf{k} \})$ taken as analytic functions of a single complex variable $\omega_{1}$, while holding all the other kinematics fixed. As we have discussed, singularities in $\psi_n ( \{ \omega \} , \{ \textbf{k} \} )$ can only exist on the negative real axis of $\omega_{1}$. This allows us to write (\ref{disrel}) as:
\begin{tcolorbox}
\begin{equation}\label{wtdisres}
    \omega_T  \psi(  \{ \omega \} ,\{\textbf{k} \}) \big|_{\omega_1 = \omega_1'} 
    =
    \int_{-\infty}^{0}\frac{d\omega_1}{2\pi i}\frac{\text{disc}\left( \omega_{T}\psi( \{ \omega \} , \{ \textbf{k} \} ) \right)}{\omega_1 -\omega_{1}' }+\underset{\omega_1 = \infty}{\text{Res}} \left(\frac{\omega_{T}\psi_n ( \{ \omega \} , \{ \textbf{k} \})}{\omega_1 - \omega_{1}' } \right).
\end{equation}
\end{tcolorbox}
\noindent This is our central application of analyticity: it allows us to connect every Wilson coefficient appearing in $\mathcal{L}_{\rm EFT}$ to an integral over the wavefunction of the underlying UV theory.

\paragraph{UV/IR sum rules.}
The idea is to expand the left-hand-side of (\ref{wtdisres}) at low energy/momenta, where it can be computed using the low-energy $\mathcal{L}_{\rm EFT}$. 
This expansion can then be matched, order-by-order, to particular high-energy integrals on the right-hand-side.
To this end, we define the UV integral,
\begin{equation}\label{IRUVsumrule2}
    \mathcal{I}_{\rm UV}^{(N)} ( \{ \omega_{a\neq1} \} , \{\textbf{k} \} ) = \int_{-\infty}^{0}\frac{d\omega_1}{2\pi i}\frac{\text{disc} \left[ \omega_{T}\psi( \{ \omega \} , \{\textbf{k}\} ) \right]}{\omega_1^{N+1}}+\underset{\omega_1 = \infty}{\text{Res}} \left( \frac{\omega_T \psi ( \{ \omega\} , \{\textbf{k} \} ) }{\omega_1^{N+1}} \right)\,.
\end{equation}
The low-energy expansion of \eqref{wtdisres} can then be written compactly as,
\begin{equation}\label{IRUVsumrule}
    \frac{1}{N!} \partial_{\omega_{1}}^{N}(\omega_{T}\psi_{\text{EFT}})  \big|_{\omega_1 = 0} =\mathcal{I}^{(N)}_{\text{UV}}( \{ \omega_{a\neq1} \} , \{\textbf{k} \} ) \; .
\end{equation}
Since the EFT does not distinguish between $\omega_1$ and the other kinematic variables, this equation should be further expanded in powers of each of the $\omega_{a \neq 1}$ and $\bfk_a$. 
In the next subsection, we explicitly construct the EFT for the wavefunction of a scalar field up to fourth order in derivatives, and show how \eqref{IRUVsumrule} relates each Wilson coefficient to a particular UV integral, $\mathcal{I}_{\rm UV}^{(N)}$.

 
\subsection{Example: a light scalar}

Let us illustrate these new sum rules using the EFT for a light scalar field $\Phi$ on a fixed Minkowski background.

\paragraph{An EFT basis for quartic interactions.}
Once total derivatives and terms proportional to the free equations of motion are included, the list of possible interactions grows rapidly with increasing mass-dimension. 
Rather than construct the most general possible EFT, we will focus on a particular subset of interactions which illustrates our sum rules simply and yet remains general enough to capture simple tree-level UV completions (of which we give an example in Sec.~\ref{sec:UV_examples}).
Firstly, we focus on quartic interactions: these would be the leading interactions in any theory with an approximate $\mathbb{Z}_2$ symmetry, $\Phi\to -\Phi$. 
We truncate the EFT at mass-dimension-8, which means we only include interactions with up to four (three) derivatives in the bulk (boundary) Lagrangian, and further assume that Lorentz symmetry is broken only by the boundary interactions. 
Finally, we focus on specific interactions of the factorised form $\mathcal{D}_1 \Phi^2 \mathcal{D}_2 \Phi^2$, where $\mathcal{D}_1$ and $\mathcal{D}_2$ are differential operators. 
Altogether, this gives the following EFT basis of interactions:
 \begin{align}\label{EFTvertices}
    \mathcal{L}_{\text{EFT}}^{\text{bulk}}[\Phi,\partial_{\mu}]&\supset\frac{\alpha_{0}}{4!}\Phi^{4}+\frac{\alpha_{2}}{4}\Phi^{2}\Box\Phi^2+\frac{\alpha_{4}}{{4} }\Phi^{2}\Box^{2}\Phi^2+\mathcal{O}(\partial^{6})\,,  \nonumber \\
    \mathcal{L}_{\text{EFT}}^{\text{boundary}}[\Phi(t_{*}),\partial_{\mu},\partial_{t}]&\supset \frac{\beta_{00}}{4!}\Phi^{4}(t_{*})-\frac{\beta_{11}}{4}\Phi^{2}(t_{*})\partial_{t}\Phi^{2}(t_{*})+\frac{\beta_{20}}{4}\Phi^{2}(t_{*})\Box\Phi^{2}(t_{*})-\frac{\beta_{22}}{4}\Phi^{2}(t_{*})\partial_{i}^{2}\Phi^{2}(t_{*})+  \nonumber \\
    &-\frac{\beta_{31}}{4}\Phi^{2}(t_{*})\partial_{t}\Box\Phi^{2}(t_{*})-\frac{\beta_{31}^{'}}{4}\Box\Phi^{2}(t_{*})\partial_{t}\Phi^{2}(t_{*})+\mathcal{O}(\partial^{4}),
 \end{align}
where the $\alpha_{a}$'s are the free Wilson coefficients of bulk interactions and the $\beta_{ab}$'s are free Wilson coefficients of boundary interactions. The first label on $\beta$ counts the total number of derivatives, while the second counts the number of derivatives that are not Lorentz invariant (such as $\partial_t$ and $\partial_i^2$). 

\paragraph{EFT wavefunction.}
We can use (\ref{EFTvertices}) to compute the tree-level four-point wavefunction coefficient up to fourth order in the momenta/energy. Even though we consider contact interactions, it is convenient to separate contributions into $s$, $t$ and $u$ ``channels'' according to the partial energies on which they depend:
\begin{equation}\label{EFTwavefunction}
    \psi_{\text{EFT}}(  \{ \omega \} , \{ \textbf{k} \} 
    )=\delta^{d}(\textbf{k}_{T})(\psi'_{\text{EFT}}(\omega_{12},\omega_{34},\textbf{k}_{s})+\psi'_{\text{EFT}}(\omega_{13},\omega_{23},\textbf{k}_{t})+\psi'_{\text{EFT}}(\omega_{14},\omega_{24},\textbf{k}_{u})) \, .
\end{equation}
An explicit calculation gives:
\begin{equation}
\begin{split} 
\psi'(\omega_{12},\omega_{34},\textbf{k}_{s}) &=\frac{1}{\omega_{T}} \left[\frac{1}{3}\alpha_{0}+\alpha_{2}(s_{12}+s_{34})+\alpha_{4}(s_{12}^{2}+s_{34}^{2}) \right]+\\
    &+\frac{i}{3}\beta_{00}+\beta_{11}\omega_{T}+i\beta_{20}(s_{12}+s_{34})+i2\beta_{22}\bfk_s^2+\\
    &+\beta_{31}(s_{12}\omega_{12}+s_{34}\omega_{34})+\beta_{31}'(s_{12}\omega_{34}+s_{34}\omega_{12})+\mathcal{O}(p^{5}).\label{EFTbasis}
\end{split}
\end{equation}
We have defined:  
\begin{align}
\textbf{k}_{s}&=\textbf{k}_{1}+\textbf{k}_{2}\,, &\textbf{k}_{t}&=\textbf{k}_{1}+\textbf{k}_{3}\,, &\textbf{k}_{u}&=\textbf{k}_{1}+\textbf{k}_{4}\,, \nonumber \\
    \omega_{ij}&=\omega_{i}+\omega_{j}\,, &\textbf{k}_{ij}&=\textbf{k}_{i}+\textbf{k}_{j}\,, & s_{ij}&=\omega_{ij}^{2}-\textbf{k}_{ij}^{2}\,.
\end{align}

\paragraph{Sum rules.}
The sum rules in \eqref{IRUVsumrule} can be used to fix each of the low-energy Wilson coefficients in \eqref{EFTvertices} in terms of an integral over the UV completion of the EFT. 
Concretely, proceeding order-by-order in derivatives:
\begin{itemize}

\item At mass-dimension-4, there is a single bulk interaction in $\mathcal{L}_{\rm EFT}$ with Wilson coefficient $\alpha_0$. The corresponding sum rule follows from evaluating \eqref{wtdisres} at all $\omega_a= k_a = 0$,
\begin{align}
\alpha_{0}=\mathcal{I}_{\rm UV}^{(0)} \left(  \{ \omega \} ,  \{\textbf{k} \} \right) \big|_{\substack{\omega_{a}=0 \\ \textbf{k}_{a}=0}},\label{0sumrules}&
\end{align}

\item At mass-dimension-5, there is a single boundary interaction in the EFT, with coefficient $\beta_{00}$. The corresponding sum rule follows from evaluating the $\partial_{\omega_1}$ of \eqref{wtdisres} at all $\omega_a = k_a = 0$,
\begin{align}
    i\beta_{00} = \mathcal{I}_{\rm UV}^{(1)}  \left(  \{ \omega \} ,  \{\textbf{k} \} \right) \big|_{\substack{\omega_{a}=0 \\ \textbf{k}_{a}=0}} ,\label{w1sumrules}&
\end{align}

\item At mass-dimension-6, there are two interactions, with Wilson coefficients $\alpha_2$ and $\beta_{11}$. 
The corresponding sum rules follow from evaluating the $\partial_{\omega_1}^2$ and $\partial_{\omega_2} \partial_{\omega_1}$ of \eqref{wtdisres} at all $\omega_a = k_a = 0$,
\begin{align}
   3(\alpha_{2}+\beta_{11}) &=\mathcal{I}_{\rm UV}^{(2)} \left(  \{ \omega \} ,  \{\textbf{k} \} \right) \big|_{\substack{\omega_{a}=0 \\ \textbf{k}_{a}=0}},\label{w1w1sumrules} \\
    2(\alpha_{2}+3\beta_{11}) &=\partial_{\omega_{2}}\mathcal{I}_{\rm UV}^{(1)}  \left(  \{ \omega \} ,  \{\textbf{k} \} \right) \big|_{\substack{\omega_{a}=0 \\ \textbf{k}_{a}=0}},\label{w2w1sumrules}
\end{align}

\item At mass-dimension-7 there are two boundary interactions, with Wilson coefficients $\beta_{20}$ and $\beta_{22}$. 
The corresponding sum rules follow
from evaluating $\partial_{\omega_1}^3$ and $\partial_{\omega_1} \partial_{k_s}^2$ of \eqref{wtdisres} at all $\omega_a = k_a = 0$,
\begin{align}
    3i\beta_{20} &= \mathcal{I}_{\rm UV}^{(3)} \left(  \{ \omega \} ,  \{\textbf{k} \} \right) \big|_{\substack{\omega_{a}=0 \\ \textbf{k}_{a}=0}} ,\label{w1w1w1sumrules}  \\
    -4i(\beta_{20}-\beta_{22}) & =\partial_{k_{s}}^{2}\mathcal{I}_{\rm UV}^{(1)} \left(  \{ \omega \} ,  \{\textbf{k} \} \right) \big|_{\substack{\omega_{a}=0 \\ \textbf{k}_{a}=0}} 
 ,\label{w1kskssumrules}
\end{align}

\item At mass-dimension-8, there are three EFT interactions, with Wilson coefficients $\{ \alpha_4, \beta_{31} , \beta_{31}' \}$. The corresponding sum rules follow from
evaluating $\partial_{\omega_1}^4$, $\partial_{\omega_1}^2 \partial_{\omega_2} \partial_{\omega_3}$ and $\partial_{\omega_1} \partial_{\omega_2} \partial_{\omega_3}  \partial_{\omega_4}$ of \eqref{wtdisres} at all $\omega_a = k_a =0$, 
\begin{align}
    3(\alpha_{4}+\beta_{31})
    &=\mathcal{I}_{\rm UV}^{(4)} \left(  \{ \omega \} ,  \{\textbf{k} \} \right) \big|_{\substack{\omega_{a}=0 \\ \textbf{k}_{a}=0}} ,\label{w1w1w1w1sumrules} \\
    6\beta_{31}+10\beta_{31}'
    &=\partial_{\omega_{2}}\partial_{\omega_{3}}\mathcal{I}_{\rm UV}^{(2)} \left(  \{ \omega \} ,  \{\textbf{k} \} \right) \big|_{\substack{\omega_{a}=0 \\ \textbf{k}_{a}=0}} ,\label{w1w1w2w3sumrules} \\
    24\beta_{31}' 
    & =\partial_{\omega_{2}}\partial_{\omega_{3}}\partial_{\omega_{4}}\mathcal{I}_{\rm UV}^{(1)} \left(  \{ \omega \} ,  \{\textbf{k} \} \right) \big|_{\substack{\omega_{a}=0 \\ \textbf{k}_{a}=0}}  . \label{w1w2w3w4sumrules} 
\end{align}

\end{itemize}

\noindent These sum rules can be similarly applied to any desired order in the EFT expansion, determining every Wilson coefficient in $\mathcal{L}_{\rm EFT}$. 
Note in particular that the boundary interactions and the bulk interactions contribute on an equal footing. 
For instance, the sum rule (\ref{w1w1sumrules}) can only unambiguously fix the bulk Wilson coefficient once supplemented with \eqref{w2w1sumrules}---generally at a given order in derivatives one requires all independent sum rules in order to solve for a particular Wilson coefficient.

 
\subsection{Comparison with amplitude sum rules}

Amplitude sum rules have been extensively studied in the  literature and have proven to be very useful in the study of EFTs. However, the amplitudes sum rules differ in two fundamental aspect from the wavefunction sum rules we have derived above:
\begin{itemize}
    \item The LSZ formula reduces the number of possible EFT interaction vertices in the bulk. For example terms like $\Phi^{2}\Box\Phi^{2}$ will not be present as they are proportional to the equations of motion.
    \item The scattering process takes place between asymptotically free past and future states. This means that one can discard total derivative interactions in the bulk, and that the EFT vertices on the time slice $t=t_{*}$ will not play a role. 
\end{itemize}
Therefore, the EFT expansion for amplitudes $S_{\text{EFT}}^{\text{amp}}[\Phi]$ only involves the bulk Lagrangian from (\ref{EFTaction}) action:
\begin{equation}
     S_{\text{EFT}}^{\text{amp}}[\Phi]=\int_{-\infty}^{+\infty}dt\int d^{3} x\,\mathcal{L}_{\text{EFT}}^{\text{bulk}}[\Phi,\partial_{\mu}].
\end{equation}
As a consequence of the above considerations, fewer operators need to be considered, namely
\begin{equation}\label{AmpEFTbasis}
     \mathcal{L}_{\text{EFT}}^{\text{bulk}}[\Phi,\partial_{\mu}]\supset\frac{\alpha_{0}}{4!}\Phi^{4}+\frac{\alpha_{4}}{4}\Phi^{2}\Box^{2}\Phi^2+\mathcal{O}(\partial^{6}) .
\end{equation}
From these interactions we can obtain the leading terms for the $2\to2$ scattering amplitude,
\begin{equation}
    A(s,t)=\alpha_{0}+2\alpha_{4}(s^{2}+t^{2}+u^{2})+\mathcal{O}(\alpha_{6}p^{6}).
\end{equation}
The analyticity for the scattering amplitude $A(s,t)$ in the complex $s$ plane allows us to write a dispersion relation for $A(s,t)$ and its derivatives. Using these we find that:
\begin{equation}
    \alpha_{0} =\int_{-\infty}^{+\infty}\frac{ds}{2\pi i}\frac{\text{disc}(A(s,t=0))}{s}+\underset{\infty}{\text{Res}}\left(\frac{\text{disc}(A(s,t=0))}{s}\right),
\end{equation}
\begin{equation}\label{ampsumrulea4}
    4\alpha_{4} =\int_{-\infty}^{+\infty}\frac{ds}{2\pi i}\frac{\text{disc}(A(s,t=0))}{s^{3}}+\underset{\infty}{\text{Res}}\left(\frac{\text{disc}(A(s,t=0))}{s^{3}}\right).
\end{equation}
Therefore, we can see that the amplitudes' sum rules only capture a reduced set of the possible EFT interactions. Not every interaction that contributes to the wavefunction may appear in the scattering amplitude, whilst all interactions that contribute to the amplitude do appear in the wavefunction coefficients\footnote{
This is necessary since in the limit $\omega_{T} \to 0$ the Minkowski wavefunction coefficients coincide with scattering amplitudes.
}. 

In the case of boundary interactions, the best case is the comparison of the sum rules for $\alpha_{4}$ from amplitudes (\ref{ampsumrulea4}) with that from wavefunction coefficients (\ref{w1w1w1w1sumrules}). Whilst the amplitude sum rule only includes information about the EFT interaction $\Phi^{2}\Box^{2}\Phi^{2}$, the wavefunction one also includes information about $\Phi^{2}\partial_{t}\Box^{2}\Phi^{2}|_{t=t_{*}}$.
For interactions proportional to the equations of motion, like $\Phi^{2}\Box\Phi^{2}$, the amplitudes sum rules are oblivious. It requires us to look at wavefunction sum rules like ($\ref{w1w1sumrules}$) and (\ref{w2w1sumrules}) to constrain the value for the Wilson coefficients of $\Phi^{2}\Box\Phi^{2}$ and $\Phi^{2}\partial_{t}\Phi^{2}|_{t=t_{*}}$.

 
\section{Example UV completions}
\label{sec:UV_examples}

In this section, we evaluate and check our proposed sum rules in a simple UV-completion of the single-scalar low-energy effective theory presented above. We do this both for a tree-level process and for a one-loop process in the UV-completion of effective theory to show cases in which both poles and branch points arise. 


\subsection{UV-completion: a tree-level example}

Consider a toy UV model of two scalars: a light field $\Phi$ and a heavy field $X$, which interact via a coupling of the form $gM\Phi^{2}X$, so that the complete renormalisable Lagragian is:
\begin{equation}\label{UVaction}
    \mathcal{L}_{\text{UV}}[\Phi,X]=-\frac{1}{2}(\partial\Phi)^{2}-\frac{1}{2}m^{2}\Phi^{2}-\frac{1}{2}(\partial X)^{2}-\frac{1}{2}M^{2}X^{2}-gM\Phi^{2}X\,.
\end{equation}
We aim to study the EFT of the light field $\Phi$ at energies well below the mass $M$ of the heavy field $X$. At tree level, it is enough to integrate out the heavy field $X$ using the classical equations of motion and substitute back into the action:
\begin{equation}\label{Xeom}
    (\Box-M^{2})X=gM\Phi^{2}\,.
\end{equation}
The solution in momentum space with boundary conditions $\bar{X}^{\text{in}}_{\textbf{k}}$ in the past and $\bar{X}^{\text{out}}_{\textbf{k}}$ in the future are:
\begin{equation}\label{heavyfield}
    X_{\textbf{k}}^\text{sol}(t)=\frac{f_{k}(t)}{f_{k}(t_{*})}\bar{X}^{\text{in}}_{\textbf{k}}+\frac{f^{*}_{k}(t)}{f^{*}_{k}(t_{*})}\bar{X}^{\text{out}}_{\textbf{k}}-gM\frac{\int_{\textbf{p}}\Phi_{\textbf{p}}(t)\Phi_{\textbf{k}-\textbf{p}}(t)}{\partial_{t}^{2}+\textbf{k}^{2}+M^{2}},
\end{equation}
 where $f_{k}(t)$ are the mode functions of the heavy scalar $X$ with $f_{k}(t)=e^{-i\Omega_k t}/\sqrt{2\Omega_{k}}$ and $\Omega_{k}^{2}=\textbf{k}^{2}+M^{2}$. The first two terms correspond to the homogeneous solution of the equation of motion. The third term corresponds accounts for the coupling to the $\Phi^2(t)$ source. 

 In order to derive the Wilson coefficients from the action we need to substitute the equations of motion for the heavy field into the original action. For generic boundary conditions, i.e. for both amplitudes and wavefunction coefficients, evaluating the action on the $X$ given in \eqref{heavyfield} gives:
 \begin{equation}\label{general}
    S_{\text{EFT}}[\Phi,X^\text{sol}]=\int dt\,d^{3}x\left[-\frac{1}{2}(\partial\Phi)^2-\frac{1}{2}m^{2}\Phi^{2}-\frac{1}{2}g M \Phi^{2}X-\frac12 \partial_\mu(X\partial^\mu X)\right].
 \end{equation}
To make progress we have to choose boundary conditions. These are different depending on whether we discuss amplitudes or wavefunction coefficients. Let's study each case in turn. 

\paragraph{Amplitudes.} 
Amplitudes are related by the LSZ reduction formula to an in-vacuum to out-vacuum Green's function. This choice corresponds to $\bar{X}^{\text{in}}_{\textbf{k}}=0$ and $\bar{X}^{\text{out}}_{\textbf{k}}=0$, and similarly for $\Phi$. This leads to
 \begin{equation}
    X_{\textbf{k}}^\text{sol}(t)=-gM\frac{\int_{\textbf{p}}\Phi_{\textbf{p}}(t)\Phi_{\textbf{k}-\textbf{p}}(t)}{\partial_{t}^{2}+\textbf{k}^{2}+M^{2}}\quad\Rightarrow \quad X^\text{sol}(t,\textbf{k})=\int_{\textbf{k}}e^{i\textbf{k}\textbf{x}}X^\text{sol}_{\textbf{k}}(t)=\frac{gM}{\Box-M^{2}}\Phi^{2}(t,\textbf{x})\,.
 \end{equation}
 When substituted back into the on-shell action the boundary term vanishes and we find
 \begin{equation}
    S^{\text{amp}}_{\text{EFT}}[\Phi,X^\text{sol}(\Phi)]=\int dt\,d^{3}x\left[-\frac{1}{2}(\partial\Phi)^2-\frac{1}{2}m^{2}\Phi^{2}-\frac{1}{2}g^{2}M^{2}\Phi^{2}\frac{1}{\Box-M^{2}}\Phi^{2}\right].
 \end{equation}
This action is clearly non-local. However, in the regime $M^{2}\gg\Box$ we can approximate it as a series of local operators. This leads to the EFT vertices that reproduce the EFT interactions in (\ref{AmpEFTbasis}):
\begin{equation}\label{lowEFT}
    S^{\text{amp}}_{\text{EFT}}[\Phi]=\int dt\, d^{3}x\left[-\frac{1}{2}(\partial\Phi)^2-\frac{1}{2}m^{2}\Phi^{2}+\frac{1}{2}g^{2}M^{2}\sum_{n=0}^{\infty}\Phi^{2} \left( \frac{\Box}{M^{2}} \right)^{n}\Phi^{2}\right].
\end{equation}
 Note $\Phi^{2}\Box\Phi^{2}$ does not contribute to any scattering amplitude as it is proportional to the equations of motion. Therefore, even though it appears in the action it cannot be captured by the amplitude sum rules. For this particular UV-completion, the Wilson coefficients in (\ref{AmpEFTbasis}) can be read off from \eqref{lowEFT}:
 \begin{equation}
     \alpha_{0}=12g^{2}\;,\;\alpha_{2}=\frac{2g^{2}}{M^{2}}\;,\;\alpha_{4}=\frac{2g^{2}}{M^{4}}\;,\;...
 \end{equation}


 \paragraph{Wavefunction coefficients.} For wavefunction coefficients the boundary conditions are different from those of amplitudes because for the ``out'' state we project onto a field eigenstate at some finite time, as opposed to a state of free particles in the infinite future. This leads to additional EFT interactions as we now show. The right boundary conditions are now $\bar  X^\text{in}_\bfk=0$ and $' X^\text{sol}_\bfk(t_\ast)=0$, and hence
\begin{align}\label{chi}
X^\text{sol}(t)&=\bco e^{i\Omega \left( t-t_{\ast} \right)}+gM\frac{\Phi^{2}(t)}{\Box-M^2}\,, & \bco&=-gM\frac{\Phi^{2}(t_{\ast})}{\Box-M^2}\,,
\end{align}
where $  \Omega=\sqrt{k^{2}+M^{2}} $ is fixed by the dispersion relation for the momentum of $  X $ or equivalently $  \Phi^{2} $. Notice that since $X(t)=0$ at $t=t_\ast$ and at $t=-\infty(1-i\e)$, the total derivative term in \eqref{general} vanishes. For the other terms in (\ref{general}) we can split the result into bulk and boundary contributions as we did in (\ref{EFTaction}). This is simplified by the fact that the on-shell action is linear in $X$:
\begin{equation}
S_{\text{EFT}}^{t_{*}}[\Phi]=S_{\text{EFT}}[\Phi,X^\text{sol}]= S^{\text{bulk}}_{\text{EFT}}[\Phi]+S_{\text{EFT}}^{\text{bdy}}[\Phi],
\end{equation}
To compute each term we notice that the relative factor between the kinetic term and the cubic interaction is different in the action \eqref{general} from what appears in the equations of motion \eqref{Xeom}. This means that we have the choice to use the equations of motion for $  X $ to eliminate either the kinetic term or the cubic interaction. Here we choose to eliminate the latter, finding
\begin{align}\label{mine}
  S_{\text{EFT}}[\Phi,X^\text{sol}] \supset \int dt\,d^{3}x\left[-\frac{1}{2}X^\text{sol} (\Box-M^2 ) X^\text{sol}\right].
\end{align} 
This has four contributions from squaring the two terms in $  X $ in \eqref{chi}. The contribution from squaring the second term in \eqref{chi} gives the bulk interactions
\begin{align}
S^{\text{bulk}}_{\text{EFT}}\supset \int d^{3}xdt\, -\frac{1}{2}g^{2}M^{2} \Phi^{2}\frac{1}{\Box-M^2} \Phi^{2}\,,
\end{align}
where $  \Box-M^2 $ cancelled out with $  (\Box-M^2)^{-1} $.
Now notice that the first term in \eqref{chi} is a solution of the homogeneous equations of motion and so it is annihilated by $  \Box-M^2 $. Hence, of the remaining three terms the only survivor is the one where $  \Box-M^2 $ hits $  \Phi^{2} $,
\begin{align}
S_{\text{EFT}}^{\text{bdy}}[\Phi]= -\int d^{3}xdt\, \frac{1}{2}\bco e^{i\Omega \left( t-t_{\ast} \right)}(\Box-M^2) gM\frac{\Phi^{2}}{\Box-M^2}\,.
\end{align}
One could choose to simplify $  \Box-M^2 $ but then one has to compute the time integral. Instead here we integrate by part twice to move $  \Box-M^2 $ onto the first factor. Since again that factor is annihilated by $  \Box-M^2 $ the only contribution comes from the boundary terms in the first and second integration by parts in time. They combine into
\begin{align}
S_{\text{EFT}}^{\text{bdy}}&= \int d^{3}x\,\frac{1}{2}g M \bco  \left(  \partial_{t}-i\Omega\right)\frac{\Phi^{2}(t_{\ast})}{\Box-M^2}\\
&= \int d^{3}x\,\frac{ig^2 M^2}{2} \frac{\Phi^{2}(t_{\ast})}{\Box-M^2} \left(  i\partial_{t}+\Omega\right)\frac{\Phi^{2}(t_{\ast})}{\Box-M^2}\,,
\end{align}
where we used \eqref{chi} to substitute for $  \bco $.

To obtain the local EFT interactions we expand in $M^{2}\gg\Box,\partial_{t}^{2},\partial_{i}^{2}$. The resulting vertices in the bulk are:
\begin{equation}\label{UVEFTbulk}
    \mathcal{L}_{\text{EFT}}^{\text{bulk}}=\sum_{n=0}\frac{g^{2}}{2M^{2n}}\Phi^{2}\Box^{n}\Phi^{2},
\end{equation}
These are the same as for amplitudes, with the difference that we cannot drop the $n=1$ term, which is proportional to the equation of motion. On the boundary, we have both Lorentz covariant and boost breaking terms:
\begin{eqnarray}    
 \mathcal{L}_{\text{EFT}}^{\text{brane}}=\frac{ig^{2}}{2M^2}\sum_{a,b}\frac{\Box^{a}\Phi^{2}(t_{\ast})}{M^{2a}}\left(i\partial_{t}+M\sum_{n=0}\begin{pmatrix}1/2 \\ n \end{pmatrix}\left(-\frac{\partial_{i}^{2}}{M^{2}}\right)^{n}\right)\frac{\Box^{b}\Phi^{2}(t_{\ast})}{M^{2b}}.
\end{eqnarray}

Therefore, we can write the first few Wilson coefficients:
\begin{equation}\label{WilsonUV}
   \begin{split}
       &\alpha_{0}=12g^{2}\;,\;\alpha_{2}=\frac{2g^{2}}{M^{2}}\;,\;\alpha_{4}=\frac{2g^{2}}{M^{4}},\\
       &\beta_{31}'=\frac{2g^{2}}{M^{4}}\;,\;\beta_{31}=\frac{2g^{2}}{M^{4}}\;,\;\beta_{11}=\frac{2g^{2}}{M^{2}}\;,\;\beta_{20}=\frac{4ig^{2}}{M^{3}}\;,\;\beta_{00}=\frac{12ig^{2}}{M}\;,\;\beta_{22}=\frac{ig^{2}}{M^{3}}.
   \end{split}
\end{equation}

The value of these coefficients have been derived from the Lagrangian of the EFT after integrating out the heavy degrees of freedom of the UV. They do not rely on the sum rules. To prove that we can use the latter to compute all the coefficients in (\ref{WilsonUV}) we will start from the UV four-point exchange wavefunction coefficient for the light scalar $\psi_{\text{UV}}$ and then we will use the sum rules (\ref{0sumrules}-\ref{w1kskssumrules}) to obtain the value of all the Wilson coefficients from  (\ref{WilsonUV}).
$\psi_{\text{UV}}$ is a exchange diagram given by a heavy internal line so it is the sum of three different channels $\psi'_{\text{UV}}$. 
An explicit computation gives,
\begin{equation}
    \psi_{\text{UV}}( \{ \omega \} , \{\textbf{k} \})= \delta \left( \bfk_T \right) \left[ 
    \psi'_{\text{UV}}(\omega_{12},\omega_{34},\textbf{k}_{s})+ \psi'_{\text{UV}}(\omega_{13},\omega_{24},\textbf{k}_{t})+ \psi'_{\text{UV}}(\omega_{14},\omega_{23},\textbf{k}_{u})
    \right]  \; , 
\end{equation}
\begin{equation}
    \omega_{T}\psi'_{\text{UV}}(\omega_{12},\omega_{34},\textbf{k}_{s})=\frac{4g^{2}M^{2}}{(\omega_{12}+\Omega_{k_{s}})(\omega_{34}+\Omega_{k_{s}})}.
\end{equation}
The object that appears on the right-hand side of the sum rules (\ref{0sumrules}-\ref{w1kskssumrules}) is the discontinuity of $\omega_{T}\psi_{\text{UV}}$ along the negative $\omega_{1}$ real axis. In this particular example, $\omega_{T}\psi_{\text{UV}}$ is a rational function of $\omega_{1}$ and therefore there are no branch cuts contributing to $\text{disc}(\omega_{T}\psi_{\text{UV}})$. However, the poles located in the negative $\omega_{1}$ plane do contribute towards  $\text{disc}(\omega_{T}\psi_{\text{UV}})$ as delta functions:
\begin{equation}
     \text{disc}(\omega_{T}\psi_{\text{UV}})=-\frac{4g^{2}M^{2}}{\omega_{34}+\Omega_{k_{s}}}2\pi i\delta(\omega_{12}+\Omega_{k_{s}})-\frac{4g^{2}M^{2}}{\omega_{24}+\Omega_{k_{t}}}2\pi i\delta(\omega_{13}+\Omega_{k_{t}})-\frac{4g^{2}M^{2}}{\omega_{23}+\Omega_{k_{u}}}2\pi i\delta(\omega_{14}+\Omega_{k_{u}}).
\end{equation}
The residue at infinity vanishes and therefore only the integral over the discontinuity contributes towards the sum rules. Now we are in position to use the sum rules to compute the Wilson coefficients and prove that they match the result from (\ref{WilsonUV}):
\begin{align}
    \mathcal{I}_{\rm UV}^{(0)} ( \{ \omega \} ,  \{\textbf{k} \} )\big|_{\substack{\omega_{i}=0 \\ \textbf{k}_{j}=0}} & = \int_{-\infty}^{0}\frac{d\omega_{1}}{2\pi i}\frac{ \text{disc}(\omega_{T}\psi_{\text{UV}})}{\omega_{1}}\bigg|_{\substack{\omega_{i}=0 \\ \textbf{k}_{j}=0}}=12g^{2} ,\\
    \mathcal{I}_{\rm UV}^{(1)}( \{ \omega \} ,  \{\textbf{k} \} )\big|_{\substack{\omega_{i}=0 \\ \textbf{k}_{j}=0}} &=\int_{-\infty}^{0}\frac{d\omega_{1}}{2\pi i}\frac{ \text{disc}(\omega_{T}\psi_{\text{UV}})}{\omega_{1}^{2}}\bigg|_{\substack{\omega_{i}=0 \\ \textbf{k}_{j}=0}}=-\frac{12g^{2}}{M} ,\\
    \mathcal{I}_{\rm UV}^{(2)} ( \{ \omega \} ,  \{\textbf{k} \} )\big|_{\substack{\omega_{i}=0 \\ \textbf{k}_{j}=0}} &= \int_{-\infty}^{0}\frac{d\omega_{1}}{2\pi i}\frac{ \text{disc}(\omega_{T}\psi_{\text{UV}})}{\omega_{1}^{3}}\bigg|_{\substack{\omega_{i}=0 \\ \textbf{k}_{j}=0}}=\frac{12g^{2}}{M^{2}} ,\\
    \partial_{\omega_{2}}\mathcal{I}_{\rm UV}^{(1)} ( \{ \omega \} ,  \{\textbf{k} \} )\big|_{\substack{\omega_{i}=0 \\ \textbf{k}_{j}=0}} &=\int_{-\infty}^{0}\frac{d\omega_{1}}{2\pi i}\frac{ \partial_{\omega_{2}}\text{disc}(\omega_{T}\psi_{\text{UV}})}{\omega_{1}^{2}}\bigg|_{\substack{\omega_{i}=0 \\ \textbf{k}_{j}=0}}=\frac{16g^{2}}{M^{2}} ,\\
    \mathcal{I}_{\rm UV}^{(3)} ( \{ \omega \} ,  \{\textbf{k} \} )\big|_{\substack{\omega_{i}=0 \\ \textbf{k}_{j}=0}}&=\int_{-\infty}^{0}\frac{d\omega_{1}}{2\pi i}\frac{ \text{disc}(\omega_{T}\psi_{\text{UV}})}{\omega_{1}^{4}}\bigg|_{\substack{\omega_{i}=0 \\ \textbf{k}_{j}=0}}=-\frac{12g^{2}}{M^{3}} ,\\
    \mathcal{I}_{\rm UV}^{(4)} ( \{ \omega \} ,  \{\textbf{k} \} )\big|_{\substack{\omega_{i}=0 \\ \textbf{k}_{j}=0}}&=\int_{-\infty}^{0}\frac{d\omega_{1}}{2\pi i}\frac{ \text{disc}(\omega_{T}\psi_{\text{UV}})}{\omega_{1}^{5}}\bigg|_{\substack{\omega_{i}=0 \\ \textbf{k}_{j}=0}}=\frac{12g^{2}}{M^{4}} ,\\
    \partial_{\omega_{2}}\partial_{\omega_{3}}\mathcal{I}_{\rm UV}^{(2)} ( \{ \omega \} ,  \{\textbf{k} \} )\big|_{\substack{\omega_{i}=0 \\ \textbf{k}_{j}=0}}&=\int_{-\infty}^{0}\frac{d\omega_{1}}{2\pi i}\frac{ \partial_{\omega_{2}}\partial_{\omega_{3}}\text{disc}(\omega_{T}\psi_{\text{UV}})}{\omega_{1}^{3}}\bigg|_{\substack{\omega_{i}=0 \\ \textbf{k}_{j}=0}}=\frac{32g^{2}}{M^{4}} ,\\
    \partial_{\omega_{2}}\partial_{\omega_{3}}\partial_{\omega_{4}}\mathcal{I}_{\rm UV}^{(1)} ( \{ \omega \} ,  \{\textbf{k} \} )\big|_{\substack{\omega_{i}=0 \\ \textbf{k}_{j}=0}} &=\int_{-\infty}^{0}\frac{d\omega_{1}}{2\pi i}\frac{ \partial_{\omega_{2}}\partial_{\omega_{3}}\partial_{\omega_{4}}\text{disc}(\omega_{T}\psi_{\text{UV}})}{\omega_{1}^{2}}\bigg|_{\substack{\omega_{i}=0 \\ \textbf{k}_{j}=0}}=\frac{48g^{2}}{M^{4}} ,\\
    \partial_{k_{s}}^{2}\mathcal{I}_{\rm UV}^{(1)} ( \{ \omega \} ,  \{\textbf{k} \} )\big|_{\substack{\omega_{i}=0 \\ \textbf{k}_{j}=0}}&=\int_{-\infty}^{0}\frac{d\omega_{1}}{2\pi i}\frac{\partial_{k_{s}}^{2}\text{disc}(\omega_{T}\psi_{\text{UV}})}{\omega_{1}^{2}}\bigg|_{\substack{\omega_{i}=0 \\ \textbf{k}_{j}=0}}=\frac{12g^{2}}{M^{3}}.
\end{align}
All of the sum rules match the result of the Wilson coefficients \eqref{WilsonUV} computed at the level of the action. 

 
\subsection{UV-completion: a one-loop example}

Now consider a different toy model:
\begin{equation} \label{eqn:4.2_LUV}
    \mathcal{L}_{\text{UV}}[\Phi,X]=-\frac{1}{2}(\partial\Phi)^{2}-\frac{1}{2}m^{2}\Phi^{2}-\frac{1}{2}(\partial X)^{2}-\frac{1}{2}M^{2}X^{2}- g \Phi^{2} X^2.
\end{equation}
Unlike the previous toy model, we can no longer use the classical equation of motion for the heavy field $X$ to obtain the effective action for $\Phi$ because the leading correction to the action is now at loop level. 
Instead, we will explicitly compute the wavefunction coefficient $\psi_{\rm UV}$, and match this to a low-energy $\psi_{\rm EFT}$ in order to fix the appropriate EFT Wilson coefficient. 

\paragraph{Computing the UV wavefunction.}
In order to compute the Wilson coefficients, we can first compute the wavefunction coefficient for the UV theory and expand the wavefunction coefficient in the low-energy limit. Since this expression needs to match the wavefunction coefficient computed from the EFT, we can read off the Wilson coefficients.
In the UV theory, the leading contribution to the two-point wavefunction coefficient is the following graph:

\begin{align}
    \includegraphics[scale=1.0]{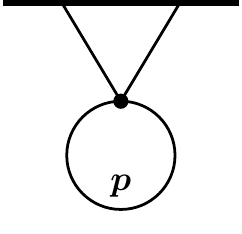}
\end{align}

\noindent This graph corresponds to the integral,
\begin{equation}
    \omega_T\psi_{\text{UV}}=  g \int_{\bfp}\frac{1}{\omega_T+2\sqrt{p^2+M^2}}.\label{looppsi1}
\end{equation}
which is computed explicitly using a hard cut-off in App.~\ref{sec:1edge}, and the full result is:
\begin{multline}  \label{eqn:psi1site_main}
    \omega_T \psi_{\text{UV}}  
    = \frac{g}{ 16 \pi^2 }\left[ 2 \Lambda (  \Lambda - \omega_T ) + M^2 + (\omega_T^2- 2M^2) \log \left( \frac{2\Lambda}{M} \right)\right. \\ + \left.
        2 \omega_T \sqrt{4M^2 - \omega_T^2} \arcsin \left( \sqrt{ \frac{ 2 M - \omega_T }{4M}} \right)
        + \mathcal{O} \left( \frac{1}{\Lambda} \right) \right] \; .
\end{multline}

\paragraph{Finding the EFT Wilson coefficients.}
Notice that this particular $\psi_{\rm UV}$ depends only on the energies of the external lines, not their spatial momenta. 
Consequently, it can be matched using an EFT in which time derivatives are treated as much larger than spatial derivatives. 
Concretely, we consider the following EFT interactions,
\begin{align}  \label{eqn:4.2_EFT}
 S_{\rm EFT} [ \Phi ; t_*] = \int_{-\infty}^{t_*} dt \int d^3 \bfx \; \frac{1}{2} \sum_{n=0} \gamma_n ( - i \partial_t )^n  \Phi^2 
\end{align}
which corresponds to a low-energy two-point wavefunction,
\begin{align}
  \omega_T \psi_{\rm EFT} = \sum_{n=0} \gamma_n \omega_T^n  \; .
  \label{eqn:4.2_psi_EFT}
\end{align}
The UV model \eqref{eqn:4.2_LUV} is one possible UV completion of this EFT, in which the first few Wilson coefficients $\gamma_n$ appearing in \eqref{eqn:4.2_psi_EFT} are fixed to be,
\begin{align}
 \gamma_0 &= \frac{g}{16 \pi^2} \left( 2 \Lambda^2 + M^2 - 2 M^2 \log \left( \frac{2 \Lambda}{M}  \right)  \right)  \nonumber \\
 \gamma_1 &= \frac{g}{16 \pi^2} \left( - 2 \Lambda + M \pi  \right) \nonumber \\
 \gamma_2 &= \frac{g}{16 \pi^2} \left( -1 + \log \left( \frac{2 \Lambda}{M} \right)\right)  \nonumber \\
 \gamma_3 &=  \frac{g}{16 \pi^2} \left( - \frac{\pi}{8 M} \right) 
\end{align}
so that $\psi_{\rm EFT}$ coincides with $\psi_{\rm UV}$ at low values of $\omega_a$. 

Notice that we have only regularized the theory, with an arbitrary cutoff, but we have not carried out renormalization. 
We will verify our sum rules at finite but arbitrary $\Lambda$: since if our rules hold for any value of the regulator, they will also hold for any renormalization scheme which is consistently implemented in both the EFT and the UV.


\paragraph{Checking the UV/IR sum rules.}
Applying the dispersion relation of Sec.~\ref{sec:dispersion}, the Wilson coefficients appearing in \eqref{eqn:4.2_EFT} are given by,
\begin{align}  \label{eqn:4.2_sum_rules}
  \gamma_n = \mathcal{I}_{\rm UV}^{(n)} \big|_{ \substack{ \omega_2 = 0 \\ \bfk = 0} }
\end{align}
where $\mathcal{I}_{\rm UV}^{(N)}$ are defined in \eqref{IRUVsumrule2} using the two-point $\psi_{\rm UV}$.

One could take the discontinuity of \eqref{eqn:psi1site_main} directly to compute the integrals $\mathcal{I}_{\rm UV}^{(N)}$, but a quicker route is to take the discontinuity of the original integrand,  
\begin{align}
        \text{disc}(\omega_T\psi_{\text{UV}})&= g \int_{\bfp} \text{disc}\left(\frac{1}{\omega_T+2\sqrt{p^2+M^2}}\right)\nonumber\\
        &= -\frac{g}{2\pi^2} \int_{0}^{\infty} dp \; p^{2} \; 2\pi i \delta(\omega_T+2\sqrt{p^2+M^2}).
\end{align}
Since $p$ is real, for $\omega_T>-2M$ there is no solution to $\omega_T+2\sqrt{p^2+M^2}=0$. 
Therefore, the discontinuity is proportional to a step function:
\begin{equation}\label{discpsi}
    \text{disc}(\omega_T\psi_{\text{UV}})=  \frac{g \, \omega_T}{16 \pi^2 } \sqrt{  \omega_T^2- 4 M^2 }  \; 2\pi i \, \Theta\left( -  \omega_T -2 M \right).
\end{equation}
Substituting \eqref{discpsi} into \eqref{IRUVsumrule2}, and again using a hard cut-off to tame any divergences, we find that in this UV model:
\begin{equation}
    \mathcal{I}^{(N)}_{\rm UV} \big|_{\substack{\omega_{2}=0\\ \bfk = 0}}=\int_{-2 \Lambda}^{-2M}d\omega \,\frac{g}{16\pi^2} \frac{ \omega \sqrt{  \omega^2 - 4 M^2 } }{ \omega^{N+1}} .
\end{equation}
These can be evaluated straightforwardly, and give,
\begin{align}
\mathcal{I}_{\rm UV}^{(0)} \big|_{\substack{\omega_{2}=0\\ \bfk = 0}} &= \frac{g}{16 \pi^2} \left( 2 \Lambda^2 + M^2 - 2 M^2 \log \left( \frac{2 \Lambda}{M}  \right)  \right)   \nonumber \\
 \mathcal{I}_{\rm UV}^{(1)} \big|_{\substack{\omega_{2}=0\\ \bfk = 0}} &= \frac{g}{16 \pi^2} \left( - 2 \Lambda + M \pi  \right) \nonumber \\
 \mathcal{I}_{\rm UV}^{(2)} \big|_{\substack{\omega_{2}=0\\ \bfk = 0}} &= \frac{g}{16 \pi^2} \left( -1 + \log \left( \frac{2 \Lambda}{M} \right)\right)  \nonumber \\
 \mathcal{I}_{\rm UV}^{(3)} \big|_{\substack{\omega_{2}=0\\ \bfk = 0}} &=  \frac{g}{16 \pi^2} \left( - \frac{\pi}{8 M} \right) 
\end{align}
in agreement with the sum rules \eqref{eqn:4.2_sum_rules}.


 
\section{Conclusions}
\label{sec:conclusions}

In this work, we have studied the analytic structure of the field theoretic Minkowski wavefunction. First, we have introduced new objects, which we dubbed \textit{off-shell wavefunction coefficients}, which are related to amputated, time-ordered in-out Green's functions as in \eqref{eqn:off_shell_psi_def}. The off-shell wavefunction coefficients are manifestly analytic in half of the complex plane because of causality (the lower-half in our conventions). We have confirmed this non-perturbative argument with an explicit study of the singularities emerging in perturbation theory. As anticipated in \cite{Arkani-Hamed:2018bjr}, the location of singularities associated to a given Feynman-Witten diagram can be deduced from a simple energy-conservation condition on each sub-diagram. We have shown how these singularities are captured by a Landau-like analysis of explicit loop integrals. In the extensive App.~\ref{App:A} we have explicitly computed a number of one-loop integrals that corroborate our general picture and allow for detailed studies of the analytic structure. In a second part of the paper, we have used analyticity to write down dispersion relations for off-shell wavefunction coefficients in the complex plane of a single off-shell frequency. These dispersion relations fix the Wilson coefficients in a low-energy effective action in terms of an appropriate integral over the discontinuities appearing in UV-complete wavefunction coefficients. These results can be thought of as new \textit{UV/IR sum rules}. An interesting difference between wavefunction sum rules and analogous relations for amplitudes is that the wavefunction is sensitive to both total derivative interactions and interactions proportional to the equations of motion. To make our general results more concrete, we have studied a simple example of a light scalar interacting with a heavy scalar. We have verified the UV/IR sum rules at tree-level, where the only singularities are poles, as well as at one-loop order, where extended discontinuities appear.

\paragraph{Outlook.}
Our results open up a number of avenues for future research, which we now discuss:
\begin{itemize}
    \item Recently, an infinite set of constraints, collectively denoted Cosmological Optical Theorem (COT), have been derived in \cite{COT} for the wavefunction as a consequence of unitary time evolution from a Bunch-Davies vacuum (see \cite{Cespedes:2020xqq} for generalizations) on any FLRW spacetime \cite{Goodhew:2021oqg} to all loop orders \cite{sCOTt}. Notice that, in spite of its name, the COT applies to the Minkowski wavefunction as well. The current formulation of the COT requires separating by hand the dependence on the norm of the external three-momenta from the dependence on their direction. The norms are analytically continued to negative values to define an ad hoc ``non-local'' discontinuity. It seems clear that the off-shell wavefunction that we introduced here is a more natural object to formulate the COT, since the off-shell frequencies appear automatically as independent variables and analyticity is ensured in the lower-half complex plane (in agreement with the COT prescription of approaching real frequencies from below). Moreover, the COT gives us a powerful non-local relation for the value of $\psi_n$ at $\{\omega \}$ and $\{-\omega\}$. It would be extremently intersting to see if this relation can be used to say something about the discontinuity across the negative real axis in the complex frequency plane, which is at the center of this paper. To build such a relation one would like to find a representation of the off-shell wavefunction that is analytic in the upper-half complex plane too (where the standard time-integral representation does not converge). Perhaps this requires invoking additional physical ingredients, such as (C)PT symmetry.
    
    \item Locality has been shown to imply a Manifestly Local Test (MLT) \cite{MLT} that constrains wavefunction coefficients in de Sitter for light fields ($m^2<H^2$) \cite{Goodhew:2022ayb} with soft interactions (not necessarily manifestly local \cite{Bonifacio:2022vwa}). The current formulation deals again with the norm of the external three momenta, but more appropriate variables would be the off-shell frequencies. After performing this reformulation, it would be interesting to investigate if one can more easily discuss local but not-manifestly-local theories such as general relativity and solids, which presented a difficulty in the standard boostless bootstrap approach \cite{BBBB}.
    
    \item It is important to develop a better understanding of the physical meaning of the different singularities appearing in the wavefunction. For example, by direct inspection one can convince oneself that wavefunction coefficients include divergences associated to one-loop diagrams with a single vertex (as in \eqref{psi1site}), but such divergences cancel out exactly from equal-time correlators of the product of fields. This is somewhat reminiscent of the cancellation of infra-red divergences for inclusive processes (the Kinoshita-Lee-Nauenberg theorem \cite{Kinoshita:1962ur,Lee:1964is}). One would like to characterize and possibly classify all divergences that disappear from (a class of) observables. 

    \item A natural next step is to extend our analysis to the wavefunction in curved spacetime and especially in de Sitter space. A few important differences will emerge. First, in dS we are usually interested in evaluating the wavefunction at the future conformal boundary ($t\to\infty$, equivalently $\eta\to0$ in conformal time). This is a conceptually important difference when one has in mind UV-completing the theory by some quantum gravity. Since there are no local, gauge-invariant observables in gravity, one expects that well defined observables in quantum gravity live at some (often conformal) boundary. Hence, in quantum gravity the boundary wavefunction in dS seems a more promising object than the finite-time Minkowski wavefunction.
    Another difference of a more technical nature is that the mode functions in dS change drastically depending on the mass and the number of spacetime dimensions. This makes it challenging to collect a sufficient number of explicit loop results, although some progress is underway \cite{Grall:2020tqc,Kristiano:2022zpn,Xianyu:2022jwk,Premkumar:2022bkm}. A final difference is the general mechanism of so-called particle production caused by the expansion in dS, which could introduce singularities of a different nature from those studied here. 

    \item One of the main motivations for constructing these UV/IR relations is to establish a set of ``positivity bounds'' for the wavefunction. Positivity bounds for amplitudes have recently been developed and applied to a wide range of effective field theories, both on Minkowski and on cosmological spacetime backgrounds \cite{Melville:2019wyy, Kim:2019wjo, Ye:2019oxx, Grall:2021xxm, Aoki:2021ffc, deRham:2021fpu, Davis:2021oce, Melville:2022ykg, Freytsis:2022aho}. However, their application on curved backgrounds has typically required taking careful subhorizon/decoupling limits in order to even define the amplitude. If analogous positivity bounds could be established directly at the level of the wavefunction, these would be more readily applicable in cosmology\footnote{
    Another interesting possibility is to consider positivity of correlation functions directly, which has recently led to robust bounds on non-trivial backgrounds in \cite{Creminelli:2022onn}.    
    }. The sum rules presented here fall short of this because, unlike their amplitude counterpart, the UV discontinuity that appears is not sign definite (which can be seen explicitly in our toy UV completions). However, as mentioned above, if unitarity can be used to constrain this discontinuity then in the future our sum rules could be used to place model-agnostic bounds on the EFT Wilson coefficients.  
    
    \item We have focussed on the analytic structure of the wavefunction when just one of the external energies is complexified (with the other kinematic variables held fixed, real and positive). It would be interesting to explore whether further dispersion relations could be constructed by analytically continuing multiple variables. These would be closer in spirit to Mandelstam's original dispersion relation, which simultaneously continues $s$, $t$ and $u$. Furthermore, using crossing symmetry to combine the dispersion relations in different channels has recently led to a number of ``null constraints'' on the integral of the UV scattering amplitude, which have been used to derive further EFT positivity bounds \cite{Bellazzini:2020cot, Tolley:2020gtv, Caron-Huot:2020cmc}. It would be interesting to explore the same possibility for the wavefunction. As a crude example: for a wavefunction coefficient in which external legs 1 and 2 are identical, one should find that $\left[ \partial_{\omega_2}^2 \partial_{\omega_1} - \partial_{\omega_1}^2 \partial_{\omega_2} \right] \psi_n = 0$, and this immediately implies that the UV integrals \eqref{IRUVsumrule} obey,
    \begin{align}
 \frac{1}{2} \partial_{\omega_2}^2 \mathcal{I}_{\rm UV} ^{(1)} |_{\bfk_1 = \bfk_2} = \partial_{\omega_2} \mathcal{I}_{\rm UV}^{(2)} |_{\bfk_1 = \bfk_2} \; . 
    \end{align}
    There are an infinite number of such constraints on the $\mathcal{I}_{\rm UV}^{(N)}$ integrals which appear in our sum rules.
    
    \item The on-shell wavefunction is a natural object to compute equal-time correlators, but its relation to unequal time correlators has until now been somewhat cumbersome. Conversely, the off-shell wavefunction we defined here is directly related to unequal time correlators. As such, it might provide a useful starting point to discuss micro-causality and the very interesting class of symmetries that involve time shifts \cite{Hui:2022dnm}.

    \item In the limit where the total energy flowing into a graph vanishes, we noticed that it contributes to the wavefunction precisely the same as the corresponding Feynman graph for a scattering amplitude. This is true at both tree- and loop-level, and the resulting Minkowski amplitude includes all finite-mass effects (unlike on cosmological backgrounds, where it is typically the high-energy limit of a corresponding amplitude). One useful direction for the future would be to further explore this connection. In particular, the dispersion relation we derived for the wavefunction coefficients contains only a single branch cut discontinuity: this is because we analytically continued one of the energies holding all others fixed. If we had first taken this amplitude limit (setting the total energy to zero), and then analytically continued, we would have found two branch cuts, reproducing the familiar $s$- and $u$-channel cuts of the scattering amplitude. Furthermore, we have focussed here on light external fields exchanging heavier fields: for the complementary process, in which sufficiently heavy external fields exchange light fields, the scattering amplitude can develop so-called ``anomalous thresholds'' on the physical sheet which are not captured by unitarity---see e.g. \cite{MartinAnomalous, Sashanotes} for a review. One starting point to investigate how these may arise in the wavefunction would be to consider the prototypical triangle diagram which mediates $2 \to 2$ scattering of a heavy particle via a loop of light particles,
    since the corresponding wavefunction coefficients must contain an anomalous threshold in the amplitude limit where the total energy vanishes. 
    
    \item Given their mathematical similarities, it seems possible to import many of the powerful techniques for evaluating Feynman loop integral to the wavefunction. For instance, in the calculation of loop corrections to scattering amplitudes, one is able to reduce the infinitely many possible integrals to a finite set of master loop integrals (bubbles, triangles and boxes). Here we have began to similarly organise and classify the structures which can appear in the wavefunction (see in particular the discussion in App.~\ref{sec:complexity}). While the number of master integrals seems to be necessarily larger for the wavefunction than for amplitudes, for a fixed number of external and internal lines it should be possible to construct an analogous basis which captures all possible non-analytic structures. Such a basis would be immensely useful for future investigations of the wavefunction beyond tree-level. Furthermore, recent progress in recasting Feynman integrals in terms of differential operators (based on canonical forms) has rendered many previously intractable diagrams now solvable, see e.g. \cite{Henn:2013pwa}. Developing an analogous technology for the wavefunction has the potential to drastically expand the range of perturbative quantum field theories that can be explicitly analysed using the sum rules developed here. 
    
    \item Finally, we have focussed on the wavefunction for scalar fields, and an important extension of these results will be to include spinning fields. For scattering amplitudes, massive spinning fields in particular can lead to additional singularities in the complex plane (so-called ``kinematical singularities'') associated with the polarisation tensors---see e.g. \cite{deRham:2017zjm} for a review. However, these arise as a consequence of the complicated crossing relation which the amplitude must satisfy, and since the wavefunction coefficients do not have this requirement they may possess a simpler analytic structure.

\end{itemize}

 
\section*{Acknowledgements} We would like to thank Harry Goodhew, Aaron Hillman,
Austin Joyce, João Penedones, Guilherme L. Pimentel, and David Stefanyszyn for useful discussions. 
S.M. is supported by a UKRI Stephen Hawking Fellowship (EP/T017481/1). 
E.P. has been supported in part by the research program VIDI with Project No. 680-47-535, which is (partly) financed by the
Netherlands Organisation for Scientific Research (NWO). 
S.A.S. is supported by a Harding Distinguished Postgraduate Scholarship. M.H.G.L. is supported by the Croucher Cambridge International Scholarship.
This work has been partially supported by STFC consolidated grant ST/T000694/1. 

 
\appendix

 
\section{One-loop computations}\label{App:A}

In this appendix, we describe in detail how to compute the momentum integrals that appear in the one-loop wavefunction, and give various examples for both massless and massive fields. 

We begin in Sec.~\ref{sec:complexity} by introducing a taxonomy for ``how complicated'' a loop integral is: this provides an organisational principle for the different $\psil{1}_n$, shown in Table~\ref{fig:genus_table_2}. 
Then we evaluate various limits of $\psil{1}_1$ in Sec.~\ref{sec:1edge}, $\psil{1}_2$ in Sec.~\ref{sec:2edge} and $\psil{1}_3$ in Sec.~\ref{sec:3edge}. 

\subsection{Classifying complexity}
\label{sec:complexity}
The majority of the loop integrals that we encounter below cannot be written in terms of elementary functions. 
It will be useful, therefore, to introduce a taxonomy for the various kinds of special functions which can arise in a given $\psi_n$. 
In particular, whenever an integral can be written in the following form,
\begin{align}
    \mathcal{I} = \int_1^{\infty} d p \, \int_{-1}^{+1} d z_1 \, ... \int_{-1}^{+1} d z_{\mathcal{T}-1} \, \frac{ R ( p , z_1, ..., z_{\mathcal{T}-1}  ) }{ \text{Poly}_{2\mathcal{G}+2} ( p , z_1, ... , z_{\mathcal{T}-1} ) },
    \label{eqn:genus_def}
\end{align}
where $R$ is a rational function of its arguments and $\text{Poly}_N$ represents a polynomial which is at most order $N$ in any one of its arguments, then we say that $\mathcal{I}$ has a \emph{degree of transcendentality} $\mathcal{T}$ and a \emph{genus} $\mathcal{G}$. 
Roughly speaking, $\mathcal{T}$ counts the number of integrals and $\mathcal{G}$ counts the number of independent square roots appearing in the integrand. 
The lower each of these numbers, the closer the integral will be to familiar elementary functions. 

In Fig.~\ref{fig:genus_table}(a) we give some examples of special functions with different $\mathcal{T}$ and $\mathcal{G}$. 
In particular, for genus $\mathcal{G} = 0$, an integral with degree of transcendentality $\mathcal{T}$ can possess polylogarithmic-type branch cuts of a polylog $ \text{Li}_{\mathcal{T}}$, where $\mathcal{T}$ determines the weight of the polylogarithm. 
For $\mathcal{T} = 1$, \eqref{eqn:genus_def} represents the so-called ``hyperelliptic integrals'', where the degree $2\mathcal{G}+2$ polynomial in the denominator is called ``hyperlliptic curve of genus $\mathcal{G}$''. 

When both $\mathcal{T} > 1$ and $\mathcal{G} > 0$ much less is known about integrals of the form \eqref{eqn:genus_def}, and the labels $\mathcal{T}$ and $\mathcal{G}$ give a useful measure of ``how complicated'' each integral is.  
For instance, we will show below that the three-point wavefunction coefficient at one-loop, $\psil{1}_3$, generically has $\mathcal{T} = 3$ and $\mathcal{G}=3$ in $d=3$ spatial dimensions, so in principle would require knowledge of the special functions which correspond to three iterated integrals over a hyperelliptic curve of genus $3$. 
A dedicated study of the properties of such functions would certainly be interesting (particularly in light of the myriad connections between amplitude Feynman integrals and pure mathematics), but here we will restrict our attention to integrals that can be written in terms of $\mathcal{G} = 0$ or $\mathcal{T} = 1$ functions only. 
In the case of $\psil{1}_3$ in $d=3$ dimensions, we will see that taking a combination of soft and massless limits can reduce $( \mathcal{T} , \mathcal{G} )$ to $( 2, 0)$, and we can therefore give an explicit expression in terms of dilogarithms (see \eqref{eqn:psi3_soft}).

\begin{table}
         \centering
         \begin{tabular}{ c | c | c | c | c}
         &  $\mathcal{T} = 1$  &   $\mathcal{T} = 2$ & $\cdots$  &  $\mathcal{T} $  \\
         \hline  
         $ \mathcal{G} = 0 $ & \begin{tabular}[c]{@{}c@{}} Logarithm \\ $\log (z)$ \end{tabular}  &  \begin{tabular}[c]{@{}c@{}} Dilogarithm \\ $\text{Li}_2 (z)$ \end{tabular} & $\cdots$ & \begin{tabular}[c]{@{}c@{}} Polylogarithm \\ $\text{Li}_{\mathcal{T}} (z)$ \end{tabular}  \\ \hline
         $ \mathcal{G} = 1$  & \begin{tabular}[c]{@{}c@{}} Elliptic integral \\ $\Pi (z)$ \end{tabular}   &  \begin{tabular}[c]{@{}c@{}} Integral of elliptic integral \\ $\int \frac{dz}{z} \Pi (z)$ \end{tabular}  &  $\cdots$ & \begin{tabular}[c]{@{}c@{}} Integrals of elliptic integral \\ $\left[ \int  \frac{dz}{z} \right]^{\mathcal{T}-1} \Pi (z)$ \end{tabular}   \\
         \hline
         $\mathcal{G} = 2$ &  Hyperelliptic integral  &  Integral of hyperelliptic integral & $\cdots$ & $\vdots$  \\
         \hline
         $\vdots$ & $\vdots$  & $\vdots$ & $\cdots$ & $\ddots$
         \end{tabular}
    \caption{To organise the special functions which arise when performing the loop integrals, we introduce a degree of transcendentality, $\mathcal{T}$, and a genus, $\mathcal{G}$, as in \eqref{eqn:genus_def}. Above we give some simple examples of functions in each class, and in Table~\ref{fig:genus_table_2} we summarise how the first three wavefunction coefficients $\psi_1, \psi_2, \psi_3$ (and their various massless and soft limits) populate these classes.}
    \label{fig:genus_table}
\end{table}

\begin{table}
\begin{subtable}[b]{0.4\textwidth}
         \centering
         \begin{tabular}{ m{1cm} | m{3.2cm} }
         & \begin{center} $\mathcal{T} = 1$ \\[-10pt]  \end{center}  \\
         \hline  
         $ \mathcal{G} = 0 $ &  \includegraphics[width=0.5\textwidth]{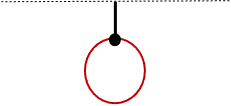}  
         \end{tabular}
         \caption{$\psil{1}_1$}
     \end{subtable} \hfill
 \begin{subtable}[b]{0.5\textwidth}
         \centering
         \begin{tabular}{ m{1cm} | m{3.2cm} | m{3.2cm} }
         & \begin{center} $\mathcal{T} = 1$ \\[-10pt] \end{center}  &  \begin{center} $\mathcal{T} = 2$ \\[-10pt]  \end{center}  \\
         \hline  
         $ \mathcal{G} = 0 $ &  \includegraphics[width=0.4\textwidth]{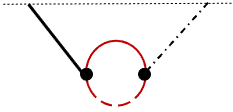}  & \includegraphics[width=0.4\textwidth]{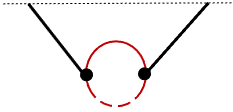}  \\ 
         $ \mathcal{G} = 1$  &  \includegraphics[width=0.4\textwidth]{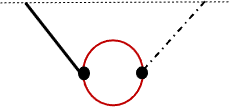}  &  \includegraphics[width=0.4\textwidth]{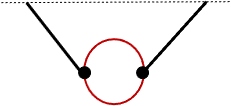}
         \end{tabular}
         \caption{$\psil{1}_2$}
     \end{subtable} \\[20pt]
     \hfill
     \begin{subtable}[b]{\textwidth}
         \centering
         \begin{tabular}{ m{1cm} | m{3.2cm} | m{3.2cm} | m{3.2cm} }
         &  \begin{center} $\mathcal{T} = 1$ \\[-10pt] \end{center}   &   \begin{center} $\mathcal{T} = 2$ \\[-10pt] \end{center}  & \begin{center} $\mathcal{T} = 3$ \\[-10pt] \end{center} \\
         \hline  
         $ \mathcal{G} = 0 $ & \includegraphics[width=0.2\textwidth]{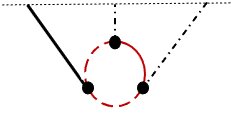}  &  \includegraphics[width=0.2\textwidth]{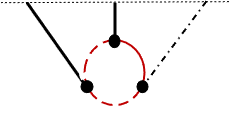}  &   \\ 
         $ \mathcal{G} = 1$  & \includegraphics[width=0.2\textwidth]{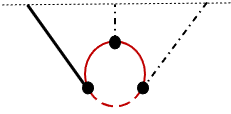}  &  \includegraphics[width=0.2\textwidth]{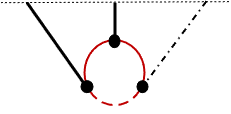}  &  \includegraphics[width=0.2\textwidth]{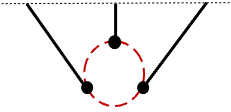}   \\
         $\mathcal{G} = 2$ &  \includegraphics[width=0.2\textwidth]{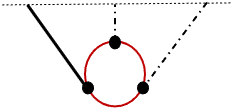}  &  \includegraphics[width=0.2\textwidth]{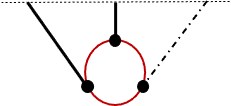}  & \includegraphics[width=0.2\textwidth]{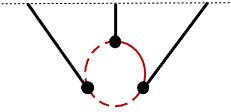}   \\
         $\mathcal{G} = 4$ &  & &  \includegraphics[width=0.2\textwidth]{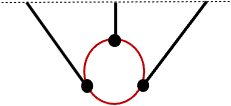}
         \end{tabular}
         \caption{$\psil{1}_3$}
     \end{subtable} 
    \caption{The degree of transcendentality $\mathcal{T}$ and genus  $\mathcal{G}$ of the different one-loop diagrams computed in App.~\ref{App:A}. A dashed internal line denotes a massless field, and a dot-dashed external line denotes a soft limit in which that field carries zero spatial momentum.}
    \label{fig:genus_table_2}
\end{table}

\paragraph{Comparison with amplitudes.}
One important point is that the one-loop wavefunction can have a much richer variety of non-analyticities than the analogous one-loop scattering amplitude (which is limited to simple poles and logarithmic branch cuts). 
This is because, once Feynman parameters are introduced, the integrand of the one-loop amplitude can be written as a rational function of the invariant $p_\mu p^\mu$ of a single $(d+1)$-dimensional momenta. 
The loop momentum integral can then be written as $d^{d+1} p = S_d d p \, p^{d-1}$ and viewed as a single integral over a rational function, which generates at most a logarithm branch cut. 
The degree of transcendentality can be used to make this precise: in the amplitude literature, it is defined\footnote{
Note that $\mathcal{T} (a b) = \mathcal{T} (a) + \mathcal{T} (b)$ can be used to count the degree of products, so $\mathcal{T} ( \log^2 ) = 2$. One often defines $\mathcal{T} ( \pi^n )  = n$ so that certain Feynman integrals have a uniform degree, e.g. $\mathcal{T} \left( \log^2 + \pi^2 \right) = 2$. 
We will instead take $\mathcal{T} ( \pi^n ) = 1$ and interpret $\mathcal{T} ( a + b)$ as $\text{max} \left( \mathcal{T} (a) , \mathcal{T} (b)  \right)$, since we will encounter sums of mixed weight and will occasionally factor out overall loop factors of e.g. $16 \pi^2$ in $d=3$.   
} to be $0$ when $f$ is any rational function, $\mathcal{T} ( \log ) = 1$ and $\mathcal{T} ( \text{Li}_n ) = n$ for an $n^{\rm th}$-order polylogarithm \cite{Henn:2013pwa} (this is the rationale for the naming of $\mathcal{T}$ in \eqref{eqn:genus_def}). 
For each loop integral we perform, the degree of transcendentality can increase by at most 1, so that\footnote{
In dimensional regularisation, this is equivalent to saying that an $L$-loop diagram diverges like $1/\epsilon^L$. Note that we are not including the additional IR divergences which can arise for massless particles and lead to $\mathcal{T} \left( \mathcal{A}^{L\text{-loop}}_n \right) = 2L$. 
},
\begin{align}
    \mathcal{T} \left( \mathcal{A}^{L\text{-loop}}_n \right) = L \; ,
\end{align}
describes how the mathematical complexity of the $n$-particle scattering amplitude grows with loop order. 

In contrast, the integrands appearing in the wavefunction coefficients are rational functions of the magnitudes, $| \bfp - \bfk_a|$, which cannot be combined using the usual Feynman parameters into the magnitude of a single vector and so will generally depend on all $d$ components of the loop momenta. Each loop integral therefore has the potential to increase the degree of transcendentality by $d$. 
However, since the $n$ external legs can carry no more than $n-1$ linearly-independent momenta $\bfk_a$ (due to momentum conservation), when $n-1 < d-1$ there are angular components of $\bfp$ which are orthogonal to every $\bfk_a$ and therefore do not appear in the loop integrand: in that case each loop integral may only increase the degree of transcendentality by $n d$.   
Overall, we expect loop integrals to produce polylogarithmic branch cuts of the following orders,
 \begin{align}
 \mathcal{T} \left( \psil{L}_n \right) = \text{min} \left( d , n  \right) \times L \; .
 \label{eqn:T_psi}
 \end{align}
This formula captures the degree to which wavefunction coefficients are more mathematically complex than their amplitude counterparts, and we will now see it borne out in several explicit examples.

\paragraph{Spurious singularities.}
Our main focus in the following will be on the analytic structure of particular loop diagrams. 
We will occasionally encounter functions that \emph{appear} to be non-analytic, and yet the residue/discontinuity of the apparent pole/branch cut is actually zero.
We refer to such non-analyticities as ``spurious''.  

The most common kind of spurious singularity is a pole of the form $f(z)/(z-a)$, where $f(z)$ in an analytic function of $z$ with a zero at $z = a$. This is also referred to as a ``removable'' singularity in the literature.
Such singularities arise whenever a rational function is decomposed into partial fractions.
For instance, suppose we define the integral,
\begin{align}
    \mathcal{I} (a) &= \int_1^{\infty} dz  \frac{ F (z) }{ (z + a ) } ,
    \label{eqn:spurious_I_eg}
\end{align}
where $F$ is some specified function and $a$ represents a particular kinematic parameter we wish to analytically continue to the complex plane. 
Further suppose that we can compute this integral, and find that it has singular points at $a = z_*$. 
The idea is that partial fractions can then be used to decompose more complicated integrals in terms of this simpler one, such as,
\begin{align}
    \int_1^{\infty} dz  \frac{ f (z) F (z ) }{ (z + f_1 (a) ) ( z + f_2 (a) ) } 
    =  \frac{ f ( f_1(a) ) \mathcal{I} ( f_1(a) ) - f ( f_2 (a) ) \mathcal{I} ( f_2(a) )  }{ f_1 (a) - f_2 (a) }  + \text{analytic} \; ,
    \label{eqn:spurious_pole_eg}
\end{align}
where $f, f_1$ and $f_2$ are any analytic functions of their arguments, and for the final term we have used that the integral of an analytic function produces an analytic function.
Looking at \eqref{eqn:spurious_pole_eg}, we would list the possible singular points of this integral as:
\begin{itemize}

\item[(i)] at $f_1 (a) = z_*$, due to the singularities of $\mathcal{I} ( f_1 (a) )$,     

\item[(ii)] at $f_2 (a) = z_*$, due to the singularities of $\mathcal{I} ( f_2 (a) )$,     

\item[{\color{red} (iii)}] {\color{red} at $f_1 (a) = f_2 (a)$}, due to the $1/(f_1 (a) - f_2(a))$ pole .

\end{itemize}
However, this apparent pole at $f_1 (a) = f_2 (a)$ is in fact spurious, since its residue vanishes at this value of $a$ and hence there is no actual non-analyticity at this point in the complex $a$ plane. 

This argument also applies with multiple factors in the denominator,
\begin{align}
    \int_1^\infty dz \, \frac{ f (z) F (z) }{ \prod_{j=1}^N (z + f_j (a) )  } \; .
    \label{eqn:spurious_pole_eg_2}
\end{align}
Such integrals can be written as a sum over $\mathcal{I} ( f_j (a) )$ using partial fractions, and one generically finds that any poles arising at $f_{j} (a) = f_{j'} (a)$ are in fact spurious. 
An analogous cancellation also takes place for certain non-analytic $f_j (a)$. For instance, performing partial fractions with a quadratic denominator can produce,
\begin{align}
    \int_1^{\infty} \frac{ F(z) }{ (z- c)^2 - a } = \frac{1}{ \bar{z}_+ (a) - \bar{z}_- (a) } \left[ \mathcal{I} ( \bar{z}_- (a) ) - \mathcal{I} ( \bar{z}_+ (a) )    \right] ,
    \label{eqn:spurious_cut_eg}
\end{align}
where $\bar{z}_{\pm} (a) = c \pm \sqrt{a}$ are the roots of the denominator $(z- c)^2 - a$. 
The factor of $\bar{z}_+ - \bar{z}_- = 2 \sqrt{a}$ gives an apparent branch point at $a=0$, but this is spurious since when expanded around this point since the square bracket contains only odd powers of $\sqrt{a}$, 
\begin{align}
    \mathcal{I} ( c - \sqrt{a} ) - \mathcal{I} ( c + \sqrt{a} )  = - 2 \sqrt{a} \sum_{n=0}^{\infty} \frac{ a^{n} }{n!} \mathcal{I}^{(n)} (c) \; , 
\end{align}
assuming that $\mathcal{I} (z)$ is analytic at $z=c$. The integral in \eqref{eqn:spurious_cut_eg} may therefore only have singularities at $\bar{z}_{\pm} (a) = z_*$, despite the appearance of an apparent branch cut at $a=0$. 

It is therefore tempting to conclude that integrals of the form \eqref{eqn:spurious_pole_eg_2} are analytic functions of $a$ modulo non-analyticities at the points given by $f_j (a) = z_*$. 
However, this still potentially overcounts, since there is a further way in which singularities can be spurious. 
Although $f (z) \mathcal{I} (z)$ may be singular at $z = z_*$, if the residue happens to be invariant under $f_1 (a) \leftrightarrow f_2 (a)$ then it will cancel out of differences like \eqref{eqn:spurious_pole_eg}.
One example of this (which we will encounter below in $\psil{1}_2$), is given by taking,
\begin{align}
 F (z ) = \log \left( a + b + z + 1 \right)
\end{align}
in \eqref{eqn:spurious_I_eg}, which gives,
 \begin{align}
     \mathcal{I} (a) = \text{Li}_2 \left( - \frac{1+b}{1+a}  \right)  \; ,
 \end{align}
and has branch points at $a = -1$ and $a = -2 - b$. 
Then the integral,
\begin{align}
\int_1^{\infty} dz \frac{ (a + b + 2 z ) F(z)}{ (z+a) ( z+ b)} =  
\text{Li}_2 \left( - \frac{1+b}{1+a}  \right) 
+
\text{Li}_2 \left( - \frac{1+a}{1+b}  \right) 
\end{align}
would seem to have singularities in the complex $a$ plane at,
\begin{itemize}

\item[(i)] $a=-1$,

\item[{\color{red} (ii)}] {\color{red} $a = -b - 2$},

\end{itemize}
from looking at each partial fraction individually, but this latter singularity is spurious since the residues of $\text{Li}_2 (z)$ and $\text{Li}_2 ( 1 /z)$ are equal and opposite at $z=1$.

So in the following, when we encounter integrals of the form \eqref{eqn:spurious_pole_eg_2}, we will perform partial fractions and focus on the singular points of the corresponding $\mathcal{I} ( f_j (a) )$, disregarding the spurious poles produced by the partial fractioning and also taking care to check for any cancellation which may happen between different values of $j$.

\subsection{One internal edge}
\label{sec:1edge}
For loop diagrams with only a single internal edge (of mass $M$), the integrand depends on the loop momenta through only a single $\Omega_p = \sqrt{p^2 + M^2}$.
Since this is a function of $p = |\bfp |$ only, the $d-1$ angular components of the loop integration can be performed immediately, giving,
\begin{align}
\int \frac{d^d \bfp}{ (2 \pi )^d } \; I ( \Omega_p ) = \frac{ S_{d-1} }{ (2 \pi )^d } \int_0^\infty d p \, p^{d-1} \; I ( \Omega_p ) ,
\end{align}
where $S_{d-1}$ is the surface area of a $(d-1)$-dimensional unit sphere,
\begin{align}
 S_{d-1} = \frac{ \pi^{d/2} }{ \Gamma \left( \frac{d}{2} \right) } \; ,
\end{align}
which takes the usual value of $S_2 = 4 \pi$ in $d=3$ spatial dimensions. 

For instance, consider the contribution from the one-loop diagram given in \eqref{eqn:I1_1loop}.
\begin{align}
    \omega \psil{1}_1 ( \omega )  =  \frac{S_{d-1}}{( 2 \pi )^d} \int_0^{\infty} d p  \, \frac{ p^{d-1} }{ \omega + 2 \Omega_p }.
\end{align}
This integral has $\mathcal{G}  ( \psil{1}_1 )= 0$ and $\mathcal{T} ( \psil{1}_1 ) = 1$ in any $d$,
The single square root in the denominator can be removed by a change of variables, and the resulting integral can be straightforwardly performed in any integer $d$ with a hard cut-off, $\int_0^\infty dp \to \int_0^\Lambda dp$. 
For instance, in $d=3$ dimensions\footnote{
Symbolic manipulation packages like Mathematica will often return a sum of $\arctan$ functions, which can be simplified using trigonometric relations like,
\begin{align}
    \arctan \left( \frac{2M + \omega}{\sqrt{ 4M^2 - \omega^2 } } \right) - \arctan \left( \frac{ \omega}{ \sqrt{4M^2 - \omega^2}} \right)  = \arcsin \left( \sqrt{ \frac{ 2 M - \omega }{4M}} \right)
\end{align}
in order to make the non-analyticity at $\omega = -2M$ manifest.
},
\begin{align}\label{md3}
    &\omega \psil{1}_1 (\omega) |_{d=3}  \nonumber \\ 
    &=\frac{1}{16 \pi ^2}\left[
    2 \Lambda (\Lambda-\omega)
    + \left(\omega^2-2 M^2\right) \log \left(\frac{2 \Lambda}{M}\right)+2 \omega \sqrt{4 M^2-\omega^2} \text{arcsin} \left( \sqrt{ \frac{2M-\omega}{4M} } \right)+ M^2 \right]\; ,
\end{align}
and in $d=1$ dimensions,
\begin{align} \label{md1}
    \omega \psil{1}_1 (\omega)  |_{d=1} = \frac{1}{4 \pi }\left[\log \left(\frac{2\Lambda}{M}\right)-\frac{2 \omega \,\text{arcsin} \left(\sqrt{\frac{2M-\omega}{4M}}\right)}{\sqrt{4 M^2-w^2}}\right] \; . 
\end{align}

In any number of dimensions, $\omega \psil{1}_1$ is an analytic function of $\omega$  except for branch points at $\omega = - 2M$ and $\omega = - \infty$, which can be connected by a cut along the negative real axis. In particular, note that \eqref{md3} and \eqref{md1} are analytic at $\omega = + 2M$, since $\sqrt{2M - \omega} \, \text{arcsin} \left( \sqrt{ \frac{ 2 M - \omega }{4 M} } \right) = \sum_j ( 2 M - \omega )^j$ for positive integer powers $j$.

\paragraph{Massless limit.}
When the mass of the internal line vanishes, $M \to 0$, the integral becomes
\begin{align}
    \omega \psil{1}_1  =  \frac{S_{d-1}}{( 2 \pi )^d} \int_0^{\infty} d p   \frac{ p^{d-1} }{ \omega + 2 p } .
\end{align}
This is an elementary logarithmic integral in any integer $d$, which can again be regulated with a hard cut-off: for instance,
 \begin{align}
    \omega \psil{1}_1 (\omega) |_{d=3}  
    =\frac{1}{16 \pi ^2}\left[ 
    2 \Lambda (\Lambda-\omega)
    + \omega^2  \log \left(\frac{2 \Lambda}{\omega}\right) \right]\; ,
\end{align}
Alternatively, using dimensional regularisation in $d = n + \epsilon$ dimensions gives the compact expression, 
\begin{align}
    \omega \psil{1}_1  =
        \frac{S_{n-1} (-\omega)^{d-1} }{ ( 16 \pi^2 )^{d/2} } \left[  - \frac{ 1 }{ \epsilon } - \log \left( \frac{\omega^2}{16 \pi} \right) +  \frac{1}{2} \psi^{(d/2)} \left( 0 \right)  \right] .
\end{align}
where $\psi^{(n)} \left( z \right)$ is the polygamma function. 
This proves that $\mathcal{T} ( \psil{1}_1 ) = 1$ in any $d$, as anticipated in \eqref{eqn:T_psi}.

\subsection{Two internal edges}
\label{sec:2edge}

For a one-loop diagram with two vertices, the integrand depends on a single loop momentum through two independent combinations, namely the energies associated with the momenta $\{ \bfq_1 , \bfq_2 \}$ of the internal lines, where momentum conservation fixes their difference to be equal to the external momentum, $\bfq_1 - \bfq_2 = \bfk$,
\begin{align}
\int \frac{d^d \bfq_1 d^d \bfq_2}{ (2 \pi )^d (2 \pi )^d} \; \mathcal{I} ( \Omega_{q_1}  , \Omega_{q_2} ) \, (2 \pi )^d \delta_D^{(d)} \left( \bfq_1 - \bfq_2 - \bfk \right) . 
\label{eqn:2edge_int}
\end{align}
Note that $\Omega_{q_1}$ and $\Omega_{q_2}$ are only independent in $d > 1$ spatial dimensions. 
We will now discuss integrals of the form \eqref{eqn:2edge_int} in $d > 1$ in some detail. We provide a convenient set of integration variables and a basis of master integrals in Sec.~\ref{sec:2edge_generalities}. Then, we evaluate $\psil{1}_2$ explicitly in the case of equal-mass internal edges (Sec.~\ref{sec:2edge_psi2}), massless internal edges (Sec.~\ref{sec:2edge_massless}) and a soft external edge (Sec.~\ref{sec:2edge_soft}). We will show in Sec.~\ref{sec:2edge_amplitude} that in the limit $\omega_{12} \to 0$ we recover the one-loop scattering amplitude on Minkowski, and finally we will describe the special case of $d=1$ dimensions in Sec.~\ref{sec:2edge_d=1}.

\subsubsection{Generalities}
\label{sec:2edge_generalities}

\begin{figure}
    \centering
    \includegraphics{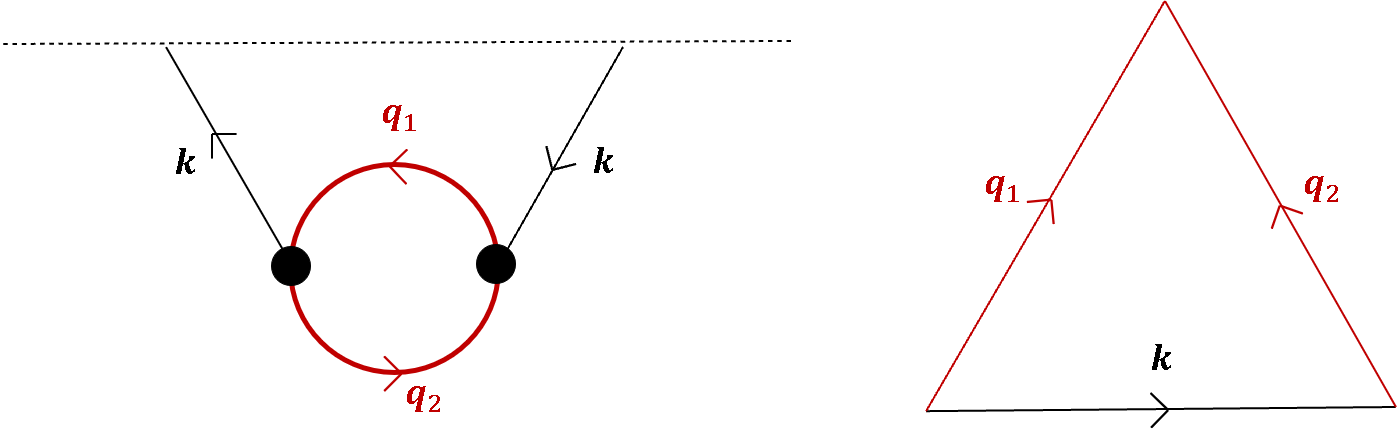}
    \caption{The one-loop contribution to $\psil{1}_2$ with two internal lines is described by three momenta which, as vectors, form the edges of a triangle due to momentum conservation.}
    \label{fig:triangle}
\end{figure}

\paragraph{Integration variables.}
Integrals of the form \eqref{eqn:2edge_int} can be simplified by choosing an appropriate set of integration variables.
When $d \geq 2$, it is possible to choose the two lengths $\{ q_1 , q_2 \}$ (together with a further $d-2$ angular co-ordinates on which the integrand does not depend). 
Given the momentum conservation constraint, the three lengths $\{ q_1, q_2, k \}$ must correspond to the edges of a triangle, as shown in Fig.~\ref{fig:triangle}. 
The domain of integration for $\{ q_1, q_2 \}$ can then be determined by the condition that such a triangle exists, namely that the three triangle inequalities are satisfied\footnote{
Given three positive numbers $\{ q_1, q_2, k\}$, \eqref{eqn:2_vertex_tri} is a necessary and sufficient condition for a triangle with these edge lengths to exist. It can also be written as a single non-linear condition, namely that the area of the formed triangle is positive, cf. \eqref{eqn:3_vertex_det}.
},
\begin{align}
    q_1 + q_2 \geq k \; , \;\; q_1 + k \geq q_2 \; , \;\; q_2 + k \geq q_1 \; . 
    \label{eqn:2_vertex_tri}
\end{align}
Note that these inequalities are only saturated when the triangle degenerates into a line. 

To change from $\{ \bfq_1, \bfq_2 \}$ to these variables, we can parametrise the two internal momenta as,
\begin{align}
    \bfq_1 = \frac{1}{2} \left( \bfp + \bfk \right) \; , \;\; \bfq_2 = \frac{1}{2} \left(  \bfp - \bfk  \right) \; ,
\end{align}
and then align one of our co-ordinate axes along $\bfk$, adopting the polar parametrisation,
\begin{align}
\bfk = \left( \begin{array}{c} k \\  \mathbf{0}_{d-1}  \end{array} \right) \; , \;\;\;\; 
\bfp =  \left( \begin{array}{c}  q_+ \cos \vartheta  \\ \sqrt{q_+^2 - k^2} \sin \vartheta  \hat{\bfp}_{d-1}  \end{array} \right) \; , 
\end{align}
where $\hat{\bfp}_{d-1}$ is a unit vector in the $(d-1)$ directions orthogonal to $\bfk$ and $\mathbf{0}_{d-1} = ( 0, ... , 0)$ is the vector of $(d-1)$ zeroes.
Integrating over the undetermined loop momentum $\bfp$ is then equivalent to integrating over $\{ q_+, \vartheta, \hat{\bfp}_{d-1} \}$, where
\begin{align}
 q_+ = q_1 + q_2 \; , \;\; k \cos \vartheta = q_1 - q_2 ,
\end{align}
and so surfaces of constant $q_+$ correspond to ellipses (with foci separated by $\bfk$) and $\vartheta$ is the eccentric anomaly of a point on each ellipse. 
We will often use the notation $q_- = q_1 - q_2$ in place of the angular $k \cos \vartheta$. 

In practice, this allows us to write integrals of the form \eqref{eqn:2edge_int} as an integral over two scalars\footnote{
One simple way to find this Jacobian is to first express $d^d \bfq_1$ in terms of spherical polars about the $\bfk$ axis (i.e. set $\bfq_1 \cdot \bfk = q_1 k \cos \theta$), and then use the explicit change of variables,
\begin{align}
q_{\pm} = q_1 \pm \sqrt{k^2 + q_1^2 - 2 q_1 k \cos \theta} \; , 
\label{eqn:2vertexJacobian}
\end{align} 
},
\begin{align}
\int \frac{d^d \bfp}{ (2 \pi )^d } \; \mathcal{I} ( \Omega_{q_1}  , \Omega_{q_2} ) = \frac{ S_{d-2} }{ (2 \pi )^d } \int_k^\infty d q_+ \int_{-k}^{+k} \frac{d q_-}{2k} \; q_1^{d-2} q_2  \;  \mathcal{I} \left( \Omega_{q_1} , \Omega_{q_2}  \right) \big|_{\substack{q_1 = \tfrac{1}{2} ( q_+ + q_- ) \\ q_2 = \tfrac{1}{2} ( q_+ - q_- ) }} \; , 
\label{eqn:2vertexMeasure}
\end{align}
where the integration region has been determined using the triangle inequalities, \eqref{eqn:2_vertex_tri}.

Finally, when considering massive fields it is often convenient to replace $q_{\pm}$ with the total/relative energy, $\Omega_{q_1} \pm \Omega_{q_2}$, which we denote by,
\begin{align}
 \Omega_{\pm} = \sqrt{  q_1^2 + M_1^2 } \pm \sqrt{ q_2^2 + M_2^2 }
 \end{align} 
The Jacobian between the two is straightforward,
\begin{align}
    \left( q_+^2 - q_-^2 \right)   \; d q_+ \; d q_- = \left( \Omega_+^2 - \Omega_-^2  \right) \, d \Omega_+ \; d \Omega_-  \; .
\end{align}
To find the integration region, we write $\bfq_1$ in polar co-ordinates about the $\bfk$ axis (as in \eqref{eqn:2vertexJacobian}), 
\begin{align}
    q_1 &= \frac{1}{2} \sqrt{ ( \Omega_+ + \Omega_- )^2 - 4 M_1^2 } \; , \;\;
    &\cos \theta &= - \frac{ \Omega_+ \Omega_- + k^2 + M_2^2 - M_1^2 }{ k \sqrt{ ( \Omega_+ + \Omega_- )^2 - 4 M_1^2} } ,
\end{align}
which makes it clear that, 
\begin{align}
    ( \Omega_+ + \Omega_- )^2 > 4 M_1^2  \;\;\;\; \text{and} \;\;\;\; | \Omega_+ \Omega_- + k^2 + M_2^2 - M_1^2 | < k \sqrt{ ( \Omega_+ + \Omega_- )^2 - 4 M_1^2} \; ,
\end{align}
and similarly for $M_1 \leftrightarrow M_2$. 
We can therefore write the allowed integration region as,
\begin{align}
    \int_{k}^{\infty} d q_+  \int_{-k}^{k} \frac{ d q_-}{k}  \left( q_+^2 - q_-^2 \right) 
    =
    \int_{\Omega_k}^{\infty} d \Omega_+  \int_{- k \, \delta_k ( \Omega_+ )}^{+ k \, \delta_k ( \Omega_+ )} \frac{ d \Omega_-}{k}  \left( \Omega_+^2 - \Omega_-^2 \right)\,,
\end{align}
where,
\begin{align}
\Omega_k &= \sqrt{k^2 + ( M_1 + M_2 )^2} \; ,   \nonumber \\
\delta_k \left( \Omega_+ \right) &= 
\frac{ \sqrt{ \Omega_+^2 - k^2 - (M_1 - M_2)^2 } \sqrt{ \Omega_+^2 - \Omega_k^2 } }{ \Omega_+^2 - k^2 } +  \frac{\Omega_+}{k} \, \frac{ | M_1^2 - M_2^2 | }{ \Omega_+^2 - k^2 } \; . 
\end{align}
In the case of equal masses, these simplify to,
\begin{align}
    \Omega_k = \sqrt{ k^2 + 4 M^2} \;\; , \;\; \delta_k \left( \Omega_+ \right) =  \frac{ \sqrt{ \Omega_+^2 - \Omega_k^2 } }{ \sqrt{ \Omega_k^2 - k^2 } } \; . 
\end{align}
To keep the following explicit expressions as succinct as possible, for the remainder of this subsubsection we are going to work with the rescaled variables,
\begin{align}
    \hat{\omega}_a = \frac{\omega_a}{\sqrt{ k^2 + 4M^2 }} \; , \;\; \hat{k} = \frac{k}{\sqrt{k^2 + 4 M^2}} \; , \;\; \hat{m} = \frac{m}{\sqrt{k^2 + 4M^2}} \; , \;\;  \hat{\Omega}_{\pm} = \frac{\Omega_{\pm}}{ \sqrt{k^2 + 4M^2} } \; . 
\end{align}
Note that $0 \leq \hat{k} \leq 1$ is bounded in these variables, and since $4 \hat{M}^2 = 1 - \hat{k}^2$ we can choose to write functions in terms of $\hat{k}$ or $\hat{M}$ as convenient. 
Altogether, \eqref{eqn:2edge_int} can therefore be written as,
\begin{align}
\int \frac{d^d \bfp}{ (2 \pi )^d } \; \mathcal{I} ( \Omega_{q_1}  , \Omega_{q_2} ) 
= 
\frac{ \Omega_k^3 S_{d-2} }{ 8 \hat{k}  (2 \pi )^d } \int_1^\infty d \hat{\Omega}_+ \int_{-\hat{k} \delta_k ( \Omega_+ )}^{+\hat{k} \delta_k ( \Omega_+ )} d \hat{\Omega}_-  \; \left( \hat{\Omega}_+^2 - \hat{\Omega}_-^2 \right)  \; q_1^{d-3} \mathcal{I} \left( \Omega_{q_1} , \Omega_{q_2}  \right) \big|_{\substack{\Omega_{q_1} = \tfrac{1}{2} k ( \hat{\Omega}_+ + \hat{\Omega}_- ) \\ \Omega_{q_2} = \tfrac{1}{2} k ( \hat{\Omega}_+ - \hat{\Omega}_- ) }} \; .
\end{align}

\paragraph{Master integrals.}
It is useful to define a small number of master integrals from which other, more complicated, integrals can be constructed. 
For $\psil{1}_2$, we define two such building blocks,
\begin{align}
\mathcal{J}_k (B) &= \int_1^{\infty} d \hat{\Omega}_+  \int_{-\hat{k} \delta_k (\Omega_+) }^{+\hat{k} \delta_k (\Omega_+)} d \hat{\Omega}_-   \frac{ 1 }{  \hat{\Omega}_+ + B  }  \; , \nonumber \\ 
\mathcal{J}_k \left( A ; B \right) &= \int_1^{\infty} d \hat{\Omega}_+  \int_{-\hat{k} \delta_k (\Omega_+) }^{+\hat{k} \delta_k (\Omega_+)} d \hat{\Omega}_-  \frac{ 1 }{ (   \hat{\Omega}_+ + \hat{\Omega}_- + A  ) ( \hat{\Omega}_+ + B  ) }   \; . \label{eqn:Jk_def}
\end{align}
Other integrals can then be written straightforwardly in terms of these using partial fractions. For instance, consider,
\begin{align}
\mathcal{I} ( A ; B_1 , B_2 ) &= \int_1^{\infty} d \hat{\Omega}_+  \int_{-\hat{k} \delta_k (\Omega_+) }^{+\hat{k} \delta_k (\Omega_+)} d \hat{\Omega}_-  \, \frac{ \hat{\Omega}_+^2 - \hat{\Omega}_-^2 }{ ( \hat{\Omega}_+ + \hat{\Omega}_- + A ) ( \hat{\Omega}_+ + B_1 ) ( \hat{\Omega}_+ + B_2 )  }  \; . 
\end{align}
Expanding the numerator in terms of the factors appearing in the denominator,
\begin{align}
 \Omega_+^2 - \Omega_-^2 &= 
   \left(   \frac{ B_1 - A }{B_1 - B_2} ( \Omega_+ + B_2) +  \frac{ B_2 - A}{B_2 - B_1} ( \Omega_+ + B_1 )  -   \Omega_- \right) ( \Omega_+ + \Omega_- + A)  \nonumber \\
  &+  
 \frac{ A ( A - 2 B_1 ) }{B_1- B_2} ( \Omega_+ + B_2 ) + \frac{ A  ( A -  2 B_2 ) }{B_2-B_1} ( \Omega_+ + B_1 )
\end{align}
immediately gives,
\begin{align}
\mathcal{I} ( A ; B_1 , B_2 ) = \frac{B_1-A}{B_1-B_2} \mathcal{J}_k (B_1)  + \frac{B_2 - A}{B_2-B_1} \mathcal{J}_k (B_2) 
+ \frac{ A (A - 2 B_1 ) }{ B_1 - B_2} \mathcal{J}_k ( A ; B_1 ) + \frac{ A ( A - 2 B_2 ) }{ B_2 - B_1 } \mathcal{J}_k ( A ; B_2 ) \; . 
\label{eqn:IB1B2}
\end{align}
As discussed in Sec.~\ref{sec:complexity}, the apparent poles at $B_1 = B_2$ are spurious: the only singularities in $\mathcal{I} (A ; B_1, B_2 )$ come from the master integrals $\mathcal{J}_k (B_j)$ and $\mathcal{J}_k (A; B_j )$.

\subsubsection{Equal masses}
\label{sec:2edge_psi2}
For the remainder of this subsection we are going to analyse the integral,
\begin{align}
 \omega_{12} \psil{1}_2  &=  \int_{\bfp}   \, \frac{ 1 }{ ( \omega_1 + \Omega_{q_1} + \Omega_{q_2} ) ( \omega_2 + \Omega_{q_2} ) } \left[ \frac{1}{ \omega_{12} + 2 \Omega_{q_1}} + \frac{1}{\omega_{12} + 2 \Omega_{q_2}} \right] \; . 
\end{align}
Changing to the $\Omega_{\pm}$ variables defined above, this can be written as,
\begin{align}
 \omega_{12} \psil{1}_2  &= \frac{1}{16 \pi^2 \hat{k}} \mathcal{I} \left( \hat{\omega}_{12} ; \hat{\omega}_1, \hat{\omega}_2 \right) .
 \label{eqn:psi2_integral_massive}
\end{align}
where $\mathcal{I} \left( A; B_1, B_2 \right)$ is given in \eqref{eqn:IB1B2} in terms of the master integrals $\mathcal{J}_k (B)$ and $\mathcal{J}_k (A ; B)$.
We will now evaluate and discuss each of these master integrals in turn---see Table~\ref{tab:Jk_singularities} for a concise summary of their singular points.

\subsubsection*{(i) The $\mathcal{J}_k (B)$ integral}

Performing the $\hat{\Omega}_-$ integral in \eqref{eqn:Jk_def}  produces a factor of $\delta_k ( \Omega_+)$ in the numerator, which we write as $( \hat{\Omega}_+^2 - 1)/\sqrt{ (\hat{\Omega}_+^2 - 1)( \hat{\Omega}_+^2 - \hat{k}^2 )  }$ in order to isolate the UV divergence at large $\hat{\Omega}_+$.
This splits $\mathcal{J}_k (B)$ into,
\begin{align}
\frac{1}{2 \hat{k}} \mathcal{J}_k ( \hat{\omega} ) = \tilde{\mathcal{J}}_k  + \tilde{\mathcal{J}}_k ( \hat{\omega} )  + \mathcal{O} (\hat{\omega})
\label{eqn:Jk1_split}
\end{align}
where,
\begin{align}
    \tilde{\mathcal{J}}_k  &=  \int_{1}^{\infty} d \hat{\Omega}_+ \; \frac{ \hat{\Omega}_+ }{ \sqrt{ \hat{\Omega}_+^2 - 1 } \sqrt{ \hat{\Omega}_+^2 - \hat{k}^2 } }  \; , \nonumber \\ 
    \tilde{\mathcal{J}}_k ( \hat{\omega} )  &=  \int_{1}^{\infty} d \hat{\Omega}_+ \; \frac{ \hat{\omega}^2 - 1 }{  \sqrt{ \hat{\Omega}_+^2 - 1 } \sqrt{ \hat{\Omega}_+^2 - \hat{k}^2 } ( \hat{\omega} + \hat{\Omega}_+ ) } \; , \label{eqn:Jk1_def}  
\end{align}
and the term linear in $\hat{\omega}$ does not contribute to $\mathcal{I} ( \omega_{12} ; \omega_1, \omega_2)$. 
The virtue of this split is that now only $\tilde{\mathcal{J}}_k$ is UV divergent, while $\tilde{\mathcal{J}}_k ( \hat{\omega} )$ (and also $\mathcal{J}_k ( \hat{\omega}_{12}, \hat{\omega}_1 )$ and $\mathcal{J}_k ( \hat{\omega}_{12}, \hat{\omega}_2 )$) are finite and responsible for the non-analytic structure. 

The UV divergent integral can be performed straightforwardly using a hard cut-off, $\int_{1}^{\hat{\Lambda}} d \hat{\Omega}_+$, which gives simply,
\begin{align}
    \tilde{\mathcal{J}}_k  &=  \log \left( \frac{\Lambda}{M}  \right) \; . 
    \label{eqn:Jk0_log}
\end{align}
This does not contribute to any non-analyticity in the complex $\omega$ planes at fixed $M$. 

Given the rational form of the finite remainder $\tilde{\mathcal{J}}_k (\hat{\omega} )$ in \eqref{eqn:Jk1_def}, we expect $\mathcal{T} \left( \mathcal{J}_k ( \hat{\omega} ) \right) = 1$ and $\mathcal{G} \left( \mathcal{J}_k ( \hat{\omega} ) \right) = 1$. 
This is the same complexity as the elliptic integrals, and indeed $\tilde{\mathcal{J}}_k ( \omega )$ integral can be written in terms of (upper\footnote{
The tilde is used to distinguish these from the more common lower incomplete integrals, which are defined using $\int_0^{\phi} dt$. 
}) elliptic integrals of the first and third kinds, which are defined by the integral representations,
\begin{align}
    \tilde{F} ( \phi , n ) = \int_\phi^1 \frac{dt}{\sqrt{1 - t^2} \sqrt{1- n^2 t^2} }  \;\; , \;\;\;\;
    \tilde{\Pi} ( \phi , n_1 , n_2 ) = \int_\phi^1 \frac{dt}{\sqrt{1 - t^2} \sqrt{1- n_1^2 t^2} ( 1 - n_2^2 t^2) } \; . 
\end{align}
Explicitly, 

\begin{align}
     \tilde{\mathcal{J}}_k ( \hat{\omega} ) &=  - \frac{ 2i }{1-\hat{k}} \left[(\hat{\omega}-1) \tilde{F} \left( \sqrt{\alpha_k} , \frac{1}{\alpha_k^2} \right) 
    + 2 \, \tilde{\Pi}\left( \sqrt{\alpha_k} , \frac{ \hat{\omega} + 1 }{\alpha_k ( \hat{\omega} - 1)} , \frac{1}{\alpha_k^2}  \right) \right] + \mathcal{O} \left( \hat{\omega} \right) \; ,
\label{eqn:Jk1_elliptic}
\end{align}
where we have discarded terms that are linear in $\omega$ (since these do not contribute to $\psil{1}_2$) and introduced the ratio,
\begin{align}
    \alpha_k = \frac{1 - \hat{k} }{1 + \hat{k} } \; .
\end{align}

\paragraph{Analytic structure.}
The representation \eqref{eqn:Jk1_elliptic} is useful because the singularities of these special functions are well-documented, and in particular \eqref{eqn:Jk1_elliptic} is an analytic function of $\omega_1$ at fixed $\{ \omega_2, k, M \}$ except for a single branch cut\footnote{
Note that, despite appearances, $\tilde{\mathcal{J}}_k (\omega)$ is analytic at $\omega = \pm 1$, since the elliptic $\tilde{\Pi}$ integral has series expansions,
\begin{align}
    \tilde{\Pi} \left( \sqrt{\alpha} , \frac{ \omega + 1 }{\alpha ( \omega - 1 ) } , \frac{1}{\alpha^2} \right) = \sum_{n=1}^{\infty} a_n (k) ( \omega - 1 )^n   
    = \tilde{F} \left( \sqrt{\alpha_k} ,  \frac{1}{\alpha_k^2} \right) + \sum_{n=0}^\infty A_n (k)  (\omega + 1 )^n \; . 
\end{align}
} at $\omega_1 = - \sqrt{k^2 + 4 M^2 }$.
This is most easily seen from a series expansion of $\tilde{\mathcal{J}}_k ( \hat{\omega})$ in $\epsilon = \hat{\omega} + 1$,  
\begin{align}
   \tilde{\mathcal{J}}_k ( -1 + \epsilon ) = \sqrt{\epsilon} \sum_{n=0}^\infty \epsilon^n B_n (k) + \sum_{n=1}^\infty \epsilon^n C_n ( k ) \; ,
    \label{eqn:Jk1_form}
\end{align}
which reveals that there is a square-root branch cut, $\tilde{\mathcal{J}}_k ( \hat{\omega} ) \sim \sqrt{ \hat{\omega} + 1}$, near the two-particle threshold. 

While the functions $B_n (k)$ and $C_n (k)$ appearing in \eqref{eqn:Jk1_form} can be found by looking up the series expansions of elliptic integrals, it is instructive to show how they can be derived from the original integral \eqref{eqn:Jk1_def} directly.
This can be done using the \emph{method of regions} familiar from Feynman integrals for amplitudes.
Changing integration variables with $\Omega_+ = 1 + z^2$ allows us to write $\mathcal{J}_k ( \omega)$ as,
\begin{align}
    \tilde{\mathcal{J}}_k ( -1 + \epsilon ) =  2 \epsilon ( \epsilon -2 ) \int_0^{\infty} d z \,  g_k ( z^2 )  \, f ( z^2 + \epsilon ) \; , 
    \label{eqn:Jk1_regions}
\end{align}
where,
\begin{align}
    g_k ( z^2 ) = \frac{1}{ \sqrt{ (z^2 + 2) ( z^2 + 1 - k) ( z^2 + 1 + k ) }} \;\; , \;\;\;\; f ( z^2 + \epsilon ) = \frac{1}{z^2 + \epsilon}  \; . 
\end{align}
The method of regions is a strategy for determining the small $\epsilon$ expansion of the integral in terms of the expansion of the integrand. 
The idea is to split the integration into two pieces, $\left( \int_0^{\Delta} + \int_{\Delta}^{\infty} \right) dz$, where $\Delta$ is chosen in the range $\epsilon \ll \Delta^2 \ll 1 - \hat{k}$ but is otherwise arbitrary. Then in the first of these integration regions we can expand $g_k (z^2)$ at small $z$ and in the second region we can expand $f (z^2 + \epsilon )$ at small $\epsilon$. 
Since $\Delta$ is an arbitrary separation that we have introduced by hand, any dependence on it will drop out of the final result (once both regions are summed). 
Applying this method to \eqref{eqn:Jk1_regions} gives,
\begin{align}
    \tilde{\mathcal{J}}_k ( -1 + \epsilon ) = 2 \epsilon ( \epsilon - 2 ) \left( I_k^{<} (\epsilon , \Delta ) + I_k^{>} ( \epsilon, \Delta ) 
    \right)\; ,
\end{align}
where the first two terms in each region are,
\begin{align}
    I_k^{<} ( \epsilon, \Delta ) 
    &=  \int_{0}^{\Delta} d z \, f ( z^2 + \epsilon ) \, \left( g_k (0 ) + g'_k (0) z^2 + \mathcal{O} \left( z^4 \right) \right)  \nonumber \\
    &=  \Delta^1 \left[  g_k' ( 0 ) + \mathcal{O} ( \epsilon )  \right] +  \Delta^0 \left[  \frac{\pi g_k (0) }{ 2 \sqrt{\epsilon}}  - \frac{ \pi g_k' (0)}{2} \sqrt{\epsilon} +  \mathcal{O} \left( \epsilon^{3/2} \right) \right]   \nonumber \\
    &+ \frac{1}{\Delta} \left[  - g_k (0) + \epsilon g_k' (0)  + \mathcal{O} \left( \epsilon^2 \right)  \right] + \frac{1}{\Delta^3} \left[  \frac{g_k(0)}{3} \epsilon  + \mathcal{O} \left( \epsilon^2 \right)    \right] + \mathcal{O} \left(  \frac{\epsilon^2}{\Delta^5 } , \Delta^3 \right)  \; , 
    \label{eqn:I_lower}
\end{align}
and (replacing $z$ with $y = z^2$),
\begin{align}
    I_k^{>} ( \epsilon, \Delta )  
    &= \int_{\Delta^2}^{\infty} \frac{d y}{2 \sqrt{y}} \, g_k ( y) \, \left(  f (y ) + \epsilon f'(y) + \mathcal{O} \left( \epsilon^2 \right)  \right) \nonumber \\
        &=  \epsilon^0 \left[ \frac{ g_k(0) }{\Delta} + c_0 ( \hat{k} )  - g_k' (0) \Delta + \mathcal{O} \left( \Delta^3 \right) \right]  
    + \epsilon \left[ 
    \frac{ - g_k (0) }{ 3 \Delta^{3} } - \frac{ g_k' ( 0)  }{ \Delta } 
    + c_1 ( \hat{k} ) + \mathcal{O} \left( \Delta \right)  \right] + \mathcal{O} \left( \epsilon^2 \right) \; . 
    \label{eqn:I_upper}
\end{align}
where the $c_n ( \hat{k})$ are particular elliptic functions of $\hat{k}$ only, the explicit form of which is unimportant for determining analyticity in $\epsilon$. 
When summed, $I_k^> + I_k^<$ is indeed independent of $\Delta$, and in fact takes the form \eqref{eqn:Jk1_form} as previously advertised. In particular, the $\Delta^0$ terms in \eqref{eqn:I_lower} determines the leading branch cut discontinuities to be\footnote{
Note that continuing to arbitrary orders in $\epsilon$ in the lower integral fully determines the branch cut discontinuity for all $n \geq 1$,
\begin{align}
    B_n (k) = \pi \left(  g^{(n)}_k ( 0 )  - 2 g_k^{(n+1)} ( 0 )  \right) \; .
\end{align}
},  
\begin{align}
        B_0 (k ) &= - 2\pi g_k (0 ) = -\frac{2\pi }{\sqrt{1-k^2}}  \; ,  \nonumber \\ 
        B_1 ( k ) &= \pi \left(   g_k (0) + 2 g_k' (0) \right)  =  \frac{B_0 (k)}{1 - k^2} \; . 
\end{align}
In summary, $\mathcal{J}_k (B)$ can be written explicitly in terms of elliptic integrals (see \eqref{eqn:Jk1_elliptic}), and has a square-root branch point at $B = -1$.

\subsubsection*{(ii) The $\mathcal{J}_k ( A; B)$ integrals}

Now we turn to the master integral $\mathcal{J}_k ( A; B)$ defined in \eqref{eqn:Jk_def}, which has $( \mathcal{T} , \mathcal{G} ) = (2,1)$. 
Performing this integral explicitly would therefore require the use of special functions such as the elliptic-dilogarithm.
Since our goal is merely to establish the analytic structure of $\mathcal{J}_k ( \hat{\omega}_1 , \hat{\omega}_2 )$ and $\mathcal{J}_k ( \omega_2, \omega_1 )$ in the complex $\omega_1$ plane, we will instead consider their first $\omega_1$ derivatives, since these share the same singular points in the complex $\omega_1$ plane but have a lower degree of transcendentality. 
These derivatives are explicitly given by,
\begin{align}
     \partial_{\hat{\omega}_1} \mathcal{J}_k ( \hat{\omega}_{12} , \hat{\omega}_2 ) &= \int_{1}^{\infty} d \hat{\Omega}_+  \; \frac{ (\hat{\Omega}_+^2 - \hat{k}^2)(\hat{\Omega}_{+}^{2}-1)  }{ \sqrt{  \hat{\Omega}_+^2 - 1 } \sqrt{ \hat{\Omega}_+^2 - \hat{k}^2 } ( \hat{\Omega}_+ + \hat{\omega}_2  ) } \;  \frac{2 \hat{k} }{ D_k ( \hat{\Omega}_+ , \hat{\omega}_{12} ) } \; , \nonumber \\
    \partial_{\hat{\omega}_1} \mathcal{J}_k ( \hat{\omega}_{12} , \hat{\omega}_1 ) &=   \int_{1}^{\infty} d \hat{\Omega}_+  \; \frac{ \hat{\Omega}_+ ( \hat{\Omega}_+ + \hat{\omega}_{12} ) }{ \sqrt{  \hat{\Omega}_+^2 - 1 } \sqrt{ \hat{\Omega}_+^2 - \hat{k}^2 } ( \hat{\Omega}_+ + \hat{\omega}_1 ) }  \;  \frac{ 8 \hat{M}^2 \hat{k} }{ D_k ( \hat{\Omega}_+ , \hat{\omega}_{12}   )   }  \; , 
    \label{eqn:Jk2_derivative}  
\end{align}
where the polynomial denominator is,
\begin{align}
    D_k ( \hat{\Omega}_+ , \hat{\omega}_{12} ) &= ( \hat{\Omega}_+ + \hat{\omega}_{12} )^2 ( \hat{\Omega}_+^2 - \hat{k}^2 ) -  \hat{k}^2 ( \hat{\Omega}_+^2 - 1 ) \; . 
\end{align}

By using partial fractions, it is possible to write \eqref{eqn:Jk2_derivative} in terms of the $\tilde{\mathcal{J}}_k ( \omega )$ integrals studied above, 
\begin{align}
 \partial_{\hat{\omega}_1} \mathcal{J}_k ( \hat{\omega}_{12} ; \hat{\omega}_2 ) 
 &= -\frac{  \hat{k}  \tilde{\mathcal{J}}_k ( \hat{\omega}_2 ) (\omega_{2}^2-k^2)}{ (\hat{\omega}_1^2 - \hat{k}^2 \delta ( \omega_2 )^2)(\hat{\omega}_2^2 - \hat{k}^2 \delta ( \omega_1 )^2) } 
    - \frac{ 1 }{ 2\sqrt{ \hat{\omega}_{12}^2 - 4 \hat{M}^2 } } 
     \frac{ \bar{z}_{+, k}^2   - \hat{k}^2 }{ \bar{z}_{+, k}  - \hat{\omega}_2} \frac{ \tilde{\mathcal{J}}_k ( \bar{z}_{+, k}  )}{ 2 \bar{z}_{+, k}  - \hat{\omega}_{12} } \nonumber \\
     &+\frac{ 1 }{ 2\sqrt{ \hat{\omega}_{12}^2 - 4 \hat{M}^2 } } \frac{ \bar{z}_{-, k}^2   - \hat{k}^2 }{ \bar{z}_{-, k}  - \hat{\omega}_2} \frac{ \tilde{\mathcal{J}}_k ( \bar{z}_{-, k}  )}{ 2 \bar{z}_{-, k}  - \hat{\omega}_{12} }  - \left( k \leftrightarrow -k  \right) 
     \; , \nonumber \\ 
 \partial_{\hat{\omega}_1} \mathcal{J}_k ( \hat{\omega}_{12} , \hat{\omega}_1 ) 
 &= - \frac{ 4 \hat{M}^2 \hat{\omega}_1 \hat{\omega_2}  }{ \left( \hat{\omega}_1 ^2 - 1 \right) \left( \hat{\omega}_1^2 - \hat{k}^2\delta(\omega_{2})^2 \right) } \frac{ 2 \hat{k} \tilde{\mathcal{J}}_k ( \hat{\omega}_1 ) }{ \hat{\omega}_2^2 - \hat{k}^2 \delta ( \omega_1 )^2 } 
-   \frac{ 4 \hat{M}^2 }{  \sqrt{ \omega_{12}^2 - 4M^2} }   
    \frac{ \bar{z}_{+, k} \bar{z}_{-, k} }{ ( \bar{z}_{+, k}^2  - 1 ) ( \bar{z}_{+,k} - \hat{\omega}_1 ) } \frac{ \tilde{\mathcal{J}}_k ( \bar{z}_{+, k} ) }{ 2 \bar{z}_{+, k} -  \hat{\omega}_{12}  } \nonumber \\
    &+\frac{ 4 \hat{M}^2 }{  \sqrt{ \omega_{12}^2 - 4m^2} }  \frac{ \bar{z}_{+, k} \bar{z}_{-, k} }{ ( \bar{z}_{-, k}^2  - 1 ) ( \bar{z}_{-,k} - \hat{\omega}_1 ) } \frac{ \tilde{\mathcal{J}}_k ( \bar{z}_{-, k} ) }{ 2 \bar{z}_{-, k} -  \hat{\omega}_{12}  }  
    - \left( k \leftrightarrow -k \right)
     \; ,
     \label{eqn:Jk2_derivative_as_Jk1}
\end{align}
where the four $\bar{z}$ are (minus) the roots of the $D_k$ polynomial,
\begin{align}
    D_k ( \hat{\Omega}_+ , \hat{\omega}_{12} ) 
    =
    ( \hat{\Omega}_+ + \bar{z}_{+, k}  )
    ( \hat{\Omega}_+ + \bar{z}_{-, k}  )
    ( \hat{\Omega}_+ + \bar{z}_{+, -k}  )
    ( \hat{\Omega}_+ + \bar{z}_{-, -k} ) \; .  
\end{align}
and can be written explicitly as\footnote{
Note that if we parameterise $\omega_{12} = 2\sqrt{p^2 + m^2}$, these roots can be written simply as $\bar{z}_{\pm, k} = \Omega_p \pm \Omega_{p + k}$.
},
\begin{align}
 \bar{z}_{\pm, k} ( \hat{\omega}_{12} ) = \frac{\hat{\omega}_{12}}{2} \pm \sqrt{ \frac{\hat{\omega}_{12}^2 }{4} + \hat{k}^2  + \hat{k} \sqrt{ \frac{\hat{\omega}_{12}^2 }{4} - \hat{M}^2   }  }   \; .
\end{align}

\paragraph{Analytic structure.}
All of the apparent poles in \eqref{eqn:Jk2_derivative_as_Jk1} introduced by the partial fractioning correspond to spurious singularities with zero residue.
To show this requires some care, since the $\bar{z}_{\pm, k} ( \hat{\omega}_{12} )$ are not analytic functions of $\omega_1$.
In particular, they have branch points at $\hat{\omega}_{12} = \pm 2M$ and at the four complex values $\hat{\omega}_{12} = 2 \sqrt{ \hat{k}} e^{i \varphi/2}$ where $\cos \varphi = \hat{k}$. 
However, these non-analyticities appear in pairs exactly analogous to the example \eqref{eqn:spurious_cut_eg} given above. 
In more detail, the spurious singularities are:
\begin{itemize}
    
    \item[(i)] at any $\bar{z} ( \hat{\omega}_{12} ) = \hat{\omega}_2$, which corresponds to $\hat{\omega}_1 = \pm k \delta_k ( \omega_2)$. This is spurious, since $\bar{z}$ is an analytic function of $\hat{\omega}_1$ near this point. To see the cancellation explicitly, fix $\omega_2$ in the interval $0 \leq \omega_2 < k$ and then expand $\hat{\omega}_1 = - k \delta_k (\omega_2 ) + \epsilon$. This gives,
\begin{align}
    \bar{z}_{-, k}  &= 
    \hat{\omega}_2 - \epsilon \; \delta_k ( \omega_2 ) \, \frac{ \hat{\omega}_2^2 - \hat{k}^2 }{ ( \hat{k} - \hat{\omega}_2 \delta_k ( \omega_2 ) ) (  k \delta_k ( \omega_2 ) + \hat{\omega}_2 ) } + \mathcal{O} ( \epsilon^2 )  \; , \;\; 
\end{align}
so we have a cancellation between the following two terms,
\begin{align}
 \left[ -\frac{  \hat{k} \, \tilde{\mathcal{J}}_k ( \hat{\omega}_2 ) (\omega_{2}-\hat{k}^2) }{ (\hat{\omega}_1^2 - \hat{k}^2 \delta ( \omega_2 )^2)(\hat{\omega}_2^2 - \hat{k}^2 \delta ( \omega_1 )^2) } 
     + \frac{1}{2\sqrt{\hat{\omega}_{12}^2 - 4\hat{M}^2 }} \frac{ \bar{z}_{-,k}^2  - \hat{k}^2 }{ \bar{z}_{-,k}  - \hat{\omega}_2} \frac{ \tilde{\mathcal{J}}_k ( \bar{z}_{-,k} )}{ 2 \bar{z}_{-,k} - \hat{\omega}_{12} }   \right] \Bigg|_{\hat{\omega}_1 = - \hat{k} \delta_k (\omega_2 ) + \epsilon} = \mathcal{O} \left( \epsilon^0 \right)  \; , 
\end{align}
which leaves a remainder that is analytic in $\epsilon$ (and hence analytic in $\hat{\omega}_1$). 

\item[(ii)] at any $\bar{z} ( \hat{\omega}_{12} ) = \hat{\omega}_1$, which corresponds to $\hat{\omega}_2 = \pm k \delta_k ( \omega_1)$. This is spurious since $\bar{z}_k$ is an analytic function of $\hat{\omega}_1$ near this point, just as (i) above. 

\item[(iii)] at $\omega_{12} = \pm 2m$, which corresponds to the roots degenerating into two pairs. 
Physically, this singularity corresponds to one of the internal edges carrying zero energy, which is why it coincides with the threshold from the one-edge loop diagram.
For instance, expanding around $\omega_{12} = + 2M + \epsilon$, 
\begin{align}
    \bar{z}_{\pm, k} ( 2m + \epsilon ) =  
    \hat{M} \pm 2 \sqrt{ \hat{M}^2 + \hat{k}^2 } - \sqrt{\epsilon} \frac{ \hat{k} \sqrt{ \hat{M} } }{ \sqrt{ \hat{M}^2 +  \hat{k}^2} } + \mathcal{O} ( \epsilon )  \; ,  \nonumber \\
    \bar{z}_{\pm, -k} ( 2m + \epsilon ) =  
    \hat{M} \pm 2 \sqrt{ \hat{M}^2 + \hat{k}^2 } + \sqrt{\epsilon} \frac{ \hat{k} \sqrt{ \hat{M} } }{ \sqrt{ \hat{M}^2 +  \hat{k}^2} } + \mathcal{O} ( \epsilon ) \; . 
\end{align}
The square brackets multiplying $1/\sqrt{\hat{\omega}_{12}^2 - 4\hat{M}^2}$ in \eqref{eqn:Jk2_derivative_as_Jk1} therefore have an odd power series in $\sqrt{\epsilon}$, so overall the product is an analytic function of $\epsilon$ (and hence $\omega_1$). 

\item[(iv)] at $\bar{z} ( \hat{\omega}_{12} ) = \hat{\omega}_{12} /2$, which corresponds to $\hat{\omega}_{12} = 2 \sqrt{ \hat{k} } e^{i \varphi /2}$ with $\cos \varphi = \hat{k}$. 
At each of these four points in the $\hat{\omega}_1$ plane, all four of the roots degenerate. 
Physically, this singularity corresponds to both of the internal edges carrying zero energy. 
But as in (iii) above, expanding in $\epsilon = \hat{\omega}_{12} - 2 \sqrt{\hat{k}} e^{i \varphi/2}$ shows that the factor of $1/( 2 \bar{z} - \hat{\omega}_{12} )$ multiplies an odd power series in $\sqrt{\epsilon}$, so overall there is no branch cut in the $\omega_1$ plane.

\end{itemize}

The only physical singularities in \eqref{eqn:Jk2_derivative_as_Jk1} therefore arise from each of the $\tilde{\mathcal{J}}_k (\omega )$ factors. 
In particular, we have shown above that $\tilde{\mathcal{J}}_k ( \hat{\omega}_1 )$ has a branch point at $\hat{\omega}_1 = -1$, and so we see that $\mathcal{J}_k ( \omega_{12}, \omega_1 )$ also has this branch point\footnote{
We also see that there can be no cancellation between the $\hat{\omega}_1 = -1$ singularities in $\tilde{\mathcal{J}}_k (\omega_1)$ and $\mathcal{J}_k (\omega_{12} , \omega_1 )$, since the latter branch cut is the integral of the former.
}. 
The only other potential singularity would be when a root $\bar{z} ( \hat{\omega}_{12} ) = -1$. 
However, for $0 < \hat{k} < 1$, the only value of $\omega_1$ for which this can happen is $\hat{\omega}_1 = -\hat{\omega}_2 - 1$, at which one of the roots has a turning point at $-1$. 
That is to say, if we expand this particular root near the potentially singular value of $\omega_1$,
\begin{align}
    \bar{z} ( -1 + \epsilon ) = -1 - \epsilon^2 \frac{2m^2 }{k^2} + \mathcal{O} ( \epsilon^3 )\; ,
\end{align}
the expansion has no linear term in $\epsilon$. 
So if we repeat the method of regions argument above, we find that,
\begin{align}
    \tilde{\mathcal{J}}_k ( \bar{z} ( -1 + \epsilon )  ) = \epsilon \sum_{n=0}^\infty \epsilon^{2n} \bar{B}_n (k) + \sum_{n=1}^{\infty} \epsilon^{2n} \bar{C}_n ( k )
    \label{eqn:Jk1_z_form}
\end{align}
and this is, in fact, an analytic function of $\epsilon$ (and hence $\hat{\omega}_1$) near $\omega_1 = - \omega_2 - 1$. 
Consequently, we can conclude from \eqref{eqn:Jk2_derivative_as_Jk1} that the only singular points in the complex $\omega_1$ plane (at fixed $\{ \omega_2, k , m\}$) arise from $\mathcal{J}_k ( \omega_{12}, \omega_1)$ at $\hat{\omega}_1 = -1$.

\begin{table}
\centering
\begin{tabular}{c|c|c|c}
     & $\mathcal{J}_k ( \hat{\omega}_1 )$ & $\mathcal{J}_k ( \hat{\omega}_{12} ; \hat{\omega}_1 )$  & $\mathcal{J}_k ( \hat{\omega}_{12} ; \hat{\omega}_2 )$  \\
     \hline 
    Finite mass ($0 < \hat{k} <1$) &   $-1$  &  $-1$ 
    & 
    \\
    Massless ($\hat{k} \to 1$) &   $-1$  &   $-1$ , \; $-\hat{\omega}_2, \;\; {\color{red} - \hat{\omega}_2 -2}$  &  $-\hat{\omega}_2, \;\; {\color{red} - \hat{\omega}_2 -2}$     \\ 
    Soft limit ($\hat{k} \to 0$) &  $-1$  & $-1$ , \; $- \hat{\omega}_2 - 1$   & $-\hat{\omega}_2 - 1$ 
\end{tabular}
\caption{
The location of singular points in the complex $\hat{\omega}_1$ plane of the three integrals from which $\psil{1}_2$ is constructed (given in \eqref{eqn:Jk1_def}). The branch point at $\hat{\omega}_{12} = -2$ which develops in the massless limit is shown in red to indicate that it cancels out in the full $\psil{1}_2$. The singularity at $\hat{\omega}_1 = -1$ (i.e. $\omega_1 = - \sqrt{k^2 + 4M^2}$) predicted by the heuristic argument is present in all cases, and in the massless and soft limits there is also the singularity at $\omega_{12} = - 2M$ predicted from the simpler one-edge loop.  
}
\label{tab:Jk_singularities}
\end{table}

\paragraph{Summary.}
Altogether, the integral \eqref{eqn:psi2_integral_massive} for $\psil{1}_2$ can be written as,
\begin{align}
    \omega_{12} \psil{1}_2  &=  
 \frac{1}{16 \pi^2} \left[ 2\tilde{\mathcal{J}}_k
+ 2\frac{\omega_1 \tilde{\mathcal{J}} ( \hat{ \omega}_2) - \omega_2 \tilde{\mathcal{J}}_k ( \hat{\omega}_1) }{\omega_1 - \omega_2} - \frac{\omega_{12}}{k} \tilde{\mathcal{J}}_k ( \hat{\omega}_1 , \hat{\omega}_2 ) - \frac{\omega_{12}}{k} \tilde{\mathcal{J}}_k ( \hat{\omega}_2, \hat{\omega}_1 ) 
 \right]
 \label{eqn:psi2_ans}
\end{align}
where $\tilde{\mathcal{J}_k}$ is a simple logarithmic divergence given in \eqref{eqn:Jk0_log}, $\tilde{ \mathcal{J}}_k ( B )$ is an elliptic integral given in \eqref{eqn:Jk1_elliptic}, and $\tilde{\mathcal{J}}_k ( A; B )$ is an integral of an elliptic integral given in \eqref{eqn:Jk2_derivative_as_Jk1}. The only singular points in the complex $\omega_1$ plane are at $\hat{\omega}_1 = -1$ (and $\hat{\omega}_1 = - \infty$).

\subsubsection{Massless limit}
\label{sec:2edge_massless}

In the massless limit, $M \to 0$, we have $\hat{k} \to 1$. 
In particular, in order to apply the method of regions in Sec.~\ref{sec:2edge_psi2}, we had to assume that $\epsilon \ll 1 - \hat{k}$.
In the massless limit this is no longer possible, and so $\psil{1}_2$ may contain additional singular points. 

Fortunately, in this limit the master integrals in \eqref{eqn:Jk_def} both have $\mathcal{G} = 0$ and can be evaluated immediately in terms of polylogarithms\footnote{
This follows immediately from the indefinite integral,
\begin{align}
\int  \frac{ d \hat{q}_+  d \hat{q}_- }{ (  A + \hat{q}_+ + \hat{q}_- ) ( B + \hat{q}_+ ) }
=
\text{Li}_2 \left( \frac{B-A - \hat{q}_-}{B+ \hat{q}_+} \right) \; . 
\end{align}
},
\begin{align}
\mathcal{J}_1 (B) &= \int_1^{\hat{\Lambda}} d \hat{q}_+  \int_{-1}^{+1} d \hat{q}_-  \frac{ 1 }{ B + \hat{q}_+  } =  2 \log \left( \frac{ \hat{\Lambda} }{B + 1} \right)   \nonumber \\ 
\mathcal{J}_1 \left( A ; B \right) &= \int_1^{\infty} d \hat{q}_+  \int_{-1}^{+1} d \hat{q}_-  \frac{ 1 }{ (  A + \hat{q}_+ + \hat{q}_- ) ( B + \hat{q}_+ ) }  =  \text{Li}_2 \left( \frac{ B - A + 1 }{B + 1} \right) - \text{Li}_2 \left( \frac{ B - A - 1 }{B + 1} \right)   
\label{eqn:Jk_massless}
\end{align}
Both have a logarithmic branch point at $B = -1$, and $\mathcal{J}_1 (A; B)$ has further dilogarithmic branch points at $A = 0$ and $A = -2$. 

In this limit, $\mathcal{J}_k ( \hat{\omega}_{12} ; \hat{\omega}_1 )$ retains the singularity at $\hat{\omega}_1 = -1$,
\begin{align}
    \lim_{k \to 1} \mathcal{J}_{k} ( \hat{\omega}_{12} , \hat{\omega}_1 ) \sim \log^2 (  \hat{\omega}_1 + 1 ) \;\; \text{near} \;\; \hat{\omega}_1 = -1\;,
\end{align}
identified above for finite masses, but now it also acquires branch points at $\hat{\omega}_1 = - \hat{\omega}_2$ and $\hat{\omega}_1 = - \hat{\omega}_2 - 2$, where the dilogarithms are finite but not smooth. These non-analyticities are due to the $\tilde{\mathcal{J}}_k ( \bar{z} )$ terms in \eqref{eqn:Jk2_derivative_as_Jk1}, since in the massless limit the four roots become,
\begin{align}
    \lim_{k \to 1} \bar{z} ( \hat{\omega}_{12} ) = \{ -1 , -1 + \hat{\omega}_{12} ,  1 + \hat{\omega}_{12} , 1 \} \;,
\end{align}
and are now linear in $\omega_1$.
However, note while $\mathcal{J}_k ( \hat{\omega}_{12}, \hat{\omega}_2 )$ was previously analytic in $\omega_1$ (at fixed finite mass), now this integral also develops branch points at $\hat{\omega}_1 = - \hat{\omega}_2$ and $\hat{\omega}_1 = - \hat{\omega}_2 - 2$. 
In particular, the residue of the $\hat{\omega}_1 = - \hat{\omega}_2 - 1$ branch points are equal and opposite, so that overall the sum of $\mathcal{J}_k ( \hat{\omega}_{12} ; \hat{\omega}_1 ) + \mathcal{J}_k ( \hat{\omega}_{12} ; \hat{\omega}_2 )$ is actually analytic there.
Concretely, the terms which are non-analytic at $\hat{\omega}_1 = - \hat{\omega}_2 -2$,
\begin{align}
\mathcal{J}_k ( \hat{\omega}_{12} ; \hat{\omega}_1 ) + \mathcal{J}_k ( \hat{\omega}_{12} ; \hat{\omega}_2 ) \supset \text{Li}_2 \left( - \frac{\hat{\omega}_1 + 1 }{ \hat{\omega}_2 + 1 }  \right) + \text{Li}_2 \left( - \frac{\hat{\omega}_2 + 1 }{ \hat{\omega}_1 + 1 }  \right)
\end{align}
partially cancel due to the dilogarithm identity,
\begin{align}
    \text{Li}_2 (z) + \text{Li}_2 \left( \frac{1}{z} \right) 
    =
    - \frac{1}{2} \log^2 ( - z) - \frac{\pi^2}{6}  \; \text{for } z > 1 \; , 
    \label{eqn:Li2_identity}
\end{align}

Altogether, substituting \eqref{eqn:Jk_massless} into \eqref{eqn:psi2_integral_massive} and using the identity~\eqref{eqn:Li2_identity}, the final result for $\psil{1}_2$ in this massless limit is,
\begin{multline}
   \omega_{12} \psil{1}_{2}(\omega_1,\omega_2,k)=\frac{1}{8\pi^2}\left[\frac{\omega_2\log\left(\frac{\omega_1+k}{\Lambda}\right)-\omega_1\log\left(\frac{\omega_2+k}{\Lambda}\right)}{\omega_1-\omega_2}\right.\\
    -\left.\frac{\omega_{12} }{2k}\left(\frac{1}{2}\log^2\left(\frac{\omega_1+k}{\omega_2+k}\right)+\frac{\pi^2}{6}+\text{Li}_2\left(\frac{k-\omega_2}{k+\omega_1}\right)+\text{Li}_2\left(\frac{k-\omega_1}{k+\omega_2}\right)\right)\right] \; . \label{psi2site}
\end{multline}

Finally, note that if we attempt the split \eqref{eqn:Jk1_split} in the massless limit, we find that both $\tilde{\mathcal{J}}_k$ and $\tilde{J}_k (B)$ in \eqref{eqn:Jk1_def} develop an additional logarithmic IR divergence. 
These can be regulated by also cutting off the lower limit of the integration, $\int_{1 + \epsilon}^{\hat{\Lambda}} d \hat{\Omega}_+$, which gives, 
\begin{align}
    \lim_{\hat{k} \to 1} \tilde{\mathcal{J}}_k  &= \frac{1}{2} \log \left( \frac{ \hat{\Lambda}^2 }{ 2 \epsilon}  \right) \; , \nonumber \\
    \lim_{\hat{k} \to 1} \mathcal{J}_{k} ( \hat{\omega} ) 
    &= \frac{1}{2} \log \left( \frac{ 2 \epsilon }{ ( \hat{\omega} + 1 )^2 }  \right) + \mathcal{O} ( \hat{\omega} ) \; . 
\end{align}
Note that the logarithmic divergence in $\epsilon$ cancels, rendering the full $\psil{1}_2$ finite as $M \to 0$.

\subsubsection{Soft limit}
\label{sec:2edge_soft}

The soft limit $\hat{k} \to 0$ is another limiting case in which the master integrals \eqref{eqn:Jk_def} become elementary,  
\begin{align}
   \mathcal{J}_0 ( \hat{\omega} )
    &= \int_1^{\infty} d \Omega_+ \frac{1}{ \sqrt{ \hat{\Omega}_+ ( \hat{\Omega}_+ - 1)} ( \Omega_+ + \omega) }   
    = \frac{2 \arcsin \left( \sqrt{-\hat{\omega}} \right) }{ \sqrt{ - \hat{\omega} (1 + \hat{\omega} ) } }   \; .  \nonumber \\
       \mathcal{J}_0 ( \hat{\omega}_{12} , \hat{\omega}_1 ) &= - \int_1^{\infty} \frac{ d \hat{\Omega}_+ }{ \hat{\Omega}_+ } \frac{2 \sqrt{ \hat{\Omega}_+^2 - 1 } }{ ( \hat{\Omega}_+ + \hat{\omega}_{12} ) ( \hat{\Omega}_+ + \hat{\omega}_{1} ) } \,  \nonumber \\ 
    &= \frac{1}{\hat{\omega}_1} \left(  \frac{\pi}{ \hat{\omega}_{12} } + \frac{ 2\sqrt{ 1 - \hat{\omega}_{12}^2}}{ \hat{\omega}_{12} }  \arccos ( \hat{\omega}_{12} )  
     + \frac{ 2\sqrt{ 1 - \hat{\omega}_{1}^2}}{ \hat{\omega}_{1} }  \arccos ( \hat{\omega}_{1} )  \right)  \; ,
     \label{eqn:Jk_soft}
\end{align}
Again we see that the only branch point in $\mathcal{J}_0 ( \hat{\omega} )$ is at the two-particle threshold $\hat{\omega} = -1$, and it is of the square-root form established by our method of regions analysis above.  
The $\mathcal{J}_0 ( \hat{\omega}_{12} ; \hat{\omega}_1)$ integral, on the other hand, has both a branch point at $\hat{\omega}_1 = -1$ and also an additional one at $\hat{\omega}_{12} = -1$ (the apparent poles at $\hat{\omega}_2=0$, $\hat{\omega}_1 = 0$ and $\hat{\omega}_{12} = 0$ are all removable singularities with zero residue). 
The analogous $\mathcal{J}_{0} ( \hat{\omega}_{12} ; \hat{\omega}_2 )$ also has a branch point at $\hat{\omega}_{12} = -1$. 
These additional branch points are due to the $\tilde{\mathcal{J}}_k ( \bar{z} )$ terms in \eqref{eqn:Jk2_derivative_as_Jk1}, since in this limit the roots again become linear in $\omega_1$, 
\begin{align}
    \lim_{k \to 0} \bar{z} ( \hat{\omega}_{12} ) = \{ \omega_{12} , \omega_{12} , 0 ,0 \} \; .
\end{align}
This branch point does not cancel out, but since $\hat{\omega}_{12} = - 1$ corresponds to $\omega_{12} = -2m$ in the soft limit we recognise this as the threshold from the simple one-edge diagram.

\subsubsection{Amplitude limit}
\label{sec:2edge_amplitude}
Note that taking $\omega_{12} \to 0$ should recover the one-loop correction to a Minkowski scattering amplitude.
Since $\mathcal{J}_k ( \hat{\omega}_{12} ; \hat{\omega}_1)$ and $\mathcal{J}_k ( \hat{\omega}_{12} ; \hat{\omega}_2)$ are both finite in that limit (for any value of the mass), these terms do not contribute to \eqref{eqn:psi2_integral_massive}.
This amplitude limit therefore corresponds to setting $(\hat{\omega}_2 + 1)/(\hat{\omega}_2 - 1 ) = (\hat{\omega}_1 - 1)/(\hat{\omega}_1 + 1 )$ in \eqref{eqn:Jk1_elliptic}, and indeed there are various identities which can simplify sums of the form $\tilde{\Pi} ( \phi, \frac{\beta}{\alpha} , \frac{1}{\alpha^2} ) + \tilde{\Pi} ( \phi, \frac{1}{\alpha \beta} , \frac{1}{\alpha^2} )$. 
A simpler route is to return to the original integral,
    \begin{align}
     \frac{\omega_1 \mathcal{J}_k ( \omega_2) - \omega_2 \mathcal{J}_k (\omega_1) }{\omega_1 - \omega_2} 
         = 
         \int_{1}^\infty d \hat{\Omega}_+ \, \delta_k ( \Omega_+ ) \frac{ \hat{\Omega}_+ + \hat{\omega}_{12}}{ ( \hat{\Omega}_+ + \hat{\omega}_1 ) ( \hat{\Omega}_+ + \hat{\omega}_2 ) } \; .
    \end{align}
    When $\omega_{12} \to 0$, we can perform this integral simply by changing variables to $y = 1/\delta_k ( \Omega_+ )$,
    \begin{align}
     \frac{1}{2} \left(  \mathcal{J}_k ( \omega) + \mathcal{J}_k ( - \omega ) \right)
         = \int_1^{\infty} \frac{d y}{y^2} \; \left[ \frac{1}{ y^2 - 1 } + \frac{2s}{ s - y^2 (s-4m^2) }   \right]\;,
    \end{align}
    where $s = \omega_1^2 - k^2$. 
    The first term is responsible for the divergence (note that it is independent of the kinematic variables), while the second term evaluates to,
    \begin{align}
        \lim_{\omega_2 \to -\omega_1} \omega_{12} \psil{1}_2 ( \omega_1, \omega_2 , k )  
         &= \text{div} + \sqrt{ \frac{4 M^2 - s }{s} } \text{arcsinh} \left( \sqrt{ \frac{s}{4M^2} }  \right)  \; ,
    \end{align}
    where $s = \omega_1^2 - k^2$. This is the familiar one-loop result\footnote{
    In fact, by changing variables to $x = s (1- 1/y^2)$, we can even write this $\text{arcsinh}$ in a familiar Feynman-parametrised form,
    \begin{align}
         \int_1^{\infty} \frac{d y}{y^2} \; \frac{2s}{ s - y^2 (s-4m^2) }    = \int_0^1 d x \log \left( \frac{ s x (1- x) +  M^2  }{4M^2}  \right)   \; . 
    \end{align}
    It would be interesting to explore whether an analogous change of variables could be used to introduce a simple Feynman parametrisation of the loop integrals for general values of $\omega_{12}$. 
    } for a scattering amplitude, with a two-particle threshold at $s = 4M^2$.

\subsubsection{In \texorpdfstring{$d=1$}{d=1} dimensions}
\label{sec:2edge_d=1}

Note that in $d = 1$ spatial dimensions, only a single integration variables is needed, and often it is most convenient to simply use $q_1$, writing \eqref{eqn:2edge_int} as
\begin{align}
    \int_{-\infty}^{\infty} \frac{d q_1}{ 2 \pi } \; \mathcal{I} (  \Omega_{q_1}  , \Omega_{q_1-k} )   \; .
\label{eqn:2vertexMeasure_d=1}
\end{align}

For instance, for the exchange of two massless edges,
\begin{align}
        \omega_{12} \psil{1}_2 &= \frac{1}{ 2 \pi } \int_{-\infty}^{\infty} d p \frac{1}{ (\omega_1 + |p| + | p - k | ) (\omega_2 + |p| + | p - k | ) } \left(  \frac{1}{\omega_{12} + 2 |p| }  +  \frac{1}{\omega_{12} + 2 | p - k |}   \right)  \nonumber \\ 
        &= \frac{1}{ \pi ( \omega_1 - \omega_2 )} \left(
        \frac{ \omega_2 \log \left( \frac{ \omega_1 + k}{\omega_1 + \omega_2} \right) }{ \omega_2^2 - k^2 } - \frac{ \omega_1 \log \left( \frac{ \omega_2 + k}{\omega_1 + \omega_2} \right) }{ \omega_1^2 - k^2 } 
 \right) \; . 
\end{align}
Note that $( \mathcal{T} , \mathcal{G} ) = (1,0)$, since despite having a dependence on both $| p|$ and $|p-k|$ there is only a single integral over a rational function\footnote{
Concretely, dividing the integration into regions $\int_{-\infty}^0 + \int_0^k + \int_k^{\infty}$ produces three separate integrals over rational functions. 
}.

\subsection{Three internal edges}
\label{sec:3edge}
For a one-loop diagram with three vertices, the integrand will depend on the loop momenta through three independent combinations, namely the energies associated with the momenta $( \bfq_{12}, \bfq_{23}, \bfq_{31} )$ of the internal lines, where momentum conservation fixes their differences,
\begin{align}
  \bfq_{12} - \bfq_{31} = \bfk_1 \; , \;\; \bfq_{23} - \bfq_{12} = \bfk_2 \; , \;\; \bfq_{31} - \bfq_{23} = \bfk_3 \;  ,
  \label{eqn:3_vertex_mom}
\end{align}
and also sets $\bfk_1 + \bfk_2 + \bfk_3 = 0$ (so only two of the three equalities in \eqref{eqn:3_vertex_mom} are independent). 
Given this constraint, the magnitudes of the internal momenta can be viewed as three edge lengths of a tetrahedron, whose triangular base is fixed by the external momenta $(\bfk_1, \bfk_2 , \bfk_3)$. 
This is shown in Fig.~\ref{fig:tetrahedron}.

\subsubsection{Generalities}

\begin{figure}
    \centering
    \includegraphics{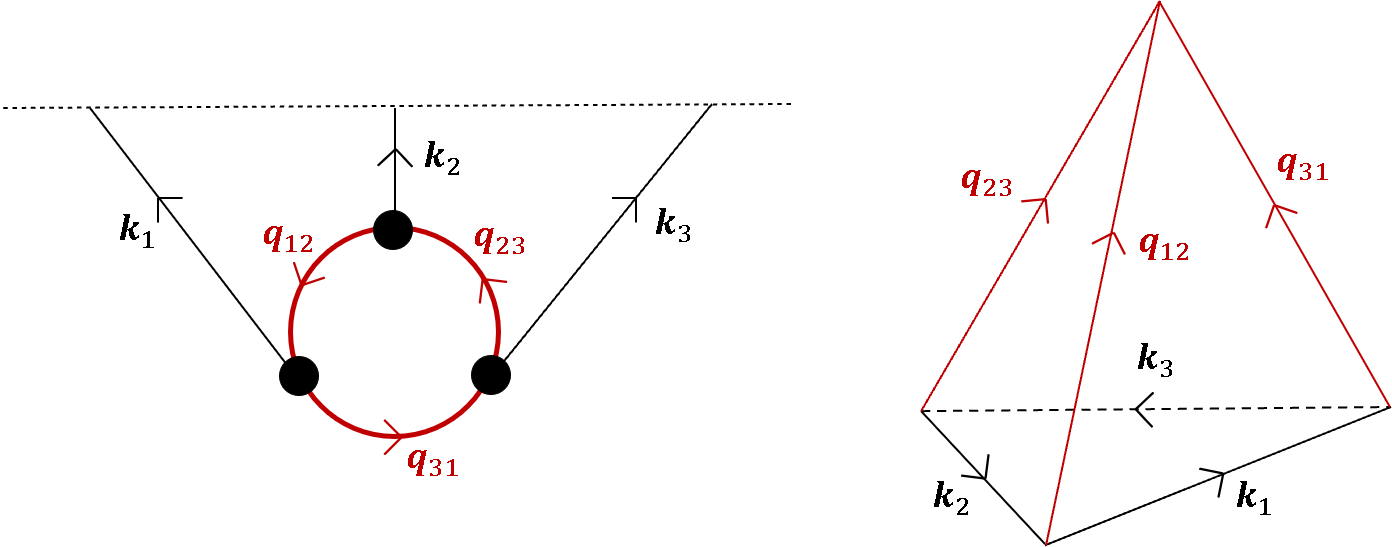}
    \caption{The one-loop contribution to $\psil{1}_3$ with three internal lines is described by six momenta which, due to momentum conservation, form the edges of a tetrahedron.}
    \label{fig:tetrahedron}
\end{figure}

Let us first consider $d \geq 3$ spatial dimensions. 
As in the two vertex case, it would seem prudent to use the edge lengths $( q_{12} , q_{23} , q_{31} )$ as our integration variables. 
The domain of integration is then determined by the condition that there exists a tetrahedron with these edge lengths. 
Clearly each of the four triangular faces of the tetrahedron lead to triangle inequalities between the internal and external momenta, for instance,
\begin{align}
    |  q_{12} - q_{31} |  < k_1 < q_{12} + q_{31} \; . 
    \label{eqn:3_vertex_tri}
\end{align}
However, while \eqref{eqn:3_vertex_tri}  and its permutations are certainly necessary conditions, they are not sufficient. 
There is one further condition which must be satisfied: the volume $V$ of the tetrahedron must be positive, 
    \begin{align}
        288 V^2 = \left| \begin{array}{c c c c c} 
        0 & 1 & 1 & 1 & 1  \\ 
        1 & 0 & k_1^2 & k_2^2 & k_3^2  \\
        1 & k_1^2 & 0 & q_{12}^2 & q_{31}^2  \\
        1 & k_2^2 & q_{12}^2  & 0 & q_{23}^2  \\ 
        1 & k_3^2 & q_{31}^2  & q_{23}^2 & 0
        \end{array} \right| \geq 0 \; , 
        \label{eqn:3_vertex_det}
    \end{align}
where this bound is saturated only for degenerate tetrahedra (for which all four faces lie in the same plane).
Taken together, \eqref{eqn:3_vertex_det} and \eqref{eqn:3_vertex_tri} (together with its permutations) are a necessary and sufficient condition\footnote{
Note that an equivalent condition is that the areas of the four tetrahedral faces obey $A_1 + A_2 + A_3 > A_4$ for all four distinct permutations of $ \{1,2,3,4\}$, which is the direct analogue of the triangle inequalities. 
The face areas are related to their side lengths $\{ a,b,c \}$ by the Cayley-Menger determinant analogous to \eqref{eqn:3_vertex_det}, 
$16 A^2 = (a+b+c) (-a+b+c)(a-b+c)(a+b-c)$. 
} for the lengths $\{ q_{12}, q_{23}, q_{31} \}$ to form a tetrahedron with triangular base $\{ k_1, k_2, k_3 \}$ as shown in~\ref{fig:tetrahedron}, and they therefore define the domain of loop integration. 
Note that \eqref{eqn:3_vertex_det} is generally a fourth order polynomial in any single $q_{ab}$, and so this condition defines a rather involved integration boundary. 
The only situation in which this nested square root simplifies 
are the degenerate cases with either $k_3 = 0$ or when $k_1 = k_2 = 0$. 

Concretely, we can orient one co-ordinate axis along $\bfk_1$ and adopt the same polar parametrisation for $\bfq_{12}$ and $\bfq_{31}$ as in the two-vertex case, namely,
\begin{align}
    \bfq_{12} = \frac{1}{2} \left(  \bfk_1 + \bfp \right) \; , \;\; \bfq_{31} = \frac{1}{2} \left( \bfk_1 - \bfp \right) ,
\end{align}
with $\bfp = ( q_+ \cos \vartheta , \sqrt{q_+^2 - k_1^2} \sin \vartheta \, \hat{\bfp}_{d-1} )$, where now,
\begin{align}
    q_+ = q_{12} + q_{31} \; , \;\; k_1 \cos \vartheta = q_{12} - q_{31} \; ,
\end{align}
and we will often also write $q_- = q_{12} - q_{31}$. 
Momentum conservation then fixes the third internal momentum as,
\begin{align}
 \bfq_{23} = \frac{1}{2} \left(  \bfp +  \bfk_1 + 2 \bfk_2 \right) \; .
\end{align}
If we then adopt the an analogous polar parametrisation of this external momenta, 
\begin{align}
 \bfk_1 + 2 \bfk_2  = \left( \begin{array}{c}  
  k' \cos \chi  \\
  k' \sin \chi  \hat{\bfk}_{d-1}'  \end{array}
 \right) \; ,
\end{align} 
we see that the only angular component of $\hat{\bfp}_{d-1}$ on which our integrand can depend is,
\begin{align}
    \hat{\bfp}_{d-1} \cdot \hat{\bfk}_{d-1}' = \cos \varphi \; . 
\end{align}
So overall, integrating over the undetermined loop momenta corresponds to integration over $\{ q_+, \vartheta, \varphi \}$ and the remaining $d-3$ angular variables in $\hat{\bfp}_{d-1}$ on which the integrand does not depend. 

In practice, this allows us to write the loop integration as,
\begin{align}
    \int \frac{d^d \bfp}{ (2 \pi )^d } \; \mathcal{I} ( \Omega_{q_{31}}, \Omega_{q_{12}}  , \Omega_{q_{23}} ) = \frac{ S_{d-3} }{ (2 \pi )^d } \int_{k_1}^\infty d q_+ \int_{-1}^{+1} \frac{d \cos \vartheta}{2} \int_{0}^{2\pi} d \varphi \; q_{12}^{d-2} q_{31}   \;  \mathcal{I} \left( \Omega_{q_{31}} , \Omega_{q_{12}}  , \Omega_{q_{23}}  \right) \big|_{\substack{q_{12} = \tfrac{1}{2} ( q_+ + k_1 \cos \vartheta ) \\ q_{31} = \tfrac{1}{2} ( q_+ - k_1 \cos \vartheta ) \\ q_{23} = \bar{q}_{23} ( q_+, \vartheta, \varphi ) }  } \; , 
\label{eqn:3vertexMeasure}
\end{align}
where,
\begin{align}
     \bar{q}_{23} ( q_+, \vartheta, \varphi ) &= \frac{1}{2} \sqrt{ q_+^2 + k'^2 - k^2 \sin^2 \vartheta + 2 q_+ k' \cos \chi \cos \vartheta + 2 k' \sqrt{q_+^2-k^2} \sin \chi \sin \vartheta \cos \varphi } \; . 
     \label{eqn:q23}
\end{align}
Since $\sin \vartheta = \sqrt{1- \cos^2 \vartheta}$, \eqref{eqn:q23} represents a nested square root: this generally leads to elliptic functions from integrals of the form \eqref{eqn:3vertexMeasure}. 

In lower dimensions, $d < 3$, we must instead integrate over only degenerate tetrahedra (confined to a plane in $d=2$ or a line in $d=1$), which can be parameterised by using $\{ q_+, \vartheta \}$ or simply $\{ q_{12} \}$. 

\subsubsection{Reducing the complexity}
Consider the correction to $\psil{1}_3$ from a loop with three internal lines, which is given by,
\begin{align}
    \omega_{123} \psil{1}_3 = \int_{\bfp} \frac{1}{ (\omega_1 + \Omega_{q_{12}} + \Omega_{q_{31}} ) ( \omega_2 + \Omega_{q_{12}} + \Omega_{q_{23}} ) ( \omega_3 + \Omega_{q_{23}} + \Omega_{q_{31}} ) } \sum_{\text{perm.}}^6  \frac{1}{ ( \omega_{123} + 2 \Omega_{q_{12}} )(\omega_{23} + \Omega_{q_{12}} + \Omega_{q_{31}}  )} \; .  
\end{align}
Transforming to the $\{ q_+ , \vartheta, \varphi \}$ variables described above, a na\"{i}ve counting of the number of integrals and the number of square roots which cannot be removed from the integrand gives $( \mathcal{T} , \mathcal{G} ) = (3,4)$. 
In more detail, we count $\mathcal{G}$ as follows. 
\begin{itemize}

\item There is one square root for each of the internal lines ($\Omega_{q_a} = \sqrt{q_a^2  + M_a^2}$) assuming the masses are unequal, 

\item Changing variables from $\varphi$ to $\cos \varphi$ introduces a square-root from the $\sqrt{1- \cos^2 \varphi}$ Jacobian,

\item There are three square roots inside $\bar{q}_{23}$ (from $\sin \vartheta = \sqrt{1- \cos ^2 \vartheta}$, $\sqrt{q_+^2 - k^2}$ and the overall square root). 

\item There is the freedom to perform an Euler substitution in each of the integration variables, which can remove three of these square roots. 

\end{itemize}
It seems that very little is known about special functions with $(\mathcal{T} , \mathcal{G} ) = (3,4)$. 
So rather than analyse this integral any further, we will look for limits which reduce this complexity.

\paragraph{Massless limit.}
In the massless limit,
\begin{align}
    \omega_{123} \psil{1}_3 = \int_{\bfp} \frac{1}{ (\omega_1 + q_{12} + q_{31} ) ( \omega_2 + q_{12} + q_{23} ) ( \omega_3 + q_{23} + q_{31} ) } \sum_{\text{perm.}}^6  \frac{1}{ ( \omega_{123} + 2 q_{12} )(\omega_{23} + q_{12} + q_{31}  )} \; .  
\end{align}
Focusing on just one of these permutations, partial fractions can be used to linearise the denominator in one of the internal momenta, say $q_{23}$.
Schematically, this produces integrals of the form,  
\begin{align}
    &\int_k^{\infty} d q_+ \int_{-1}^{+1} d \cos \vartheta   \int_{0}^{2 \pi} d \phi  \frac{ R ( q_+, \cos \vartheta ) }{ \bar{q}_{23} ( q_+, \vartheta , \varphi )  + q_+ + k_1 \cos \vartheta + \omega_2  } \nonumber \\ 
    &= \int_k^{\infty} d q_+ \int_{-1}^{+1} d \cos \vartheta \;  R (q_+ , \cos \vartheta ) \; \Pi \left( q_+ + k_1 \cos \vartheta , \bar{p}_{23} ( q_+ , \vartheta , 0 ) , \bar{p}_{23} ( q_+ , \vartheta , \pi ) \right) ,
\end{align}
where $R(x,y)$ represents a rational function of $\{ x, y \}$ and $\Pi ( x , y , z )$ represents a combination of elliptic integrals whose arguments are rational functions of $\{ x, y, z\}$ (note that a rational dependence on the external $\{ \omega_a \}$ is implicit in both of these functions).
Performing the remaining two integrals will produce special functions with $( \mathcal{T} , \mathcal{G} ) = (3,1)$.

\paragraph{Soft limit.}
Note that if we further take one of the external edges to be soft, e.g. $k_3 \to 0$, then integration variables can be chosen so that the integrand no longer depends on $\varphi$. 
This leads to an integral with $( \mathcal{T} , \mathcal{G} ) = (2,0)$, which is the same complexity as the massless $\psil{1}_2$ integral computed in \eqref{psi2site} above.
We will focus on this limit, in which all three external lines are massless and one external line carries zero spatial momentum, for the remainder of the subsection, since in this case we are able to give closed form expressions for $\psil{1}_3$ using (at most) dilogarithms.

\subsubsection{Soft limit}

Taking the soft limit $k_3 \to 0$ removes the need to integrate over $\varphi$, since the tetrahedron of momenta degenerates into a triangle and we can proceed as in the two-edge case of Sec.~\ref{sec:2edge}. 
Explicitly, there is now only one independent external momenta which we write as $\bfk_1 = - \bfk_{2} = \bfk$, and the three internal momenta are now constrained by,
\begin{align}
    \bfq_{23} = \bfq_{31} \;, \;\; \bfq_{12} - \bfq_{31} = \bfk \; . 
\end{align}
This limit therefore lowers both the transcendentality and the genus of the loop integral, which can be written in terms of $q_{\pm} = q_{12} \pm q_{31}$, 
\begin{align}
\omega_{123} \psil{1}_3 = \int_{\bfp} &\frac{1}{ (\omega_1 + q_+ ) ( \omega_2 + q_+ ) ( \omega_3 + q_+ - q_- ) } \Bigg\{   \frac{1}{  \omega_{123} + q_+ + q_- } \left( \frac{1 }{  \omega_{13} + q_+ } + \frac{1}{ \omega_{23} + q_{+} }  \right) \hfill \nonumber \\
&\qquad\qquad\qquad\qquad\qquad+
 \frac{1}{  \omega_{123} + q_+ - q_- } \left( \frac{1 }{  \omega_{13} + q_+ } + \frac{1}{ \omega_{23} + q_{+} } + \frac{2}{ \omega_{12} + q_+ - q_-  }  \right)  \Bigg\} \; .
 \label{eqn:psi3_soft_int}
\end{align}

To express \eqref{eqn:psi3_soft_int} compactly, it will be useful to generalise \eqref{eqn:IB1B2} to an arbitrary number of factors in the denominator, defining
\begin{align}
\mathcal{I} ( A ; B_1, ... , B_n )  &=  \int_1^{\infty} d \hat{q}_+ \int_{-1}^{+1} d \hat{q}_- \, \frac{ \hat{q}_+^2 - \hat{q}_-^2 }{ ( \hat{q}_+ + \hat{q}_- + A ) \prod_{j=1}^n ( \hat{q}_+ + B_j ) }  \; .  \\
\end{align}
Using partial fractions, integrals of this kind can be written in terms of the two master integrals \eqref{eqn:Jk_def} of Sec.~\ref{sec:2edge},
\begin{align}
\mathcal{I} ( A ; B_1, ... , B_n ) &= \sum_{j=1}^n \frac{A (A - 2 B_j )}{ \prod_{i \neq j} (B_j - B_i ) } \mathcal{J}_1 ( A ; B_j )  +  \sum_{j=1}^n \frac{ B_j - A  }{ \prod_{i \neq j} ( B_j - B_i ) }  \mathcal{J}_1 ( B_j ) \; . 
\label{eqn:IAB1...Bn}
\end{align}
where in the massless limit the $\mathcal{J}_1$ are given in \eqref{eqn:Jk_massless} in terms of logs and dilogs.
Note that the apparent poles when two of the $B_j$ coincide are spurious---the only branch points are at each $B_j = -1$ and at $A=0$ or $A=-2$. 

In terms of these \eqref{eqn:IAB1...Bn} integrals, the three-point coefficient \eqref{eqn:psi3_soft_int} is given by, 
\begin{align}
\omega_{123} \psil{1}_3 
&=
\frac{1}{2} \mathcal{I} \left( \omega_{3} ; \omega_1, \omega_2 , \omega_{13} , \omega_3 + \frac{\omega_{12}}{2}   \right)
+ \frac{1}{2} \mathcal{I} \left( \omega_{123} ; \omega_1, \omega_2 , \omega_{13} , \omega_3 + \frac{\omega_{12}}{2}   \right) 
 \nonumber \\
 &+
 \frac{1}{2} \mathcal{I} \left( \omega_{3} ; \omega_1, \omega_2 , \omega_{23} , \omega_3 + \frac{\omega_{12}}{2}   \right)
 +\frac{1}{2} \mathcal{I} \left( \omega_{123} ; \omega_1, \omega_2 , \omega_{23} , \omega_3 + \frac{\omega_{12}}{2}   \right)  \nonumber \\
&+ \frac{1}{\omega_{12}} \mathcal{I} \left( \omega_3  ; \omega_1, \omega_2 , \omega_{13}  \right)  -  \frac{1}{\omega_{12}} \mathcal{I} \left( \omega_{123}  ; \omega_1, \omega_2 , \omega_{13}  \right)  \label{eqn:psi3_soft} \\
&+ \frac{1}{\omega_{12}} \mathcal{I} \left( \omega_3  ; \omega_1, \omega_2 , \omega_{23}  \right)  -  \frac{1}{\omega_{12}} \mathcal{I} \left( \omega_{123}  ; \omega_1, \omega_2 , \omega_{23}  \right)   \nonumber \\
&+ 
\frac{2 \, \mathcal{I} \left( \omega_{3} ; \omega_1 , \omega_2  \right)}{\omega_{12} (\omega_{12} - \omega_3) } 
+ \frac{2\,  \mathcal{I} \left( \omega_{123} ; \omega_1 , \omega_2  \right) }{\omega_3 \omega_{12}} 
- \frac{2 \, \mathcal{I} \left( \omega_{12} ; \omega_1 , \omega_2  \right) }{\omega_3 ( \omega_{12} - \omega_{3} )} \; . \nonumber 
\end{align}
The apparent poles at $\omega_{12} = 0$, $\omega_1 = 0$ and $\omega_2 = 0$ are all spurious. 
Only some of the branch points have non-zero residues.
In the complex $\omega_1$ plane (at fixed $\omega_2, \omega_3$), there are four physical singularities,
\begin{itemize}

\item[(i)] $\hat{\omega}_1 = - 1$,

\item[(ii)] $\omega_{12} = 0$, 

\item[(iii)] $\hat{\omega}_{13} = -1$, 

\item[(iv)] $\omega_{123} = 0$,

\end{itemize}
which agree with the energy-conservation condition of Sec.~\ref{ssec:analytic_heuristic}. 
There are three spurious branch points, 
\begin{itemize}

\item[{\color{red} (v)}] {\color{red} $\hat{\omega}_{12} = -2$}. This singularity is spurious since the only term with a possible branch point there is the $\mathcal{I} ( \omega_{12} ; \omega_1, \omega_2 )$, but this is precisely the $\psil{1}_2 (\omega_1, \omega_2, k)$ integral analysed above.

\item[{\color{red} (vi)}] {\color{red} $\hat{\omega}_{123} = -2$}.
This branch point occurs in every term separately, but cancels out of the total sum. 

\item[{\color{red} (vii)}] {\color{red} $\hat{\omega}_3 + \frac{\hat{\omega}_{12}}{2} = -1$}.
It is only the first line of \eqref{eqn:psi3_soft_int} that could contain such a branch point, and when we compare the prefactors in \eqref{eqn:IAB1...Bn} (in particular the $1/(B_j - B_i)$ product), we find that they are proportional to,
\begin{align}
\frac{1}{ ( \omega_3 + \frac{\omega_{12}}{2} ) - \omega_{13} } + \frac{1}{ ( \omega_3 + \frac{\omega_{12}}{2} ) - \omega_{23} } = 0 \; , 
\end{align}
and consequently the branch point at $\omega_3 + \omega_{12}/2$ has zero residue and is spurious.

\end{itemize}

The complex $\omega_2$ plane is analogous, since \eqref{eqn:psi3_soft} is symmetric in $\omega_1 \leftrightarrow \omega_2$. 
In the complex $\omega_3$ plane (at fixed $\omega_1, \omega_2$), there are five physical singularities,
\begin{itemize}

\item[(i)] $\omega_3 = 0$,

\item[(ii)] $\hat{\omega}_{3} = -2$, 

\item[(iii)] $\hat{\omega}_{13} = -1$,

\item[(iv)] $\hat{\omega}_{23} = -1$,

\item[(v)] $\omega_{123} = 0$.

\end{itemize}
which again agree with the energy-conservation condition of Sec.~\ref{ssec:analytic_heuristic}. 
There are two spurious branch points, at $\hat{\omega}_{123} = -2$ and $\hat{\omega}_3 + \frac{\hat{\omega}_{12}}{2} = -1$, which have vanishing residue for the reasons described above.

 
\section{Feynman rules for the on- and off-shell wavefunction}\label{A:rules}

In this appendix, for completeness we briefly review the Feynman rules to compute the Minkowski on- and off-shell wavefunction at time $t=0$. We never actually use these rules in the paper because the handy recursion relations of \cite{Benincasa:2019vqr} already give us the desired results. 
\begin{itemize}
\item For a desired $  n $-point off-shell wavefunction coefficient $ \psi_{n}(\omega_{a})$ draw a connected diagram with $  V $ vertices, $  I $ internal lines, and $  n $ external lines going from a vertex to the spatial hypersurface at time $t=0$. It is conventional to represent time as running vertically from the infinite past at the bottom to $t=0$ at the top.
\item The $  n $ external lines have ingoing spatial momenta $ \bfk_{a}  $ with $  a=1,\dots,n $, which is then conserved at every vertex. 
\item  External lines are associated to bulk-boundary propagators $K_{\omega_{a}}$ satisfying  the boundary conditions
\begin{align}
\lim_{\eta\rightarrow\eta_0} K_{\omega_{a}}  ( \eta, \eta_0 )&=1 \,, & \lim_{\eta\to -\infty(1-i\epsilon)}K_{\omega_{a}}   ( \eta , \eta_0 )&=0\,,
\label{eqn:propagator_def}
\end{align}
In the on-shell case (where $\omega_{a}=\Omega_{k_{a}}$) the bulk-to-boundary propagators also obeys the equations of motion of the free theory,
\begin{align}
(\partial_{t}^{2}+ k^2+m^2) K_{\Omega_k}(t)=0  \;\, .
\end{align}
The appropriate solution for a mode with momentum $\bfk$ and mass $m$ is then
\begin{align}
\text{On-shell propagator:}\;\;K_{\Omega_k} ( t )  &= e^{i\Omega_{k} t}\,,
\end{align}
with $\Omega_{k}=\sqrt{k^2+m^2}$. To obtain the off-shell wavefunction coefficient we analytically extend the bulk-to-boundary propagator to be a function of the frequency $\omega_{a}$:
\begin{equation}
 \text{Off-shell propagator:}\;\;K_{\omega_{a}} ( t )  = e^{i\omega_{a} t}\,.
\end{equation}
\item The $  I $ internal lines carry spatial momenta $ \bfp_{i}  $ with $  i=1,\dots,I $, which is fixed by momentum conservation at each vertex (up to running loop momenta). There can be a number $ L $ of loop momenta that should be integrated over. 
\item Internal lines are associated with bulk-bulk propagators $G_{p_{i}}$ satisfying the equations of motion with (minus) a delta source and the boundary conditions:
\begin{align}
(\partial_{t}^{2}+\textbf{p}_{i}^2+m^2)G_{p_{i}}(t_{1},t_{2})=-\delta(t_{1}-t_{2}),\;\;\lim_{\eta \rightarrow\eta_0} G_{{p}_{i}}  (\eta,\eta')&=0,&  \lim_{\eta \to -\infty(1-i\epsilon)} G_{{p}_{i}}   ( \eta,\eta')&=0\,.\label{Gbound}
\end{align}
The resulting expression for a propagator connecting vertices at times $t_1$ and $t_2$ is
\begin{align} \label{G}
G_{p}(t_1, t_2 ) &= -i \frac{e^{i \Omega_{p} t_2 } }{ \Omega_p } \sin ( \Omega_{p} t_1 ) \theta ( t_1 - t_2 ) + \left( t_1 \leftrightarrow t_2 \right)\;. 
\end{align}
Notice that for both the on- and off-shell wavefunctions the bulk-bulk propagator $G$ is always on-shell.
\item For each vertex at time $t_A$, with $A=1,\dots,V$, one should integrate in $dt_A$ over the range
\begin{align}
 -\infty(1-i\e)<t_A \leq 0 \,.
\end{align}
For polynomial interactions such as $  \mathcal{H}_{\rm int} = \lambda \Phi^{n}/(n!) $, one should add a factor $i \lambda $.
\end{itemize}

\section{A non-perturbative definition of the off-shell wavefunction}\label{App:C}

In this appendix we provide a pedagogical discussion and derivation of the the non-perturbative definition of wavefunction coefficients given in \eqref{eqn:off_shell_psi_def}. The argument essentially proceeds by analogy with how the off-shell scattering amplitude is defined using the time-ordered Green's function, a connection which we will now briefly review.

\paragraph{Off-shell amplitudes.}
The starting point for defining an off-shell extension of the scattering amplitude is the (time-ordered) Green's function,
\begin{align}
 \langle \Omega_{\rm out} | T \hat{\Phi}_{\bfk_1} ( t_1)  ... \hat{\Phi}_{\bfk_n} (t_n)   | \Omega_{\rm in} \rangle,
 \label{eqn:TOC}
\end{align}
where $| \Omega_{\rm in} \rangle$ and $| \Omega_{\rm out} \rangle$ are the states in the interacting theory which coincide with the vacuum in the asymptotic past/future. 
This object can be computed in perturbation theory in the usual way, for instance via a diagrammatic expansion of the corresponding path integral, where each edge corresponds to the Wick contraction,
\begin{align}
 \langle \Omega_{\rm out} |  \; T \; \hat{\Phi}_{\bfk} (t) \hat{\Phi}_{\bfk'} (t') \;  | \Omega_{\rm in} \rangle
 =
 G_{\bfk}^F ( t, t') \, \delta_D^{(3)} ( \bfk - \bfk' ).
\end{align}
where $G^F_{\bfk} (t, t')$ is the Feynman propagator. 

This correlator \eqref{eqn:TOC} is related to an $S$-matrix element by the LSZ reduction procedure, pedagogical accounts of which can be found in most introductory QFT textbooks. 
There are two steps to this procedure,
\begin{itemize}

\item[(i)] \emph{Amputate the external legs}. This amounts to replacing each $\hat{\Phi}_{\bfk} (t)$ with its classical equation of motion, $\mathcal{E} \hat{\Phi}_{\bfk} (t)$, since this replaces the external propagators with $\delta$ functions,
\begin{align}
 \mathcal{E} G_{\bfk} ( t , t' )=  i \left( \partial_{t}^2 + k^2 + m^2 \right)G_{\bfk} ( t , t' ) = \delta_D^{(1)} ( t - t' ) \; .
 \label{eqn:EG_def}
\end{align}

\item[(ii)] \emph{Put the external momenta on-shell}. 
This amounts to transforming to the frequency domain and then setting $\omega_a = \pm \sqrt{ k_a^2 + m_a^2}$ for each ingoing/outgoing particle. 

\end{itemize}

\noindent Concretely, if we write the amputated Green's function in the frequency domain as,
\begin{align}
 \left[ \prod_{j=1}^n \int_{-\infty}^{+\infty} d t_j \, e^{ i \omega_j t_j}  \mathcal{E}_j\right]\;  \langle \Omega_{\rm out} |  \; T \;  \hat{\Phi}_{\bfk} ( t_1 ) ...  \hat{\Phi}_{\bfk_n} ( t_n ) \;  | \Omega_{\rm in} \rangle 
 =
 \mathcal{S}_n \left( \{ \omega \} , \{ \bfk \} \right) \; \delta_D^{(3)} \left( \bfk_1 + ... + \bfk_n \right) 
 \; , 
\label{eqn:off_shell_S_def}
\end{align}
then LSZ reduction corresponds to the statement that $\mathcal{S}_n$ becomes an $n$-particle scattering amplitude when each $\omega$ is put on-shell. 

To illustrate this, consider the simple interaction, $\mathcal{L}_{\rm int} = \frac{\lambda}{4!} \Phi^4$. 
The corresponding four-point Green's function is,
\begin{align}
 \langle \Omega_{\rm out} | T \hat{\Phi}_{\bfk_1} ( t_1)  ... \hat{\Phi}_{\bfk_4} (t_4)   | \Omega_{\rm in} \rangle
 &= \left[ \int_{-\infty}^{+\infty} d t \,  \prod_{j=1}^4 G_{\bfk_j}^F ( t - t_j )  \right]  \; \delta_D^{(d)} \left( \bfk_1 + \bfk_2 + \bfk_3 + \bfk_4 \right)  \; . 
\end{align} 
The amputated Green's function in the time domain is therefore,
\begin{align}
  \left[ \prod_{j=1}^{4}\mathcal{E}_j  \right]\langle \Omega_{\rm out} | T  \hat{\Phi}_{\bfk_1} ( t_1)  ...  \hat{\Phi}_{\bfk_4} (t_4)   | \Omega_{\rm in} \rangle
 &= \lambda \left[ \int_{-\infty}^{+\infty} d t \,  \prod_{j=1}^4 \delta ( t - t_j )  \right]  \; \delta_D^{(d)} \left( \bfk_1 + \bfk_2 + \bfk_3 + \bfk_4 \right)  \; ,
\end{align}
and so transforming to frequency space gives,
\begin{align}
 \mathcal{S}_4 \left( \{ \omega \} , \{ \bfk \} \right) 
 &= 
\lambda \left[  \int_{-\infty}^{+\infty} d t \, \prod_{j=1}^4 e^{i \omega_j t}  \right]  
= \lambda \, \delta_D^{(1)} \left( \omega_1 + \omega_2 + \omega_3 + \omega_4 \right)  \; .
\end{align}
The physical (on-shell) $S$-matrix is then obtained by fixing each of the energies as $\omega_a = \pm \sqrt{ k_a^2 + m_a^2}$. 

In general, the off-shell $\mathcal{S}_n$ can be computed in perturbation theory via a diagrammatic expansion in which internal edges correspond to $G_{\bfk}^F$ propagators and external edges correspond to factors of $e^{i \omega_a t}$ (which are the amputated propagators written in frequency space). 
Schematically, each diagram then contributes to $\mathcal{S}_n$ via an integral of the form, 
\begin{align}
 \mathcal{S}_n (  \{ \omega \} , \{ \bfk \}  ) =  \left[ 
 \prod_{v=1}^V \int_{-\infty}^{+\infty} d t_v
 e^{i \omega_v t_v} \right]  \left[ \prod_{ \ell =1}^L \int \frac{d^{3} \bfp_{\ell} }{ (2\pi )^d} \right] \prod_{i = 1}^I G_i^F ( \{ \bfk \} , \{ \bfp \} ) ,
 \label{eqn:S_schematic}
\end{align}
where $\omega_a$ is the total energy flowing into each of the $V$ vertices, and $G_i^F$ are the Feynman propagators for each of the $I$ internal lines, which (once 3-momentum conservation is imposed at each vertex) depend only on the external momenta $\bfk$ and $L$ loop momenta $\bfp$.

\paragraph{Off-shell wavefunction coefficients.}
In Sec.~\ref{offshell}, we defined the off-shell wavefunction coefficients perturbatively via a very similar set of Feynman rules: they are given by a diagrammatic expansion in which each internal line corresponds to a factor of the bulk-to-bulk propagator $G_{\bfk}$ and each external line corresponds to a factor of $e^{i \omega_a t}$. 
Schematically, each diagram then contributes to $\psi_n$ an integral of the form, 
\begin{align}
 \psi (  \{ \omega \} , \{ \bfk \}  ) =  \left[ 
 \prod_{v=1}^V \int_{-\infty}^0 d t_v
 e^{i \omega_v t_v} \right]  \left[ \prod_{ \ell =1}^L \int \frac{d^{3} \bfp_{\ell} }{ (2\pi )^d} \right] \prod_{i = 1}^I G_i ( \{ \bfk \} , \{ \bfp \} )  \; , 
 \label{eqn:off_shell_psi_schematic}
\end{align}
where $\omega_a$ is the total energy flowing into each of the $V$ vertices, and $G_i$ are the bulk-to-bulk propagators for each of the $I$ internal lines, which (once 3-momentum conservation is imposed at each vertex) depend only on the external momenta $\bfk$ and $L$ loop momenta $\bfp$. 
Comparing with \eqref{eqn:S_schematic}, we see that this perturbative definition of $\psi_n$ is almost identical to the perturbative definition of an off-shell scattering amplitude. The only differences are,
\begin{itemize}

\item the domain of the time integration is $-\infty < t < 0$ (rather than $-\infty < t < +\infty$), 

\item the use of bulk-to-bulk propagators (rather than Feynman propagators). 

\end{itemize}

\noindent To implement these differences, we consider the modified Green's function, 
\begin{align}
\langle \phi (t=0)  = 0 | \; T \; \Phi_{\bfk_1} ( t_1 )  ... \hat{\Phi}_{\bfk_n} (t_n)  \;  | \Omega_{\rm in} \rangle \; , 
\label{eqn:TOC_2}
\end{align}
where each $t_a$ argument is restricted to lie in the range $-\infty < t < 0$. 
This differs from \eqref{eqn:TOC} in that $| \Omega_{\rm out} \rangle$ at $t \to +\infty$ has been replaced by the zero-field eigenstate at $t \to 0$. 
This effectively changes the Feynman rules, so that in the perturbative expansion of \eqref{eqn:TOC_2} each edge corresponds to a factor of,
\begin{align}
\langle \phi (t=0)  = 0 |  T \Phi_{\bfk} ( t )  \hat{\Phi}_{\bfk'} (t')   | \Omega_{\rm in} \rangle = G_{\bfk} (t, t') \; \delta_D^{(3)} \left( \bfk + \bfk' \right) \; , 
\end{align}
where $G_{\bfk} (t, t')$ is the bulk-to-bulk propagator introduced in the main text (and which differs from $G_k^F$ by a particular boundary term due to the different bra boundary condition). 
It also changes the range of $t$ over which interaction vertices should be inserted, since now the corresponding path integral is defined using the finite-time action $S_{t=0} [\phi ]$ described in Sec.~\ref{ssec:analytic_review} above.

These difference aside, we can then proceed by analogy with the usual LSZ procedure and define the amputated Green's function in the frequency domain using \eqref{eqn:off_shell_psi_def}. 
To illustrate this, consider again the simple interaction, $\mathcal{L}_{\rm int} = \frac{\lambda}{4!} \Phi^4$. 
The corresponding four-point Green's function is,
\begin{align}
 \langle \phi (0) = 0 | \; T \; \hat{\Phi}_{\bfk_1} ( t_1)  ... \hat{\Phi}_{\bfk_4} (t_4)  \;  | \Omega_{\rm in} \rangle
 &= \left[ \int_{-\infty}^{0} d t \,  \prod_{j=1}^4 G_{\bfk_j} ( t - t_j )  \right]  \; \delta_D^{(d)} \left( \bfk_1 + \bfk_2 + \bfk_3 + \bfk_4 \right)  \; . 
\end{align} 
and the amputated Green's function is therefore,
\begin{align}
  \left[ \prod_{j=1}^{4}\mathcal{E}_j  \right]\langle \phi (0) = 0 | T  \hat{\Phi}_{\bfk_1} ( t_1)  ... \hat{\Phi}_{\bfk_4} (t_4)   | \Omega_{\rm in} \rangle
 &= \lambda \left[ \int_{-\infty}^{0} d t \,  \prod_{j=1}^4 \delta ( t - t_j )  \right]  \; \delta_D^{(d)} \left( \bfk_1 + \bfk_2 + \bfk_3 + \bfk_4 \right)  \; ,
\end{align}
since $G_{\bfk} (t, t')$ also satisfies \eqref{eqn:EG_def}.
Transforming to frequency space then gives,
\begin{align}
 \psi_4 \left( \{ \omega \} , \{ \bfk \} \right) 
 &= 
\lambda \left[  \int_{-\infty}^{0} d t \, \prod_{j=1}^4 e^{i \omega_j t}  \right]  
= \frac{ \lambda }{ i ( \omega_1 + \omega_2 + \omega_3 + \omega_4 ) }  \; ,
\end{align}
which indeed agrees with our perturbative tree-level result in the main text. 
For general interactions, it is not difficult to show that each diagram produces precisely the integral \eqref{eqn:off_shell_psi_schematic} considered in the main text, so in that sense \eqref{eqn:off_shell_psi_def} is the non-perturbative completion of this diagrammatic series.

\paragraph{Going on-shell.}
One benefit of the non-perturbative definition~\ref{eqn:off_shell_psi_def} is that it shows how our $\psi_n$ reduce to the usual wavefunction coefficients in the on-shell limit, $\omega_a \to \Omega_{k_a}$.
 This follows almost immediately from the following observation:
\begin{align}
  e^{i \Omega_k t} \,     \mathcal{E}    \hat{\Phi}_{\bfk} ( t )     
 =  \partial_t \left[   e^{i \Omega_k t} \overset{\leftrightarrow}{  \partial_{t}  } \hat{\Phi}_{\bfk} ( t )    \right] \; . 
\end{align}
This allows the time integrals to be performed exactly, each giving boundary contributions at $t = 0$ and $t= - \infty$. 
These boundary terms can then be written as,
\begin{align}
\langle \phi (0) = 0 |  \, \lim_{t \to 0} e^{i \Omega_k t} \overset{\leftrightarrow}{  \partial_{t}  } \hat{\Phi}_{\bfk} ( t ) 
&=  \langle \phi (0) = 0 | \, \hat{\Pi}_{\bfk} ( 0 ) \; ,  \nonumber \\
 \lim_{t \to - \infty} e^{i \Omega_k t} \overset{\leftrightarrow}{  \partial_{t}  } \hat{\Phi}_{\bfk} ( t )  | \Omega_{\rm in} \rangle 
&\propto  \hat{a}_{\bfk}  | \Omega_{\rm in} \rangle = 0  \; ,
\end{align}
where $\hat{\Pi}_{\bfk} (t) = \partial_t \hat{\Phi}_{\bfk} (t)$ is the momentum conjugate to $\Phi (t)$ in the free theory, and we have used the usual decomposition $\hat{\Phi}_{\bfk} (t) \propto e^{-i \Omega_k t } \hat{a}_{\bfk} + e^{+ i \Omega_k t} \hat{a}_{\bfk}^{\dagger}$ of the field into annihilation and creation operators in the asymptotic past.  
Performing this rewriting for each of the fields in \eqref{eqn:off_shell_psi_def} gives\footnote{
In writing \eqref{eqn:psi_on_shell} we have assumed that no two $\bfk_a$ coincide, so that $[ \hat{\Phi}_{\bfk}, \hat{\Pi}_{\bfk'} ] = 0$ for every pair of operators. When two $\bfk_a$ coincide, there can be additional disconnected contributions to \eqref{eqn:off_shell_psi_def}. In that case, one could have instead used the connected Green's function in \eqref{eqn:off_shell_psi_def} and arrived at \eqref{eqn:off_shell_psi_def} for the connected part of the wavefunction coefficient.
},
\begin{align}
\psi_n \left( \{ \Omega_{k} \}  , \{ \bfk \} \right)
 &= \langle \phi (0) = 0 | \; i \hat{\Pi}_{\bfk_1} ( 0 ) \; ... \; i \hat{\Pi}_{\bfk_n} ( 0 ) \; | \Omega_{\rm in} \rangle \;  \nonumber \\ 
 &= \frac{\delta}{\delta \phi_{\bfk_1} } ... \frac{\delta}{\delta \phi_{\bfk_n} }  \; \ln \Psi_0 [ \phi ] \bigg|_{\phi = 0},
 \label{eqn:psi_on_shell}
\end{align}
where $\Psi_0 [\phi ] \propto \langle \phi (0) | \Omega_{\rm in} \rangle$ is the usual equal-time wavefunction at $t=0$. 


~\\
In summary, we have shown that the matrix element \eqref{eqn:off_shell_psi_def} indeed coincides with our perturbative definition of the off-shell wavefunction coefficients, and moreover in the on-shell limit it matches the usual definition of the wavefunction coefficients. 
Although our analysis in the main text has focused on the analyticity of $\psi_n$ in perturbation theory, the non-perturbative representation \eqref{eqn:off_shell_psi_def} will be a useful starting point for future investigations of the analytic structure of the wavefunction beyond perturbation theory.

\bibliographystyle{JHEP}
\bibliography{refs}
\end{document}